\documentclass[
reprint,
superscriptaddress,
 amsmath,amssymb,
 pre
]{revtex4-1}

\usepackage{probability}
\usepackage{graphicx}
\usepackage{dcolumn}
\usepackage{bm}
\usepackage{bbm}
\usepackage[colorlinks=true,allcolors=blue]{hyperref}
\usepackage[mathscr]{eucal}
\usepackage{tikz}
\usetikzlibrary{shapes,arrows,backgrounds,patterns,decorations.markings}
\usepackage{longtable}
\usepackage{booktabs}
\usepackage{hhline}
\usepackage{mathtools}
\usepackage{balance}

\newcommand{\N}{\mathbb{N}}

\newcommand*{\swap}[2]{\hspace{-0.5ex}#2#1}

\DeclareMathOperator*{\argmin}{arg\,min}

\LTcapwidth=\textwidth

\definecolor{red}{RGB}{227,26,28}
\definecolor{orange}{RGB}{255,127,0}
\definecolor{lightblue}{RGB}{31,120,180}
\definecolor{green}{RGB}{51,160,44}
\definecolor{tablered}{RGB}{213,94,0}
\definecolor{tableorange}{RGB}{240,228,66}
\definecolor{tableblue}{RGB}{86,180,233}
\definecolor{tablegreen}{RGB}{0,158,115}

\newcommand{\xiThreshold}           {1/4}
\newcommand{\gammaThreshold}        {5}
\newcommand{\undirectedPL}          {$49\%$ }
\newcommand{\undirectedDSM}          {$29\%$}
\newcommand{\directedBothDSM}        {$6\%$  }
\newcommand{\directedBothPL}        {$24\%$ }
\newcommand{\directedAtLeastOnePL}  {$82\%$ }
\newcommand{\directedAtLeastOneDSM}  {$45\%$ }
\newcommand{\bipartiteBothDSM}       {$13\%$ }
\newcommand{\bipartiteBothPL}       {$35\%$ } 
\newcommand{\bipartiteAtLeastOnePL} {$74\%$ }
\newcommand{\bipartiteAtLeastOneDSM} {$55\%$ }

\usepackage{xcolor}

\begin{document}

\title{Scale-Free Networks Well Done}
\author{Ivan Voitalov}
\affiliation{Department of Physics, Northeastern University, Boston, Massachusetts 02115, USA}
\affiliation{Network Science Institute, Northeastern University, Boston, Massachusetts 02115, USA}
\author{Pim van der Hoorn}
\affiliation{Department of Physics, Northeastern University, Boston, Massachusetts 02115, USA}
\affiliation{Network Science Institute, Northeastern University, Boston, Massachusetts 02115, USA}
\author{Remco van der Hofstad}
\affiliation{Department of Mathematics and Computer Science, Eindhoven University of Technology, Postbus 513, 5600 MB Eindhoven, Netherlands}
\author{Dmitri Krioukov}
\affiliation{Department of Physics, Northeastern University, Boston, Massachusetts 02115, USA}
\affiliation{Network Science Institute, Northeastern University, Boston, Massachusetts 02115, USA}
\affiliation{Department of Mathematics, Northeastern University, Boston, Massachusetts 02115, USA}
\affiliation{Department of Electrical \& Computer Engineering, Northeastern University, Boston, Massachusetts 02115, USA}

\begin{abstract}
We bring rigor to the vibrant activity of detecting power laws in empirical degree distributions in real-world networks. We first provide a rigorous definition of power-law distributions, equivalent to the definition of regularly varying distributions that are widely used in statistics and other fields. This definition allows the distribution to deviate from a pure power law arbitrarily but without affecting the power-law tail exponent. We then identify three estimators of these exponents that are proven to be statistically consistent---that is, converging to the true value of the exponent for any regularly varying distribution---and that satisfy some additional niceness requirements. In contrast to estimators that are currently popular in network science, the estimators considered here are based on fundamental results in extreme value theory, and so are the proofs of their consistency. Finally, we apply these estimators to a representative collection of synthetic and real-world data. According to their estimates, real-world scale-free networks are definitely not as rare as one would conclude based on the popular but unrealistic assumption that real-world data comes from power laws of pristine purity, void of noise and deviations.
\end{abstract}

\pacs{}

\maketitle

\section{Introduction}

\emph{Scale-free} and \emph{power-law} are sacral words in network science, a mature field that studies complex systems in nature and society by representing these systems as networks of interacting elements~\cite{barabasi2016network,newman2018networks,barrat2008dynamical,bornholdt2002handbook}. The most basic property of any network, second only to the network size and average degree, is the degree distribution, and the early days of network science were filled with the surprising and exciting news that degree distributions in many real-world networks of completely different origins are scale-free, i.e., ``close to power laws.'' This property means that the node degrees in a network are highly variable and lack a characteristic scale, with a multitude of profound and far-reaching implications for a wide spectrum of structural and dynamical properties of networks~\cite{barabasi2016network,newman2018networks,barrat2008dynamical,bornholdt2002handbook,hofstad2016random,dorogovtsev2008critical,arenas2008synchronization,dallasta2006dynamical}. These implications are the reason why these scale-free findings were extremely impactful, and why they steered the whole field of network science in the direction it has followed for nearly two decades. They impacted essentially all the key aspects of network science, from the basic tasks of network modeling, all the way down to concrete applications, such as prediction and control of the dynamics of real-world complex systems, or identifying their vulnerabilities~\cite{barabasi2016network,newman2018networks,barrat2008dynamical,bornholdt2002handbook}.

Yet there is one glaring problem behind all these exciting developments. The problem is that scale-free networks do not have any widely agreed-upon rigorous definitions. Specifically,
it is quite unclear what it really means for a degree sequence in a given real-world network to be power-law or ``close'' to a power law. This lack of rigor has led and still leads to confused controversy and never-ending heated debates~\cite{willinger2002scaling,li2005theory,noauthor2006network,krioukov2007workshop,willinger2009mathematics,mitzenmacher2004brief,khanin2006scale,stumpf2012critical,corral2011noncharacteristic,corral2018power,clauset2009power,broido2018scale,klarreich2018scant,holme2019rare,lee2018bayesian,drees2018minimum,gerlach2019testing,serafino2019scale,charpentier2019extended}. This controversy has culminated in the recent work~\cite{broido2018scale} that concluded that ``scale-free networks are rare.'' Here we arrive at quite a different conclusion based on a state-of-the-art statistical analysis and a more general definition of power laws.

Faced with the question whether a given real-world network is scale-free or not, one first has to decide how much the data can be trusted---how well does the measured degree sequence reflect the actual degree sequence in the network? We do not address this question here, and assume that we can trust the data. Under this assumption, the next questions are:
\begin{enumerate}
  \item What exactly does it mean that a distribution is approximately a power law?
  \item What are the correct, i.e., statistically consistent, methods to estimate the tail exponent of this power law from the measured degree sequence?
  \item How likely is it that the measured sequence comes from a power law with the estimated exponent?
\end{enumerate}
Here we address all these three questions.

One of the most frequently seen formula in the early days of network science was
\begin{equation}\label{eq:power_law_approx_pdf}
	P(k) \sim k^{-\gamma}.
\end{equation}
It intended to say that the fraction $P(k)$ of nodes of degree $k$ in a network under consideration decays with $k$ approximately as a power law with exponent~$\gamma$. The symbol `$\sim$' could mean anything, but usually its intended meaning was something like ``roughly proportional.''
The literature was also abundant with plots of empirical probability mass/density functions (PMFs/PDFs) $P(k)$ and complementary cumulative distribution functions (CCDFs) $\overline{F}(k)$ of degrees~$k$
drawn on the loglog scale to illustrate that these functions are ``roughly straight lines,'' so that the network is power-law, thus deserving a publication.

The first attempt to introduce some rigor into this vibrant activity, which became overwhelmingly popular in network science, came in~\cite{clauset2009power}, when network science was about a decade old. In~\cite{clauset2009power}, Eq.~\eqref{eq:power_law_approx_pdf} was taken literally to mean that $P(k)$ for $k \geq k_{\min}$ is \emph{exactly} proportional to $k^{-\gamma}$, i.e.,
\begin{equation}\label{eq:power_law_equal_pdf}
	P(k) = c\,k^{-\gamma},
\end{equation}
where $c$ is the normalization constant.

But complexly mixed stochastic processes driving evolution of many different real-world networks are of different origins and nature. Worse, they all are prone to different types and magnitudes of noise and fluctuations. Therefore, basic common sense suggests that these processes can hardly produce beautifully clean power-law dependencies void of any deviations from~\eqref{eq:power_law_equal_pdf}. This is similar to how one cannot expect Newton's laws on Earth with friction to yield results as beautiful as Newton's laws in the empty space without friction. That is why it is not surprising that if one looks for such idealized power-law dependencies in real-world networks, one is doomed to find them quite rare~\cite{broido2018scale}. And as far as power-law network \emph{models} are concerned, even the most basic such model, preferential attachment, is known to have a degree distribution with a power-law tail, but the exact expression for the degree distribution in preferential attachment networks is not a pure power law~\eqref{eq:power_law_equal_pdf}, as shown in~\cite{dorogovtsev2000structure,krapivsky2000connectivity,bollobas2001degree}.
In fact, power-law network models with the pristine purity of~\eqref{eq:power_law_equal_pdf} are an exception rather than a rule.

For all these reasons, in statistics one considers the class of \emph{regularly varying distributions}~\cite{resnick2007heavy,bingham1989regular,foss2011introduction,beirlant2006statistics} instead of the pure power laws~\eqref{eq:power_law_equal_pdf}. Compared to the rather restrictive distribution class~\eqref{eq:power_law_equal_pdf}, the class of regularly varying distributions is much larger. In particular, it contains all the distributions whose PDFs are given by
\begin{equation}\label{eq:power_law_slowly_pdf}
	P(k) = \ell(k)k^{-\gamma},
\end{equation}
thus allowing for deviations from pure power laws by means of a slowly varying function $\ell(k)$, i.e., a function that varies slowly at infinity, classic examples including functions converging to constants or $\log^ak$ for any constant~$a$. The exact definition of regularly varying distributions requires their CCDFs to be of the form
\begin{equation}\label{eq:power_law_slowly_cdf}
	\overline{F}(k) \coloneqq \sum_{k'>k}P(k') = \ell^\prime(k)k^{-\alpha},
\end{equation}
where $\alpha=\gamma-1$, and $\ell^\prime(k)$ is also a slowly varying function. The class of distributions that satisfy~\eqref{eq:power_law_slowly_cdf} is even more general than~\eqref{eq:power_law_slowly_pdf}: if \eqref{eq:power_law_slowly_pdf} holds for a distribution, then so does~\eqref{eq:power_law_slowly_cdf}, but not necessarily the other way around.

Compared to~\eqref{eq:power_law_equal_pdf}, any distribution in the class~\eqref{eq:power_law_slowly_cdf} has the same power-law tail exponent~$\gamma$, but it can have drastically different shapes for finite degrees. The exact shape of $\ell(k)$ is of much less significance than the value of the tail exponent~$\gamma$, because it is $\gamma$, and not $\ell(k)$, that is solely definitive for a number of important structural and dynamical properties of networks in the limit of large network size~\cite{hofstad2016random,dorogovtsev2008critical,arenas2008synchronization,dallasta2006dynamical,stegehuis2017clustering,van2007distances,van2018typical,boguna2009navigability,delre2010will,doerr2012rumors,bringmann2016greedy}. As the simplest example, the value of $\gamma$ determines how many moments of the degree distribution remain bounded in the large-graph limit, affecting many important network properties.
Yet we also note that some properties of finite-size networks may and usually do depend on a specific form of~$\ell(k)$.

For all these reasons, and following the well-established tradition in statistics, in Section~\ref{sec:definitions} we \emph{define} a distribution to be \emph{power-law} if it is \emph{regularly varying}, i.e., if its CCDF satisfies~\eqref{eq:power_law_slowly_cdf}.

The next question, that we address in Section~\ref{sec:estimating_tails}, is how to properly estimate the value of $\gamma$ under the assumption that a given degree sequence comes from a regularly varying distribution. This question has attracted extensive research attention in
probability, statistics, physics, engineering, and finance~\cite{resnick2007heavy,bingham1989regular,foss2011introduction,beirlant2006statistics,ameraoui2016bayesian,embrechts2013modelling,boucheron2015tail,mcneil2000estimation,embrechts2013modelling,jansen1991frequency,mcculloch1996financial,kotulski1995asymptotic,metzler2000random,lu2001well,resnick1997heavy,nikias1995signal}, where a variety of estimators have been developed for this task, all based on extreme value theory. We identify the maximal subset of such estimators that, to the best of our knowledge, are the only currently existing estimators that
\begin{enumerate}
  \item are applicable to any regularly varying distribution;
  \item are statistically consistent, i.e., have been proven to converge to the true $\gamma$, if applied to increasing-length sequences sampled from any regularly varying distribution; and
  \item can be fully automated by the means of the double bootstrap method that has been proven to yield the optimal estimation of $\gamma$ for any finite sequence of numbers sampled from any regularly varying distribution.
\end{enumerate}

It is important to stress here that~\eqref{eq:power_law_equal_pdf} is just one representative of the extremely wide class of regularly varying distributions~\eqref{eq:power_law_slowly_cdf}. Therefore, as opposed to the methods in~\cite{clauset2009power,broido2018scale} that are consistent only under the assumption that a given degree sequence comes from a pure power law~\eqref{eq:power_law_equal_pdf} above a certain minimal degree threshold, the estimators that we discuss in Section~\ref{sec:estimating_tails} are proven to be consistent under the much more general assumption that the sequence comes from any impure power law, including any distribution that satisfies~\eqref{eq:power_law_slowly_pdf} or even~\eqref{eq:power_law_slowly_cdf} with any nontrivial slowly varying functions~$\ell(k)$, $\ell^\prime(k)$.

In Section~\ref{sec:validation} we evaluate these estimators by applying them to a wide range of synthetic sequences sampled from a variety of regularly varying distributions, as well as to degree sequences in paradigmatic network models---the configuration model, preferential attachment, and random hyperbolic graphs. In all the considered cases, all the considered estimators converge as expected. We also compare their performance to that of the PLFit algorithm from~\cite{clauset2009power,broido2018scale}, which is believed to represent the state of the art in network science. We find that PLFit tends to show much worse performance when applied to distributions with nontrivial slowly varying functions. Remarkably, one example of such nontrivial distributions is the degree distribution in the ``harmonic oscillator'' of power laws---the preferential attachment model.

The key strength behind the estimators considered in this paper is that most of them have been proven to be consistent not only under the assumption that the sampling distribution $P(k)$ is regularly varying, but also under the even more general assumption that it is \emph{any}\/ distribution belonging to the maximum domain of attraction of \emph{any}\/ extreme value distribution with \emph{any}\/ index $\xi$, which is the main parameter of an extreme value distribution. The extreme value distributions are the $n\to\infty$ limit distributions of rescaled maximum values among $n$ samples from any given distribution $P(k)$. If $P(k)$ is regularly varying, then $\xi$ is strictly positive, and the tail exponent~$\gamma$ and extreme value index~$\xi$ are related by
\begin{equation}
\xi=\frac{1}{\gamma-1}.
\end{equation}
If $P(k)$ is not regularly varying, then $\xi$ is either negative or zero, in which case the tail exponent~$\gamma$ is undefined. None of the considered estimators estimates~$\gamma$ directly. They all are based on extreme value theory, and estimate the index~$\xi$ instead.

The last question from the list of the three questions above is about hypothesis testing. Given any degree sequence, one can always apply to it any $\xi$-estimator that will always return some $\xi$-estimate $\hat{\xi}$. How likely is it that this sequence comes from a regularly varying distribution with exponent $\gamma=1+1/\hat{\xi}$? Clearly, if $\hat{\xi}$ is either negative or zero, then this question is ill-posed since one cannot even tell what the $\gamma$ is. But what if $\hat{\xi}$ is positive?

Section~\ref{sec:power-law-networks} is dedicated to the explanation that even in this case one cannot devise any hypothesis test to answer the above question. The popular $p$-value approach used often in hypothesis testing is deeply problematic and should be avoided, as has been long known and recently well documented in a statement article by the American Statistical Association~\cite{wasserstein2016asa}, followed by a special issue of {\it The American Statistician}~\cite{wasserstein2019moving}. But it is not that $p$-values are bad, and there is a better way. \emph{Hypothesis testing is} \emph{simply impossible with regularly varying distributions.} Intuitively, the main reason for this impossibility is the infinite number of ``degrees of freedom'' contained in the space of slowly varying functions $\ell(k)$ that make the space of regularly varying distributions nonparametric. In particular, there is an infinite number of regularly varying distributions such that for any \emph{finite} sequence length, degree sequences of this length sampled from these distributions do not appear to be regularly varying, or the other way around, there is an infinite number of distributions that are not regularly varying, but such that random sequences of any \emph{finite} length sampled from these distributions appear as regularly varying.  

In view of this extremely important but badly misunderstood observation, which is one of the key points in this paper, the best strategy one can follow is to consult as many consistent $\gamma$-estimators as possible to see whether they agree on the ranges of their $\gamma$-estimates on a given sequence~\cite{resnick2007heavy}. And this is indeed the strategy we follow in Section~\ref{sec:power-law-networks} to \emph{define} what it means for a given degree sequence to be power-law. If at least one of the considered estimators returns a negative or zero value of $\hat{\xi}$, then we call the degree sequence \emph{not power-law}, but if all the estimators agree that $\hat{\xi}>\xiThreshold$, then we say that the sequence is \emph{power-law}. If neither of these conditions are satisfied, then we call the degree sequence \emph{hardly power-law}. The threshold $\hat{\xi}=\xiThreshold$ between the power-law and hardly power-law ranges is completely arbitrary, and one is free to choose any nonnegative value of $\xi$ for this threshold, determining the value of $\gamma$ above which one can hardly call a network power-law. We chose this value to be~$\gamma=1+1/\xi=\gammaThreshold$ for the reasons discussed in Section~\ref{sec:power-law-networks}.

Finally, in Section~\ref{sec:real_networks}, we implement all the considered estimators in a software package~\cite{githubcode} available to the public, and apply them to the degree sequences of $115$ real-world networks with more than $1,000$ nodes collected from the KONECT database~\cite{kunegis2013konect}. The collection contains many paradigmatic networks from different domains. Some of them were found to be power-law in the past (the Internet, for instance), while others were documented not to be power-law (road networks are a classic example). We find that the considered consistent estimators mostly agree with this classification, while overall, according to the definitions above, these estimators report that \undirectedPL of the considered undirected networks have degree sequences that are power-law. Among the considered directed networks, \directedBothPL have both in- and out-degree sequences that are power-law, while \directedAtLeastOnePL have either in- or out-degree sequence that is power-law. The bipartite networks exhibit a similar picture according to the estimators: 
\bipartiteBothPL of them have power-law degree sequences for both types of nodes, while in \bipartiteAtLeastOnePL of them at least one type of nodes has a power-law degree sequence.

In summary, if we relax the unrealistic requirement that degree distributions in real-world networks must be pure power laws, and allow for real-world impurity via regularly varying distributions, then upon the application of the state-of-the-art
methods in statistics to detect such distributions in empirical data, we find that one can definitely not call scale-free networks ``rare.''

\section{Power-Law Distributions}\label{sec:definitions}

We \emph{define} a distribution to be \emph{power-law} if it is \emph{regularly varying}. A distribution with PDF $P(k)$ is called \emph{regularly varying~\cite{bingham1989regular,foss2011introduction} if its CCDF
\begin{align}\label{eq:power_law_approx_cdf}
	\overline{F}(k) &\coloneqq 1 - F(k) = \sum_{k'>k}P(k')
\end{align}
satisfies}
\begin{equation}\label{eq:def_power_law}
	\overline{F}(k) = \ell(k)k^{-\alpha},
\end{equation}
where $\alpha>0$, and $\ell(k)$ is a slowly varying function. A function $\ell(x)$ is called \emph{slowly varying} if
\begin{equation}
  \lim_{x \to \infty} \frac{\ell(tx)}{\ell(x)} = 1
\end{equation}
for any $t>0$.
If the PDF of a distribution satisfies~\eqref{eq:power_law_slowly_pdf} with some slowly varying function, then the distribution is regularly varying, i.e., its CCDF satisfies~\eqref{eq:def_power_law} with some other slowly varying function. The converse may or may not be true, as discussed in Appendix~\ref{sec:heavy_tails}.

If a distribution is regularly varying, but its slowly varying function $\ell(k)$ in~\eqref{eq:def_power_law} does not vary at all, i.e., if it is constant, then we call such a distribution a \emph{pure power law}. If $k$ is integer-valued, $k=k_{\min}, k_{\min}+1, \ldots$, where $k_{\min}$ is a natural number, then this pure power law is known as the generalized zeta distribution with PDF
\begin{equation}\label{eq:zeta_def}
  P(k) = \frac{k^{-\gamma}}{\zeta(\gamma,k_{\min})},
\end{equation}
where $\gamma$ is the PDF tail exponent, and $\zeta(\gamma,k_{\min})$ is the Hurwitz zeta function. If $k=x$ is real and $x\geq x_{\min}>0$, then this pure power law is known as the Pareto distribution whose PDF is
\begin{equation}\label{eq:pareto_def}
  P(x) = \alpha x_{\min}^\alpha x^{-\gamma},
\end{equation}
where $\alpha=\gamma-1$. In both cases the constant slowly varying functions are simply the normalization constants. Clearly, pure power laws form a small subset of general power laws, i.e., regularly varying distributions.

The definition of power-law distributions as regularly varying distributions formalize the point that the distribution exhibits a power-law tail at high degrees, but has an arbitrary shape at small degrees.
They follow the well-established convention in probability, statistics,
physics, engineering, and finance~\cite{resnick2007heavy,bingham1989regular,foss2011introduction,beirlant2006statistics,ameraoui2016bayesian,embrechts2013modelling,boucheron2015tail,mcneil2000estimation,embrechts2013modelling,jansen1991frequency,mcculloch1996financial,kotulski1995asymptotic,metzler2000random,lu2001well,resnick1997heavy,nikias1995signal}, where regularly varying distributions are the best studied subclass of much larger classes of distributions, such as \emph{heavy-tailed} and others, see Appendix~\ref{sec:heavy_tails}.

We also note that the rigorous definition of regularly varying distributions in~\eqref{eq:def_power_law} perfectly formalizes the common traditional intuition behind the `$\sim$' sign in the non-rigorous ``scale-free formula''~\eqref{eq:power_law_approx_pdf}. Indeed, if the regularly varying functions $\log(ck)k^{-\alpha}$ and $C k^{-\alpha}$, for example, are drawn on the loglog scale, one would see nothing but straight lines at large $k$ in both cases, even though the first case is not a pure power law. This observation is formalized by Potter's Theorem~\cite[Theorem~1.5.6]{bingham1989regular}, stating that $\lim_{k \to \infty} \ell(k) k^{-\delta} = 0$ for any slowly varying function $\ell(k)$ and any $\delta > 0$. Therefore, in both cases one would be tempted to write $\overline{F}(k) \sim k^{-\alpha}$, so that the power-law definition~\eqref{eq:def_power_law} is indeed a perfect way to hide any distributional peculiarities that do not asymptotically influence the power-law shape of the distribution tail.

We emphasize here that due to the nature of slowly varying functions, 
definition~\eqref{eq:def_power_law} is intrinsically asymptotic, dealing with the $k \to \infty$ limit. In particular this implies that a distribution satisfying~\eqref{eq:def_power_law} can take any form for all degrees $k<K$ below an arbitrarily large but fixed threshold $K>0$. This observation, and more generally, the asymptotic nature of power laws is the key factor responsible for the impossibility of hypothesis testing with regularly varying distributions, Section~\ref{sec:power-law-networks}.

The simplest and most frequently seen examples of regularly varying distributions can be found in Appendix~\ref{sec:heavy_tails}.

\section{Consistent Estimators of the Tail Exponent}\label{sec:estimating_tails}

We now turn to the question of how to estimate the tail exponent of a regularly varying distribution given a finite collection of samples (e.g., node degrees) from it. We employ three estimators---\emph{Hill}~\cite{hill1975simple}, \emph{Moments}~\cite{dekkers1989moment} and \emph{Kernel}~\cite{groeneboom2003kernel}---that have been long proven to be statistically consistent at this task. \emph{Consistency} means that as the number of samples increases, the estimated values of the exponent $\hat{\xi}$ are guaranteed to converge to the true exponent value $\xi$ \emph{regardless} of the slowing-varying function~$\ell(k)$.

We note that all the considered estimators are consistent under the assumption that the data that they are applied to is a collection of i.i.d.\ (independent, identically distributed) samples from a regularly varying distribution. There is no, and cannot be any, hypothesis testing procedure that will tell whether a given sequence (of degrees in a (real-world) network) is an i.i.d.\ sequence from a regularly varying distribution, as we explain in detail in Section~\ref{sec:power-law-networks}. Therefore the application of these estimators to degree sequences of real-world networks can be justified only indirectly. In particular, their consistency has been recently proven for a wide range of preferential-attachment models, in which degree sequences are not exactly i.i.d.~\cite{wang2018consistency}. In case of the configuration model~\cite{bender1978asymptotic,bollobas1980probabilistic,wormald1980some}, it is known that a degree sequence sampled i.i.d.'ly from a distribution with finite variance is graphical with positive probability~\cite[Theorem 7.21]{hofstad2016random}. This probability is very close to $1/2$ for any distribution with a finite mean that takes odd values with positive probability, a surprising fact proven in~\cite{arratia2005likely}. This means that random graphs with a power-law degree distribution can be sampled by first sampling i.i.d.'ly a degree sequence from the distribution, and then constructing a graph with this degree sequence using known techniques~\cite{del2010efficient}. Such a graph exists with non-zero probability because the degree sequence is graphical with this probability. Applied to graphs constructed this way, the estimators are consistent because the degree sequences in these graphs are i.i.d. Yet proving the consistency of these estimators applied to other network models is an open research area, which is only tangentially related to justifying their application to real-world networks, since there cannot be any ``ultimately best'' model for any real-world network. We also note that these estimators are actively employed in practice, in particular in financial mathematics~\cite{chan2006using,danielsson2000value,gilli2006application,embrechts2013modelling,mcneil2000estimation}, where regularly varying distributions are abundant, where the estimation of rare events is of key importance (e.g., for portfolio or fund management), and where the i.i.d.\ assumption cannot be checked to hold in real-world data either.

All the considered estimators do not estimate either the PDF or CCDF tail exponents $\gamma$ or $\alpha=\gamma-1$ directly. They are all based on extreme value theory, so that instead of estimating $\gamma$ or $\alpha$, they estimate the extreme value index $\xi$ of the distribution. Given a sequence of $n$ i.i.d.\ samples $x_1, \dots, x_n$ from a distribution, extreme value theory is concerned with the behavior of the maximum value $m_n=\max_{i=1}^n x_i$ in this sample. In particular, one is typically interested in finding $n$-dependent constants $c_n$ and $d_n$ such that the distribution of $(m_n - d_n)/c_n$ has a non-degenerate limit. This limit distribution, if it exists, is called an \emph{extreme value distribution (EVD)}, and the distribution of $x$'s is then said to belong to the \emph{maximum domain of attraction (MDA)} of this EVD. One of the key results in extreme value theory~\cite{fisher1928limiting} is that there are only three families of EVDs. They are parameterized by a real number $\xi$, called the \emph{extreme value index}. The three families are Weibull with $\xi < 0$, Gumbel with $\xi = 0$, and Fr\'{e}chet with $\xi > 0$. 

The reason why extreme value estimators are the standard tool in statistics to infer the tail exponent of regularly varying distribution, is the fundamental fact proven in~\cite{gnedenko1943distribution}. It states that the class of all distributions that belong to the Fr\'{e}chet MDA with $\xi > 0$ is exactly the class of all regularly varying distributions, i.e., those distributions whose CCDFs satisfy~\eqref{eq:def_power_law}. Moreover, the PDF and CCDF tail exponents $\gamma$ and $\alpha$ are related to the extreme value index~$\xi$ in this case by
\begin{equation}
  \xi = \frac{1}{\gamma-1}=\frac{1}{\alpha}.
\end{equation}
It is this intimate relation between regularly varying distributions and extreme value theory that provides a rigorous and well-explored framework to analyze regularly varying distributions and make inferences concerning them.  

We note that while the Hill estimator is consistent under the assumption that a given sequence is sampled only from a regularly varying distribution, i.e., that it is in the Fr\'{e}chet MDA, the other considered estimators---that is, the Moments and Kernel estimators---are consistent for degree sequences sampled from \emph{any} distribution belonging to the MDA of \emph{any} extreme value distribution. This means that if these estimators are applied to increasing-length sequences sampled from distributions belonging to the Fr\'{e}chet, Gumbel, or Weibull MDAs, then in all these three cases the estimates of these estimators are guaranteed to converge to the true values of $\xi$ that are positive, zero, and negative, respectively. As a side note, while the Fr\'{e}chet MDA is exactly all the regularly varying distributions, the Weibull MDA consists of distributions with upper-bounded supports, while the Gumbel MDA contains all other distributions that can be either light-tailed or heavy-tailed, but not regularly varying. Appendix~\ref{sec:estimators} contains all the relevant details.

The key point here, which we rely upon in the next section, is that if the estimators, applied to a particular degree sequence, return either negative values of $\xi$, or values of $\xi$ close to zero, then this sequence is quite unlikely to come from the Fr\'{e}chet MDA, i.e., from a regularly varying distribution. Yet again, there is no way to quantify this unlikeliness rigorously, as explained in Section~\ref{sec:power-law-networks}.

Applied to $n$ data samples $x_1,x_2,\ldots,x_n$, the estimators operate by first sorting the data in non-increasing order, $x_{(1)} \ge x_{(2)} \ge \ldots \ge x_{(n)}$, and then limiting their consideration only to the $\kappa$ largest data samples $x_{(1)}, x_{(2)}, \ldots, x_{(\kappa)}$, where $\kappa$ is a free parameter. Since the $\kappa$-th
order statistic is a random variable representing the $\kappa$-th
largest element among $n$ i.i.d.\ samples from a distribution, the $\kappa$ parameter is known as the \emph{number of order statistics}. The estimators thus operate only 
on the $\kappa$-tail of the empirical distribution represented by the $\kappa$ order statistics. Given this tail, different estimators provide different expressions, documented in Appendix~\ref{sec:estimators}, for the estimated value $\widehat{\xi}_{\kappa,n}$ of $\xi$, which depends on $\kappa$. These expressions rely on different aspects of the order statistics contained in the tail, but all these expressions are consistent, meaning that
\begin{equation}\label{eq:def_consistency}
	\widehat{\xi}_{\kappa,n} \to \xi \quad \text{as } \kappa, n \to \infty, \, \kappa/n \to 0,
\end{equation}
for all the estimators. The convergence above is usually in probability, although in some cases some stronger results, such as almost sure convergence or asymptotic normality, are available under additional assumptions on the data.

It is important to note here that in proving this convergence, the number of order statistics $\kappa$ cannot be fixed, it must diverge with the number of samples $n$ to incorporate more and more data in the tail, so that the estimated value of $\xi$ is less and less affected by the fluctuations in the tail. Yet $\kappa$ cannot be equal to $n$ either, since in this case the estimated $\xi$ would be affected by the slowly varying function $\ell(k)$. This implies that if applied to finite-size data samples, these estimators will not give a good estimate of $\xi$ for either small or large values of $\kappa$. One option to deal with this problem in practice is to investigate the plot of $\widehat{\xi}_{\kappa,n}$ as a function of $\kappa$ in order to find the value of $\kappa$ where this function is ``most flat/constant.'' This subjective approach can clearly not be rigorous. Worse, on real-world data, these functions can behave violently, see for instance the figures in Chapter~4 of~\cite{resnick2007heavy} or in~\cite{matsui2013estimation}, so that finding such a flat region of $\widehat{\xi}_{\kappa,n}$ may be quite problematic.

Fortunately, for the three estimators that we consider, the double bootstrap method documented in Appendix~\ref{sec:real_estimation} is proven to find the optimal value $\kappa^\ast$ of $\kappa$. \emph{Optimality} means here that the error between the estimated and true values of $\xi$ is minimized, Appendix~\ref{sec:real_estimation}. The double bootstrap method is also proven not to break consistency, meaning that as a function of $n$, the value of $\kappa^\ast_n$ diverges sublinearly, so that in view of~\eqref{eq:def_consistency}, the estimated value of~$\xi$, $\widehat{\xi}_{\kappa^\ast_n,n}$, converges to the true~$\xi$: 
\begin{equation}
	\widehat{\xi}_{\kappa^\ast_n,n} \to \xi \quad \text{as } n \to \infty.
\end{equation}

In addition to the Hill, Moments, and Kernel estimators, the Pickands estimator~\cite{pickands1975statistical} and its generalized versions~\cite{segers2005generalized} are also often considered. However, only for one of these generalizations has the double bootstrap method been proven to be consistent, Appendix~\ref{sec:estimators}.
Worse, in application to real-world data, the Pickands estimator has been shown to be unstable and volatile~\cite{segers2005generalized, shinyie2013semi} and to have poor efficiency~\cite{groeneboom2003kernel,muller2009smooth}. Many other $\xi$-estimators exist, see~\cite{gomes2015extreme} for a review, but the proofs of consistency of the double bootstrap method are available only for the Hill, Moments, Kernel, and Pickands estimator.

Therefore, to the best of our knowledge, the Hill, Moments, and Kernel estimators are the maximal subset of consistent, stable, and efficient estimators, for which the double bootstrap method that automatically determines the optimal value of $\kappa$, is proven to be both optimal and consistent. The reason we consider not one but all such estimators is mentioned above: since as we explain in Section~\ref{sec:power-law-networks} there can be no hypothesis test to tell whether a given degree sequence is an i.i.d.\ sequence sampled from a regularly varying distribution, the best one can do is to consider as many consistent estimators as possible, testing as many different aspects of the degree sequence as possible, and see whether they agree in their estimations~\cite{resnick2007heavy}.

\section{Evaluation of Estimator Performance}\label{sec:validation}

In Appendix~\ref{sec:experiments} we perform an in-depth evaluation of all the three estimators based on extreme value (EV) theory from the previous section. We apply them to a collection of random sequences sampled from various distributions, as well as to degree sequences in three popular network models---the configuration model, preferential attachment, and random hyperbolic graphs. We also juxtapose these validation results against the performance of the PLFit algorithm from~\cite{clauset2009power,broido2018scale}.

The results of these experiments are as expected: all the EV estimators converge to the true value of $\xi$ if the distribution is regularly varying, and they do not converge if it is not. They also converge even in the case where we sample not from a fixed regularly varying distribution, but from a sequence of distributions that are not regularly varying but that converge to a regularly varying distribution---the case with a Pareto distribution with the diverging natural exponential cutoff. On degree sequences in network models where individual degrees are not i.i.d.\ samples from a fixed degree distribution, the estimators converge as well, even though the i.i.d.\ assumption no longer holds.

For PLFit we find in Appendix~\ref{sec:experiments} that if the sample distribution is sufficiently ``nice,'' then the estimation accuracy and convergence rates of the PLFit are comparable to those of the EV estimators. However, in cases where the distribution is not so nice and is further from a pure power law, the EV estimators perform significantly better than the PLFit. This is the case, for example, with distributions that can be fitted by power laws with wrong exponents in the region of small degrees. Remarkably, one example of such a distribution is the degree distribution in the preferential attachment model, a ``harmonic oscillator'' of power laws in network science~\cite{dorogovtsev2000structure,krapivsky2000connectivity,bollobas2001degree}.
For these and a number of other lower-level technical reasons, all documented in Appendix~\ref{sec:experiments} and fully supported in a more recent and detailed focused study~\cite{drees2018minimum}, we exclude the PLFit from the subsequent considerations here.

\section{Power-Law Degree Sequences}\label{sec:power-law-networks}

There is no way to test the hypothesis that any given number was sampled from any given distribution that contains the number in its support. Yet if one has a long sequence of numbers, then there is a multitude of hypothesis testing procedures to measure how likely it is that this sequence was sampled from the distribution. The longer the sequence, the more reliable such procedures are, and any good procedure will give a definitive answer as the sequence length approaches infinity. This statistical methodology is widely known to work not only for a fixed distribution, but also for many parametric families of distributions. In the latter case, the testing involves one additional step: the parameters of the distribution are first to be estimated from the sequence using a consistent estimator.

A variation of this standard approach is at the core of~\cite{clauset2009power,broido2018scale}, where the parametric family of distributions consists of pure power laws---the zeta or Pareto distributions. Their parameters, the tail exponents, are estimated using a combination of the likelihood maximization and Kolmogorov-Smirnov (KS) distance minimization techniques documented in Appendix~\ref{sec:experiments}. Finally, the hypothesis testing procedure is the KS test, yielding a popular $p$-value number reflecting roughly how likely a given sequence comes from the pure power law with the estimated exponent.

We now come to the key point that this or any other hypothesis testing approach is not, and cannot be, applicable to regularly varying distributions, simply because these distributions do \emph{not} form a parametric family of distributions. Instead, they are a nonparametric class of distributions of an asymptotic nature with an uncountably infinite number of ``degrees of freedom'' contained in the slowly varying functions~$\ell(k)$ (Appendix~\ref{sec:heavy_tails}).
Testing whether a given \emph{finite} collection of numbers was sampled from such an infinite-dimensional family of distributions is akin to testing whether a given number was sampled from a given distribution, which clearly is impossible as mentioned above.

Situations of this type are quite familiar for a physicist or network scientist. Phase transitions are a classic example: true phase transitions occur only in the thermodynamic limit, while for any finite system we can only observe their signs. The simplest example in network science is graph sparsity. The definition of sparse graphs applies only to family of graphs whose size tends to infinity, and one cannot say anything at all about how sparse any given finite-size graph is, even if this is an empty graph of $n=10^{10}$ nodes, simply because this graph can be considered as a typical Erd\H{o}s-R\'enyi graph with the connection probability $p=10^{-10^{10}}$, which is dense.

Yet, for a variety of reasons, these matters, including the impossibility of hypothesis testing with nonparametric families of distributions, as well as various consequences of this impossibility, are routinely overlooked and misunderstood. For these reasons, we first discuss the general picture behind this impossibility, and then illustrate it with a collection of examples.

First, the general picture is as follows. Recall that the consistency of an exponent estimator means that if we sample i.i.d.'ly increasingly larger numbers $n$ of random numbers $k_i$, $i=1,\ldots,n$, from a fixed regularly varying distribution with exponent $\gamma$ and any slowly varying function $\ell(k)$, then the estimates $\hat{\gamma}_n$ that the estimator returns are guaranteed to converge to $\gamma$. Observe that while $\gamma$ is a fixed number, $\hat{\gamma}_n$ is a random number, i.e., a random variable, because the $k_i$s are random. That is why one has to be careful with statements concerning in what particular sense the random variable $\hat{\gamma}_n$ converges to number $\gamma$. As stated in Section~\ref{sec:estimating_tails}, the convergence is usually in probability, but in some cases one can prove that $\gamma_n$ converges to a normally distributed random variable with mean $\gamma$ and some vanishing variance. For different definitions of convergence of random variables, we refer to any textbook on probability, such as~\cite{billingsley2013convergence}.

It is crucially important to recognize that the convergence in probability does not mean that for any \emph{finite} $n$ there are any guarantees on how close the estimate $\hat{\gamma}_n$ will be to the true $\gamma$.
To see why, observe that the slowly varying function $\ell(k)$
can be arbitrarily bad, breaking pure power laws for any arbitrarily large number of degree samples or range of degrees, while the true tail of the distribution can be inferred only in the limit of infinitely long sampled sequences, which one never has in practice.

We thus see that this general picture is very different from the one with hypothesis testing with a parametric family of distributions, such as the normal distributions or pure power laws. If we employ MLE, for instance, to estimate the parameters of such distributions, we usually know all we need to tell how close our estimates are expected to be to the true values for any given sample of size $n$. We often even know the full distribution of these estimates as random variables, and we then have the luxury to employ any reasonable hypothesis test of our choice, or to compute $p$-values to quantify chances if we wish. With regularly varying distributions, the situation is very different because if we do not know $\ell(k)$, we simply do not know how large $n$ must be so that our estimators and hypothesis tests start showing any signs of convergence, simply because $\ell(k)$ can be arbitrarily bad.

To illustrate this extremely important point, we consider several examples next. The first two are of artificial/adversarial nature, while the last one is a well-studied network model.

\begin{figure}[t]
 \hspace*{-20pt}\includegraphics[width=.3\textwidth]{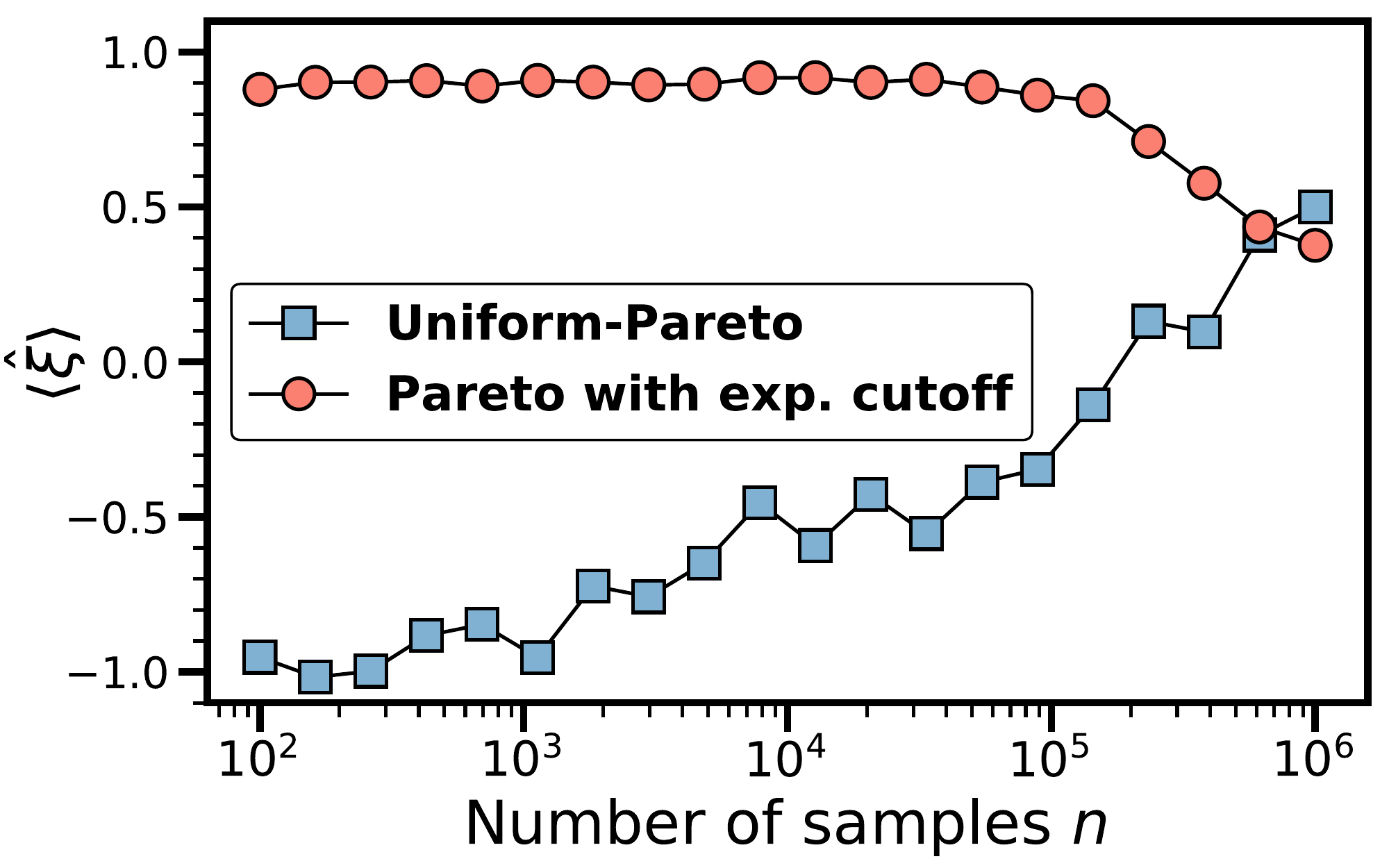}
  \caption{
  {\bf The extreme value index estimates for the two adversarial examples in Section~\ref{sec:power-law-networks}.}
  The sequences of varying length $n$ are sampled from the distributions defined by Eq.~\eqref{eq:adversarial1} with $c_1 = 1$, $c_2 = 2$, $f = 2\cdot10^{-4}$, and $\gamma = 2$ (blue squares), and by Eq.~\eqref{eq:adversarial2} with $c = 500$ and $\gamma = 2$ (red circles). The data shown are the estimates of the Moments estimator with the double bootstrap procedure applied to these samples. The results are averaged over 100 random sequences for each data point. In the case of blue squares, the distribution is regularly varying with $\gamma=2$, so that $\xi=1$. However, if $n$ is not sufficiently larger than $1/f$, the sequences sampled from this distribution appear as if sampled from the uniform distribution that belongs to the Weibull MDA with $\xi=-1$. In the case of red circles, the distribution is not regularly varying. It belongs to the Gumbel MDA with $\xi=0$. However, if $n$ is not sufficiently larger than $c^{\gamma-1}$, the sequences sampled from this distribution appear as if sampled from a regularly varying distribution with $\gamma=2$ and $\xi=1$. The examples illustrate that for any finite $n$ there is absolutely no way to tell what distribution class (regularly varying or not) the samples are coming from.
  }
  \label{fig:adv-examples}
\end{figure}

The first example is a regularly varying distribution with support on $[c_1, \infty)$ and PDF with $\gamma > 1$ and constants $c_2 > c_1 \geq 0$ and $f \in [0,1]$:
\begin{align}\label{eq:adversarial1}
  P(x) &= \ell(x) x^{-\gamma}, \text{ where} \\
  \ell(x) &=
  \begin{cases}
    \frac{1-f}{c_2 - c_1} x^{\gamma}, & \mbox{if } x\in[c_1, c_2], \nonumber\\
    (\gamma - 1) f c_2^{\gamma-1}, & \mbox{if } x\in(c_2, \infty).
  \end{cases}
\end{align}
In words, this distribution is uniform on the interval $[c_1,c_2]$, and a pure power law (Pareto) for $x>c_2$. The parameter~$f$ is the fraction of the distribution mass falling within the Pareto region. This distribution is regularly varying for any constants $c_2 > c_1 \geq 0$, $f \in (0,1]$, and $\gamma>1$ because for $x > c_2$ the slowly varying function of its CCDF is constant, or in simpler terms, because it has a pure power law tail. However, if we sample $n < 1/f$ random numbers from this distribution, then there is no way to infer from these samples that the distribution is regularly varying with exponent $\gamma$ because the expected number of samples in the Pareto region is below 1, so that all samples are expected to be from the uniform part of the distribution. Only if the number of samples $n$ is sufficiently larger than $1/f$, can we expect to start seeing signs of the presence of a power-law tail. Figure~\ref{fig:adv-examples} confirms that this is indeed the case. Clearly, one can replace the uniform part of the distribution with an arbitrary function, thus reflecting the reality of degree sequences observed in many real-world networks much more closely.

As another example, consider the Pareto distribution supported on $[1, \infty)$ with a fixed exponential cutoff at $c>1$:
\begin{align}\label{eq:adversarial2}
  P(x) &= \frac{c^{\gamma - 1}}{\Gamma(1-\gamma, 1/c)} x^{-\gamma}{\mathrm e}^{-x/c},
\end{align}
where $\Gamma$ denotes the upper incomplete gamma function. This distribution is not regularly varying, yet if our sample size~$n$ satisfies $n<c^{\gamma - 1}$, then we will be tempted to conclude that the distribution is regularly varying, and that the exponent is $\gamma$, because almost all samples will be from the Pareto part of the distribution. Only if the number of samples $n$ is sufficiently larger than $c^{\gamma - 1}$, will we see signs of that this distribution does not really have any power-law tail, as confirmed in Fig.~\ref{fig:adv-examples}.

To see that such deceiving situations can occur in quite reasonable network models we refer to superlinear preferential attachment. In this model of growing networks, new nodes join a network one at a time, and connect to existing nodes of degree~$k$ with probability proportional to $k^\delta$, where $\delta>1$. For any such $\delta$ the limit degree distribution is not regularly varying: the number of nodes with degrees exceeding a certain fixed threshold is finite~\cite{krapivsky2008scale}. Yet this threshold becomes larger if $\delta$ approaches~$1$. The threshold is also a growing function of the average degree $\bar{k}$, i.e., the number of links that new nodes establish. More importantly, the larger this threshold, the more slowly the degree distribution approaches its limit, appearing as a reasonably ``clean'' power law in its vast preasymptotic regime. For example, for $\delta=1.15$ and $\bar{k}=4$, there are no noticeable deviations from this seemingly pure power-law behavior until the network size reaches about $10^{17}$~\cite{krapivsky2008scale}.

All these examples illustrate the point that based on any given \emph{finite} degree sequence (of a real-world network), there is absolutely no way to tell how likely the hypothesis is that this sequence was sampled from a regularly varying distribution. In view of this impossibility, the best strategy one can follow is to simply rely on the estimates of the consistent estimators discussed in the previous section~\cite{resnick2007heavy}. If the estimates of $\xi$ that these estimators report on a given sequence are all positive, then it might be the case that this sequence comes from a regularly varying distribution. Yet if these estimates are negative or close to zero, then the chances of that are slim. However, \emph{there is absolutely no, and cannot be any, rigorous way to quantify these chances, using $p$-values or any other methods,} for the reasons above. This is the key point in our paper.

In view of these considerations, we take a conservative approach, and propose the following definition of a power-law degree sequences, based on the values of $\xi$ that the three estimators from the previous section return upon their application to the sequence:
\begin{itemize}
  \item[$\rhd$] {\bf A degree sequence is not power-law (NPL)} if at least one estimator returns a negative or zero value of $\xi$;
  \item[$\rhd$] {\bf A degree sequence is hardly power-law (HPL)} if all the estimators return positive values of $\xi$, and if at least one estimator returns a value of $\xi\leq\xiThreshold$;
  \item[$\rhd$] {\bf A degree sequence is power-law (PL)} if all the estimators return values of $\xi>\xiThreshold$.
\end{itemize}

In purely intuitive and non-rigorous terms, the larger the $\xi$, the more likely it is that the degree sequence comes from a distribution with a power-law tail. These chances are the smaller, the closer the positive $\xi$ is to zero, and we take a conservative approach to doubt that the degree sequence is power-law if $\xi\leq\xiThreshold$. If $\xi\leq0$, these chances are really slim. Unfortunately, as discussed above, it is principally impossible to attach any rigorous quantifiers to this intuition.

Yet we note that one important advantage of this classification scheme is that it tries to make a decision based on information from several estimators that are known to be consistent, instead of just one of unknown consistency. It is also possible to include other consistent estimators to collect more information about a degree sequence. We reiterate that we employ the Hill, Moments and Kernel estimators here because they are the only three consistent estimators that are known to be stable on real-world data, and for which the double bootstrap procedure has been proven to be consistent.

We also note that the choice of the hardly-power-law $\xi=\xiThreshold$ threshold is completely arbitrary, and in view of the considerations above we should not have defined any \emph{hardly} power-law regime, and call a degree sequence power-law if all $\xi$s are positive. Yet if $\xi=0.01$, for instance, then $\gamma=101$. To call a degree sequence with such $\gamma$ a power law is an unsatisfactory stretch of terminology. Another reason to define this threshold is that it is very difficult to tell whether a very small value $\hat{\xi}>0$ that an estimator returns is an estimation of $\xi=0$ or of a very small $\xi>0$. In the latter case, the sequence was sampled from a regularly varying distribution, while in the latter case it was sampled from a distribution in the Gumbel MDA. This MDA consists of all kinds of distributions, including both light-tailed and heavy-tailed, but not regularly varying. The lognormal distribution, for example, is not regularly varying, but it is heavy-tailed and belongs to the Gumbel MDA, see Appendix~\ref{sec:estimators}. Yet if the task is to tell whether a sequence was sampled from a regularly varying distribution or not, then classifying the sequence as regularly varying based on a small value $\hat{\xi}$ increases the chances of false positives because this small $\hat{\xi}$ may be an estimate of $\xi=0$, in which case the source distribution is not regularly varying. To minimize the chances of such false positives, we do define the hardly-power-law threshold $\xi=\xiThreshold$, so that if at least one estimator thinks that $\xi\leq\xiThreshold$, we doubt that the sequence is coming from a regularly varying distribution. We set this threshold to $\xi=\xiThreshold$ here by selecting the largest value of $\gamma$ that is known to us to still matter. That is, we are unaware of any value of $\gamma$ that would correspond to any critical point, and that is larger than $\gamma=1+1/\xi=5$ in the Ising model on random graphs in~\cite{dommers2014ising}.

Power-law degree sequences whose distributions have divergent second moments, meaning $\gamma<3$ and $\xi>1/2$, are of particular interest to network science for a variety of reasons. For example, networks with such degree sequences are particularly robust thanks to the absence of the percolation threshold~\cite{dorogovtsev2008critical}, they are ultrasmall worlds versus small worlds~\cite{cohen2003scale}, the degree correlations in them are unavoidable due to structural constraints~\cite{boguna2004cut}, etc. Therefore, we also define a subclass of power-law degree sequences with divergent second moments of their distributions:
\begin{itemize}
  \item[$\rhd$] {\bf A power-law degree sequence has a divergent second moment (DSM)} if all the estimators return values of $\xi>1/2$.
\end{itemize}

We note that we do not put any restriction on how close to each other the estimated values reported by the different estimators must be in the definitions above. The main reason for that is that the speeds of convergence of these estimators are not known. They may converge to the true $\xi$ at different rates. However, as discussed above, if the data size is relatively large, and all the estimators report values $\hat{\xi} > \xiThreshold$, the chances that the degree sequence does not come from a regularly varying distribution ought to be slim. The power-law sequence definitions above represent one of many possible classification schemes. But if a degree sequence is classified as PL or DSM according to this scheme, and if all the estimators report values that are close to each other, then one can be confident about the true values of $\xi$ and $\gamma$. Unfortunately, there is no way to quantify this confidence. Since the convergence speeds are unknown, one cannot attach any rigorous bounds on how close the estimated values must be to yield any given accuracy in the estimation of the true $\gamma$. It is also important to recognize that these considerations apply not only to the estimators considered here, but also to any other estimator, including the PLFit~\cite{clauset2009power}, whose convergence speed on regularly varying distributions is not known.

Finally, if a network is simple unweighted undirected unipartite single-layer and static, then it has only one degree sequence associated with it, so that it is straightforward to call such a network power-law if its degree sequence is power-law. However, in more complicated situations, such as directed, multipartite, multilayer, multiplex, and/or temporal networks, there are not one but many degree sequences associated with the network. To call such a network power-law based only on one, or all, or some percentage (as in~\cite{broido2018scale}) of the total number of its degree sequences, is purely a matter of taste. What usually does matter is a specific question, e.g., the spread of a disease, posed for the network, and different degree sequences, e.g., in- versus out-degrees, are of interest for different questions. Therefore, we do not propose to classify such networks as power-law or not, and instead report the data for each degree sequence separately in the next section.

\section{Real-World Networks}\label{sec:real_networks}

\begin{figure*}[ht!]
  \centering
  \includegraphics[width=.8\textwidth]{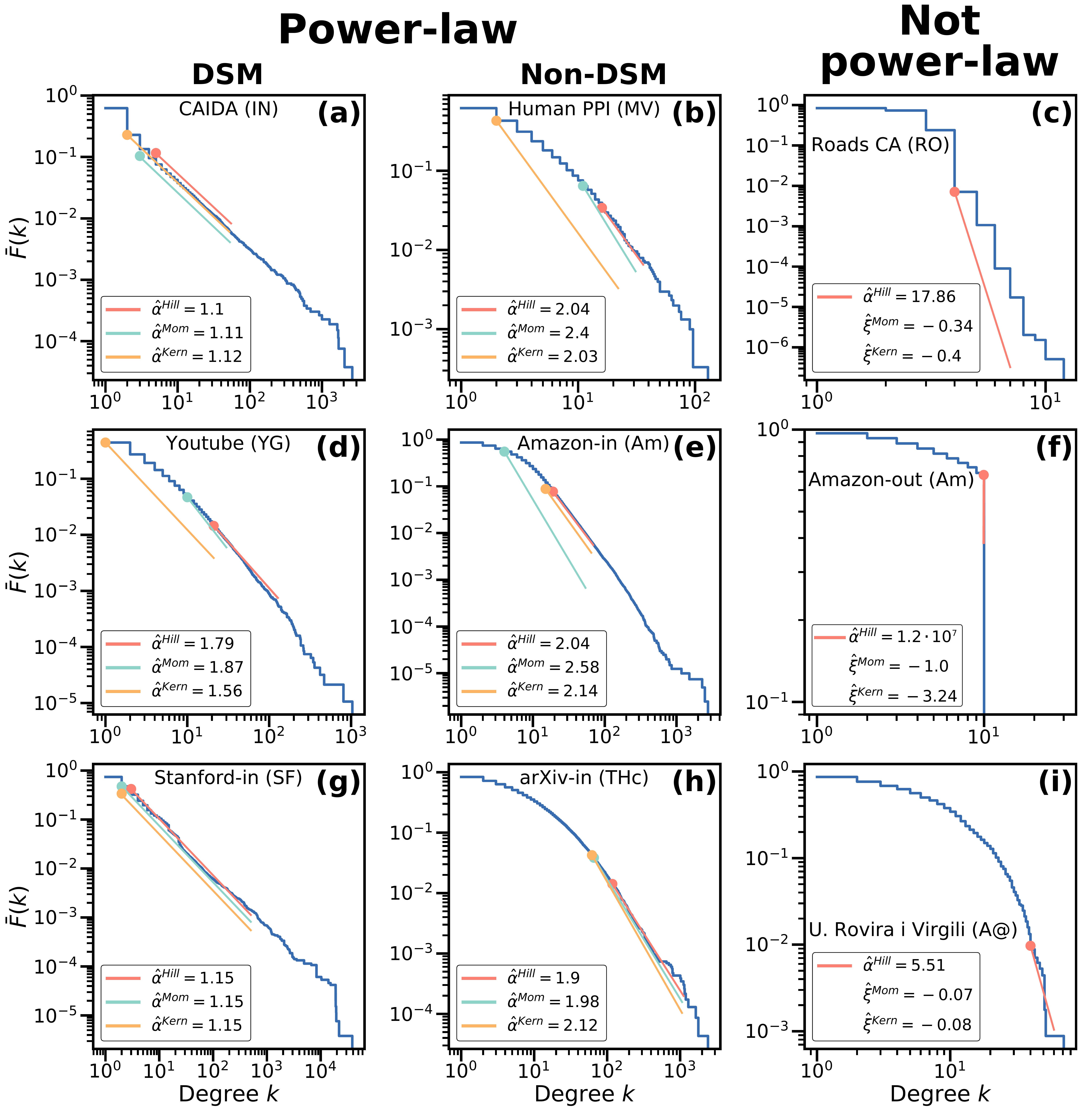}
  \caption{
  {\bf The results of the Hill, Moments, and Kernel estimators applied to the degree sequences of nine real-world networks.}
  The degree sequences belong to different classes defined in Section~\ref{sec:power-law-networks}: \textbf{(a,d,g) \textit{power-law DSM:} (a)~CAIDA}, the undirected network of the Internet at the autonomous system level; {\bf (d)~Youtube}, the user degree distribution of the bipartite network of Youtube users and their group memberships; {\bf (g)~Stanford}, the in-degree distribution of the directed network of hyperlinks between the WWW pages at the Stanford University website; \textbf{(b,e,h) \textit{power-law non-DSM:} (b)~Human PPI}, the undirected network of human protein-protein interactions; {\bf (e)~Amazon}, the in-degree distribution of the directed network of product recommendations at 
  Amazon; {\bf (h)~arXiv}, the in-degree distribution of the directed citation network of publications on High Energy Physics Theory at arXiv; \textbf{(c,f,i) \textit{not power-law:} (c)~Roads CA}, the undirected network of road intersections in the state of California; {\bf (f)~Amazon}, the out-degree distribution of the same network as in~(e); {\bf (i)~U.~Rovira i Virgili}, the undirected email communication network at the University of Rovira i Virgili.
  The shown network names are their codenames used in the KONECT database~\cite{kunegis2013konect}, and they also appear in Tables~\ref{tab:real-undir}-\ref{tab:real-bip}.
  Each panel shows the empirical complementary cumulative distribution functions (CCDFs) $\overline{F}(k)$ of the degree sequences on loglog scale. The straight lines visualize the estimated values of the CCDF exponents $\alpha=1/\xi$. The filled circles are the optimal values of the number of order statistics $\kappa^{\ast}$ found by the double bootstrap method. The estimators operate only on degrees larger than $\kappa^{\ast}$. The estimated values of $\alpha$ are $\hat{\alpha}=1/\hat{\xi}(\kappa^\ast)$, where $\hat{\xi}(\kappa)$ is the estimated value of the tail index $\xi$ as a function of $\kappa$. For non-positive values of $\hat{\xi}(\kappa^\ast)$, the $\hat{\alpha}$ is undefined, so that the legends in panels~(c,f,i) show $\hat{\xi}=\hat{\xi}(\kappa^\ast)$ instead. Hardly power-law examples are not shown as they are not particularly interesting, lying somewhere in between power-law and not power-law examples.
  }
  \label{fig:examples}
\end{figure*}

\begin{figure*}[ht!]
  \centering
  \includegraphics[width=.99\textwidth]{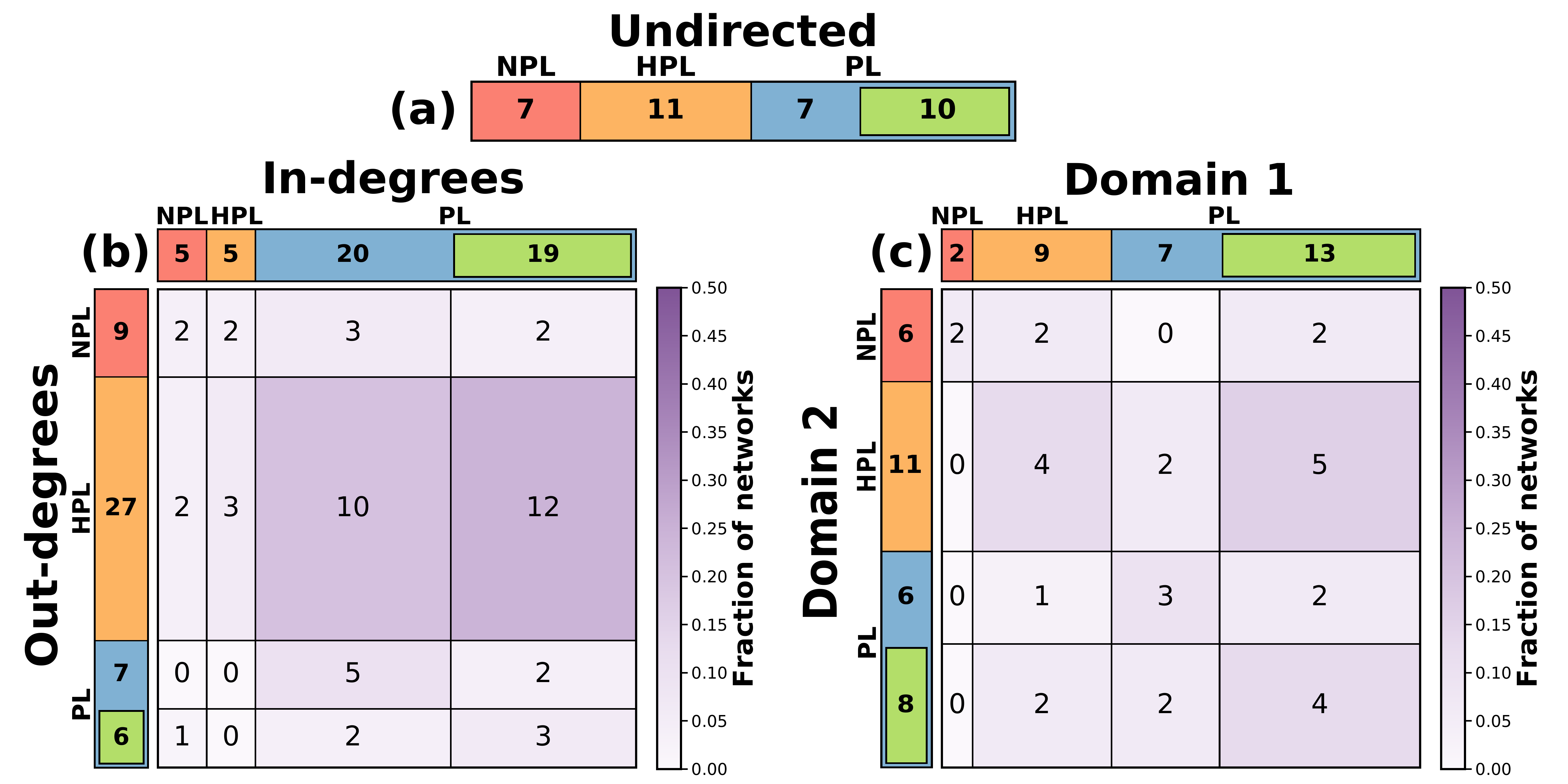}
  \caption{{\bf The breakdown of the degree sequences of the considered real-world networks into the three classes defined in Section~\ref{sec:power-law-networks}:} \textit{not power-law (NPL)}, \textit{hardly power-law (HPL)}, and \textit{power-law (PL)}, the latter containing the \textit{divergent second moment (DSM)} subclass, shown in green. The full data appears in Tables~\ref{tab:real-undir}-\ref{tab:real-bip}. The data is shown for \textbf{(a)}~undirected, \textbf{(b)}~directed, and \textbf{(c)}~bipartite networks. The numbers in the boxes are the numbers of networks falling within the corresponding (sub)class. In directed networks, the in- and out-degree sequences are classified separately. Similarly, in bipartite networks, the degree sequences of nodes of types 1 and 2 (domains 1 and 2) are classified separately as well. The numbers of networks with all possible combinations of the two classifications are then shown in the squares, along with the marginals outside of the squares. The color shading in the squares shows the fraction of directed and bipartite networks in each category.
  }
  \label{fig:real-nets}
\end{figure*}

Here we apply the Hill, Moments, and Kernel estimators to a collection of degree sequences in real-world networks from the KONECT database~\cite{kunegis2013konect}. The database is a curated collection of real-world networks categorized by several network attributes such as size, (un)directedness, (un)weightedness, etc. The database uses a unified edge list format to store the data, which simplifies the automation of data processing. Better yet, the database allows one to sort networks by their properties, and to filter out networks with possibly incomplete information. This is in contrast to other large network collections, such as ICON~\cite{icondata}, that link their entries to third-party databases of various formats and origins, which makes it quite difficult  to collect and process the data in an automated manner.

To streamline data processing, we do not consider networks in the database that are not downloadable in the KONECT edge list format. We also ignore temporal networks to avoid arbitrariness in selecting the temporal scale for data aggregation. Among database entries that possibly represent the same real-world network (for example, \href{http://konect.uni-koblenz.de/networks/dbpedia-link}{Wikipedia (EN) hyperlinks} and \href{http://konect.uni-koblenz.de/networks/wikipedia_link_en}{Wikipedia (EN) links}, both representing the English Wikipedia), we select only one entry. We also ignore networks that are marked as \emph{incomplete} in the database. Finally, since the estimation of $\xi$ cannot be reliable for networks of a small size, we only consider networks consisting of at least $n=1000$ nodes.

The KONECT database contains not only undirected networks, but also directed and bipartite. For the latter two classes, we obtain not one, but two degree sequences for each network: the in- and out-degree sequences for directed networks, and one degree sequence for each of the two types of nodes in bipartite networks. We also remove all self-loops and multi-edges from each collected network. After these filtering steps, we are left with $115$ networks of three different types: undirected (35), directed (49), and bipartite (31). The degree sequences of these networks are available at the software package repository~\cite{githubcode}.\\

We then feed the obtained degree sequences to the three estimators. Figure~\ref{fig:examples} shows the exponent estimation results that the estimators produce on some paradigmatic real-world networks in different domains, while Tables~\ref{tab:real-undir}, \ref{tab:real-dir} and \ref{tab:real-bip} contain the full lists of these estimations for the undirected, directed and bipartite networks respectively. We see that many networks that were found to be power-law in the past have degree sequences that are classified as such by these estimators as well. These include the Internet, WWW, human protein interactions, social group memberships, citations, and product recommendation networks. The other way around, degree sequences of networks that are known not to be power-law are classified as not power-law---the California road network or the out-degree distribution in the directed network of Amazon product recommendations, for instance.

We emphasize again the importance of using as many consistent estimators as possible: on any finite degree sequence, different estimators are not guaranteed to return the same $\xi$-estimation, as they may explore different parts of the distribution, especially if the slowly varying function $\ell(k)$ is nontrivial, Appendix~\ref{sec:real_estimation}. That is why we use the maximal subset of stable and efficient estimators for which the double bootstrap method to determine the optimal number of order statistics $\kappa^\ast$ is proven to be consistent.

Finally, Figure~\ref{fig:real-nets} summarizes the estimation results for $\hat{\gamma}=1+1/\hat{\xi}$ in Tables~\ref{tab:real-undir}-\ref{tab:real-bip} by classifying the degree sequences of all the considered networks into the not power-law (NPL), hardly power-law (HPL), and power-law (PL) classes, the latter containing the subclass of power-law networks with divergent second moments (DSM), defined in the previous section. We see that the percentages of power-law and DSM degree sequences in undirected networks are \undirectedPL and \undirectedDSM, respectively. Among the considered directed networks, \directedBothPL and \directedBothDSM have both in- and out-degree sequences that are power-law and DSM, while \directedAtLeastOnePL and \directedAtLeastOneDSM of these networks have either in- or out- degree sequence which is power-law and DSM, with a majority of those being in-degree sequences. The bipartite networks exhibit a similar picture: \bipartiteBothPL and \bipartiteBothDSM of them are power-law and DSM according to both types of nodes, while \bipartiteAtLeastOnePL and \bipartiteAtLeastOneDSM are power-law and DSM according to at least one type of nodes.

While one cannot directly compare these results to the ones in~\cite{broido2018scale}, they present quite a different picture than painted there.

\section{Conclusion and Discussion}

In summary, we call a distribution power-law if it is regularly varying. The pure power laws---the Pareto and zeta distributions---are a small subset of this more general, realistic, and well-studied class of distributions. This class constitutes the most inclusive theoretical framework capable of formalizing all the aspects of the ``\emph{straight line on log-log scale}'' intuition behind power-law observations in real-world networks. Utilizing the connection between this class of distributions and the maximum domain of attraction of the Fr\'{e}chet distribution in extreme value theory, we identify state-of-the-art statistical tools to estimate the tail exponent $\gamma$ in a given degree sequence. These are then deployed to design a classification scheme for degree sequences. The application of this scheme to a representative collection of degree sequences in real-world networks reveals that significant fractions of these networks have power-law degree sequences.

We note that the problem of classifying a given degree sequence as power-law or not has nothing to do with possible mechanisms that may lead to the emergence of power-law distributions in real data, and that are of great interest to network science in general. The reason why such mechanisms are a completely different subject altogether is simple. We can think of different mechanisms as different network models approximating stochastic processes that drive the evolution of real-world networks, and it is quite well known that completely different network models and thus completely different network formation mechanisms may lead to networks that have exactly the same degree distribution. That is, these networks may certainly be very different in all respects other than the degree distribution~\cite{orsini2015quantifying}. Therefore the question of what mechanism causes this or that degree distribution is completely irrelevant and ill-posed, as it is impossible in principle to infer it based only on the degree distribution.

The impossibility of hypothesis testing for regularly varying distributions is the reason why one cannot attach any statistical weight, such as a $p$-value, to the statement that a given finite sequence is regularly varying or not. Yet many other aspects of the current state of affairs in statistics related to detecting power laws in empirical network data do allow for improvement, so we comment on some of them here.\\

{\bf Fundamental limitations of estimators based on extreme value theory.} The existing consistent estimators of tail exponents are based on extreme value (EV) theory. These estimators cannot generally differentiate between heavy-tailed and light-tailed distributions, simply because the maximum domain of attraction of the Gumbel EV distribution contains distributions of both types---the light-tailed normal and heavy-tailed lognormal distributions, for example, Appendix~\ref{sec:estimators}. Since for many applications in network science an important question is whether a degree distribution is heavy- or light-tailed, versus regularly varying or not, it is of particular interest to devise other estimators, not based on EV theory, that would be capable of differentiating between these two types of distributions. Some initial steps in this direction have recently been made~\cite{hill2019measure}. Even more generally, it is often of interest whether a given degree sequence comes from a distribution with an infinite or finite second moment, versus power-law or not, so that it would be desirable to develop statistically consistent methods to test the infiniteness of the second moment. Such tests cannot be based on EV theory either.

Yet even for EV-based estimators there are many paths to improve their applicability and rigorous guarantees, which we discuss next.

{\bf The i.i.d.\ assumption.} First, it would be nice to relax the i.i.d.\ assumption for these estimators, and to prove their consistency in application to network models. The first step in this direction was made in~\cite{wang2018consistency}. We saw in our experiments in Appendix~\ref{sec:experiments} that all the considered estimators converge in all the considered network models, but there are no proofs for this convergence for any network model other than preferential attachment, to the best of our knowledge.

{\bf Convergence speed.} Another important open problem is the convergence speed. All we currently know is that the considered estimators converge to the true value of the power-law exponent $\gamma$ on sequences of random numbers of increasing length $n$ sampled i.i.d.'ly from any regularly varying distribution with this $\gamma$, but we do not know how quickly this convergence occurs, so that, for instance, there is no way to tell how close the estimates of different estimators on the same finite-$n$ sequence are supposed to be, even if this sequence is sampled i.i.d.'ly from a regularly varying distribution. The speed of this convergence depends not only on $\gamma$ but also on the slowly varying function $\ell(k)$. Thus, the problem is to obtain bounds, as functions of $n$, on the error of estimation of $\gamma$ for a given $\gamma$ and $\ell(k)$. Can such bounds be obtained for certain classes of $\ell(k)$s?

{\bf Not one sequence but many sequences.} More pertinent to networks, and also closely related to the convergence speed, is the question of dealing with not one sequence but with sequences of sequences. For some real-world networks there exist data not only on one snapshot of the network but also on a historical series of such snapshots. In this case, we have not one degree sequence but a series of degree sequences. One can then apply the estimators to these series, obtaining a series of estimates. Given such an estimate series and the length of the sequence attached to each element of the series, i.e., the network size, can one extract any additional information about the convergence of the series, and possibly devise some tests of the hypothesis that the series comes from a regularly varying generative process? To the best of our knowledge, these questions are wide open.

{\bf Integer-valued sequences.} Another network-specific issue is that degree sequences are integer-valued, while the considered EV estimators were designed with real-valued data in mind. As a consequence, these estimators are known to be unstable and to converge quite slowly in the case of integer-valued regularly varying distributions, Appendix~\ref{sec:real_estimation}. We circumvent this issue in our experiments by adding symmetric uniform noise, but it would be nice to design estimators that work reliably on integer-valued data directly.

{\bf The second order condition.} Another down-to-earth issue is the second order condition needed to prove the consistency of the double bootstrap method, Appendix~\ref{sec:real_estimation}. This condition is violated by pure power-law distributions, the Pareto and zeta distributions. We saw in our experiments in Appendix~\ref{sec:experiments} that the estimators equipped with the double bootstrap method converge in these cases as well, but there are no proofs of the consistency of the double bootstrap method in these cases.\\

{\bf Cutoffs.} Finally, we comment on the important issue of cutoffs that often causes much confusion. Here we have to differentiate between many possibilities of what a \emph{cutoff} might mean. Two classes of such possibilities are finite-size effects and true cutoffs. In the first case, a cutoff is just an illusion due to a finite sample size. 
If one samples an insufficiently large number of i.i.d.\ samples from a regularly varying distribution, the empirical distribution of these samples may appear to have a cutoff, even though the distribution we are sampling from does not have any cutoffs by definition of it being regularly varying. In simple terms, the tail of the empirical distribution may bend downwards, but this effect is simply due to the insufficient number of samples. In such cases, if one explores the empirical distribution tail, one finds only a few data points there. We note that EV theory gives not only the expected value of the maximum among these samples, but also the exact distribution of this properly rescaled maximum in the limit, Appendix~\ref{sec:estimators}.

In networks, however, this maximum can simply not be greater than the network size $n$ which is equal to the degree sequence length, and there are other kinds of degree correlations and degree sequence constraints that are forced by the network structure, many documented in~\cite{boguna2004cut}, for instance. These constraints can be such that the degree distribution does have true cutoffs. More generally, it may very well happen that the process driving the evolution of a given network is such that its degree distribution does converge to a distribution with true cutoffs, sharp or soft. Examples are the preferential attachment model with a preference kernel which is constant above a certain degree threshold~\cite[Section 4]{berger2005degree}, or the causal set of the universe~\cite{krioukov2012network}.

In these cases, one has to further differentiate between the following two possibilities. First, the cutoff can be constant, that is, independent of the network size/degree sequence length. In this case, the distribution is not regularly varying by definition, so that one cannot call it power-law. If one still wishes to estimate~$\gamma$ in samples from, for example, the distribution class $P(k)=\ell(k)k^{-\gamma}{\mathrm e}^{-k/c}$ where $\ell(k)$ is a slowly varying function, and $c>0$ a constant, then it is yet another open problem since EV-based estimators can clearly not be employed for this estimation, simply because the distributions in this class are not regularly varying. Neither are we aware of any consistent estimators that can do this estimation. In fact, such estimators are quite unlikely to exist, simply because this task appears to be ill-defined. Indeed, $\ell(k)$ can be arbitrarily bad for any finite~$k$. All we know about this function is that it varies slowly at infinity. But we also know from the shape of the distribution that it is exponential at infinity.

The other possibility is that the cutoff diverges with the network size. In this case we have a scenario that can possibly be modeled by random sequences of varying length $n$ sampled from a sequence of distributions parameterized by $n$. If their cutoff diverges with $n$, then the latter sequence may or may not converge to a regularly varying distribution in the $n\to\infty$ limit. In Appendix~\ref{sec:experiments} we considered an example of this sort, diverging natural exponential cutoffs, where the $n$-dependent distributions $P_n(k)=C_nk^{-\gamma}{\mathrm e}^{-k/n^\xi}$ do converge to the regularly varying Pareto distribution. We saw there that even in this case, the considered estimators converge to the true values of $\gamma$, even though the key assumptions behind the proofs of their convergence are violated. Proving the consistency of these and other estimators for sequences of random numbers sampled from sequences of distributions converging to regularly varying distributions, is thus yet another open problem.\\ 

Notwithstanding these open problems, the consistent estimators considered in this paper represent the current state of the art in the rigorous detection of power laws in empirical data. Their implementation is available in~\cite{githubcode}, and their application to a representative collection of degree sequences in real-world networks confirms that scale-free networks are not rare.

\begin{acknowledgments}
We thank S.~Resnick, S.~Foss, A.~Vespignani, A.~Broido, and J.~Kelley for useful discussions and suggestions. This work was supported by NSF Grant No.~IIS-1741355, ARO Grant Nos.~W911NF-16-1-0391 and W911NF-17-1-0491, NWO VICI grant No.~639.033.806, and the Gravitation {\sc Networks} grant No.~024.002.003.
\end{acknowledgments}

\newpage

\appendix

\section{Classes of distributions with heavy tails}\label{sec:heavy_tails}

Here we briefly review the taxonomy of distributions with heavy tails and provide the definition of the simplest and most frequently seen regularly varying distributions. All the distribution classes mentioned here are characterized by the key property that their tails decay more slowly than exponentially. The most general class is that of the heavy-tailed distributions. We note that ``fat-tailed'' distributions are also mentioned sometimes in the literature, but do not appear to have any rigorous definition. We focus on distributions with support on $\mathbb{R}_+$. Chapters~2 and~3 in~\cite{foss2011introduction} contain further details.

\subsection{Heavy-tailed distributions}

A distribution with cumulative distribution function (CDF) $F(x)$ is said to be \emph{heavy-tailed}~\cite[Theorem 2.6]{foss2011introduction} if its complementary CDF (CCDF) $\overline{F}(x)$ satisfies, for any $t > 0$,
\[
	\limsup_{x \to \infty} \mathrm{e}^{tx} \, \overline{F}(x) = \infty.
\]
In words, this definition literally says that the tail of the distribution $\overline{F}(x)$ decays more slowly than exponentially.

\begin{figure}
\centerline{\includegraphics{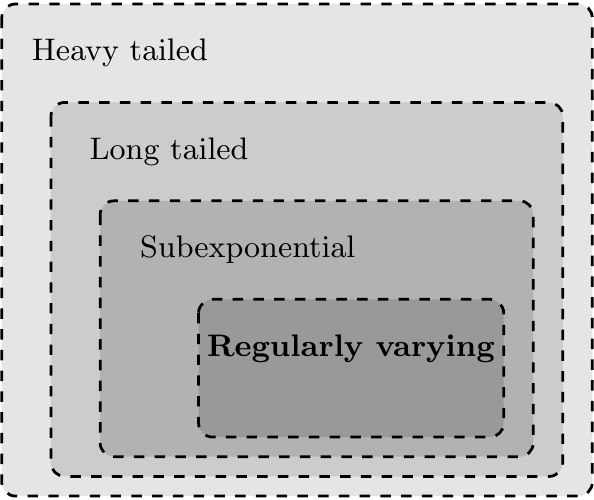}}
\caption{Schematic overview of the landscape of heavy-tailed distributions, containing regularly varying distributions.}
\label{fig:heavy_tailed_landscape}
\end{figure}

The class of heavy-tailed distributions is quite vast and general which makes it rather difficult to work with them in their full generality. Therefore, many different narrower and more tractable subclasses of heavy-tailed distributions have been defined and studied, see Figure~\ref{fig:heavy_tailed_landscape} for an overview of the landscape of heavy-tailed distributions. For completeness, we briefly discuss two important subclasses that encapsulate regularly varying distributions, which are our main interest.

\paragraph{Long-tailed distributions.}

A distribution with CDF $F(x)$ is called long-tailed~\cite[Definition 2.21]{foss2011introduction} if its CCDF satisfies, for any fixed $y > 0$,
\begin{equation}
	\lim_{x \to \infty} \frac{\overline{F}(x+y)}{\overline{F}(x)} = 1,
\end{equation}
meaning that any finite shift does not asymptotically affect the tail of the distribution. This property is nice and useful as, for instance, if $X$ is a random variable which has a long-tailed distribution, and $Y$ a random variable that only takes values on a finite set, then the tail of the distribution of $X + Y$ is asymptotically equivalent to that of $X$~\cite[Corollary 2.32]{foss2011introduction}.  

Long-tailed distributions are heavy-tailed~\cite[Lemma 2.17]{foss2011introduction}, but not all heavy-tailed distributions are long-tailed. A simple example of a heavy-tailed function which is not long-tailed is
\[
	f(x) = \sum_{k = 1}^\infty 2^{-k} \mathbbm{1}\{2^{(k-1)} \le x < 2^k\},
\]
where $\mathbbm{1}$ is the indicator function. Indeed, for any $t > 0$
\[
	\limsup_{x \to \infty} {\mathrm e}^{tx} f(x) \ge \limsup_{k \to \infty} {\mathrm e}^{t 2^k} 2^{-k} = \infty, 
\]
so that $f$ is heavy-tailed, but
\[
	\liminf_{x \to \infty} \frac{f(x+1)}{f(x)} \le \liminf_{k \to \infty} \frac{f(2^k + 1)}{f(2^k)} = \frac{1}{2} \ne 0,
\]
so that $f$ is not long-tailed.

\paragraph{Subexponential distributions.}

Let $(F\ast F)(x)$ be the convolution of CDF $F(x)$ with itself. That is, $F\ast F$ is the CDF of $X + X^\prime$, where $X$ and $X^\prime$ are independent random variables with CDF $F$. A distribution with CDF $F(x)$ is said to be subexponential~\cite[Definition 3.1]{foss2011introduction} if
\begin{equation}
	\lim_{x \to \infty} \frac{\overline{(F \ast F)}(x)}{\overline{F}(x)} = 2.
\end{equation}
This definition means that if $X$ and $X^\prime$ are independent samples from a subexponential distribution, then the CCDF of $X + X^\prime$ is asymptotically twice as large as the CCDF of the original distribution. This property implies, for instance, that if the sum $\sum_{i = 1}^n X_i$ of $n$ independent samples from a subexponential distribution exceeds some large threshold, then it is because just one $X_i$
has exceeded this threshold. This is in contrast to independent samples from a Poisson distribution, for instance, as their 
sums exceeding a large threshold do not contain, with high probability, any terms exceeding this threshold.

The class of subexponential distributions is contained in that of long-tailed distribution~\cite[Lemma 3.2]{foss2011introduction}, hence they are heavy-tailed. In fact, it is strictly contained. However, unlike the case for heavy-tailed versus long-tailed distributions, examples of long-tailed distributions that are not subexponential are more involved, see Section~3.7 in~\cite{foss2011introduction}.

Our main interest is in regularly varying distributions, which form a subclass of subexponential distributions~\cite[Theorem 3.29]{foss2011introduction}. This hierarchy endows regularly varying distributions with all the nice theoretical properties of the subexponential and long-tailed ones, but in contrast to these more general classes, regularly varying distributions are equipped with a concise and tractable representation that makes them very convenient to work with in statistical inference settings.

\subsection{Regularly varying distributions}

A function $f(x)$ is said to be \emph{regularly varying at infinity with index $\alpha$}~\cite{bingham1989regular,foss2011introduction} if there exists a slowly varying function $\ell(x)$, such that
\begin{equation}\label{eq:def_regularly_varying} 
	f(x) = \ell(x) x^{-\alpha},
\end{equation}
where a slowly varying function $\ell(x)$ is defined to be a function satisfying, for any $t > 0$,
\[
	\lim_{x \to \infty} \frac{\ell(tx)}{\ell(x)} = 1.
\]
The simplest examples of slowly varying functions are functions converging to constants or $\log^a(bx)$ for any $a\in\mathbb{R}$ and $b > 0$.

The full class of slowly varying functions is of course much richer, and it is fully characterized by Karamata's Representation Theorem~\cite[Corollary~2.1]{resnick2007heavy} stating that
\[
	\ell(x) = c(x) \exp\left\{\int_1^x t^{-1}\varepsilon(t) \, dt\right\},
\]
for some functions $c, \varepsilon : \mathbb{R}_+ \mapsto \mathbb{R}_+$ satisfying
\[
	\lim_{x \to \infty} c(x) = c \in (0,\infty), \quad \lim_{x \to \infty} \varepsilon(x) = 0.
\]
The theory of regular variations is a rich and well-developed one, and for further details we refer to~\cite{bingham1989regular}.

A distribution is \emph{defined} to be regularly varying if its CCDF $\overline{F}(x)$ is a regularly varying function. In Section~\ref{sec:definitions} we also define a \emph{power-law distribution} to be a regularly varying distribution.

We note that if the PDF of a distribution is regularly varying, then so is its CCDF with another slowly varying function~$\ell^\prime(x)$,
\[
	P(x) = \ell(x) x^{-\gamma}\Rightarrow\overline{F}(x) = \ell^\prime(x) x^{-\alpha}, \text{ where } \alpha=\gamma-1,
\]
according to Karamata's theorem~\cite[Theorem 2.1]{resnick2007heavy} in the case of continuous distributions, and to~\cite[Lemma~9.1]{hoorn2018average} in the case of discrete ones.
The converse is not generally true, and depends on the exact form of the slowly varying function $\ell^\prime(x)$. A simple example of a distribution whose PDF is not regularly varying but whose CCDF is, is given by the PDF $P(x) = c\,\sin(x)^2\,x^{-3}$ with support $x\ge1$, and $c$ the normalization constant. This PDF is not regularly varying since $\ell(x)=c\,\sin^2x$ is not slowly varying. However, the CCDF of this distribution $\overline{F}(x)=\ell^\prime(x)\,x^{-2}$, where $\ell^\prime(x)=(c/2)\left(\sin^2x + x\sin(2x) - 2x^2 \mathrm{Ci}(2x)\right)$ and $\mathrm{Ci}(x) = - \int_x^{\infty} \cos(t)/t \, dt$ is the cosine integral, is regularly varying because $\ell^\prime(x)$ is slowly varying: it converges to the constant $c/4$ at $x\to\infty$.

Another important property of regularly varying distributions is that if the sum $X+Y$ of two random variables is regularly varying, and $\lim_{z\to\infty}\overline{F}_Y(z)/\overline{F}_{X+Y}(z)=0$, then $X$ is regularly varying as well, and the tail exponents~$\gamma$ of $X+Y$ and $X$ are the same~\cite[Lemma~3.12]{jessen2006regularly}. In application to directed networks, this means for instance that if the total degree distribution is regularly varying and either the in-degree or out-degree distribution is not heavy-tailed, then the other distribution must be regularly varying with the same exponent as the total degree distribution.

As a subclass of heavy-tailed distributions, regularly varying distributions can model data with high variability, yet here we stress again that they are far from being as general as heavy-tailed distributions, which means, in particular, that if a given data fails to be regularly varying, it does not necessarily mean that it is not heavy-tailed or even subexponential. The simplest example of a subexponential distribution which is not regularly varying is the lognormal distribution. Yet on the other hand, regularly varying distributions are a vast generalization of pure power laws exclusively considered in~\cite{clauset2009power,broido2018scale}, i.e., of the Pareto distribution~\eqref{eq:pareto_def} if $x$ is continuous, or of the generalized zeta distribution~\eqref{eq:zeta_def} if $x$ is integer-valued.

\subsection{Simplest examples of regularly varying distributions}\label{ssec:examples_regularly_varying}

To make the definition \eqref{eq:def_regularly_varying} more concrete, here we give the simplest examples of regularly varying distributions, both continuous and integer-valued ones.

The simplest example is the continuous Pareto distribution with scale $x^\ast$ and shape $\alpha$, or exponent $\gamma=\alpha+1$:
\[
	P_{\mathrm{Pareto}}(x) = \begin{cases}
		\alpha (x^\ast)^\alpha x^{-\gamma} &\mbox{if } x \ge x^\ast, \\
		0 &\mbox{else.}
	\end{cases}
\]
Here the slowly varying function is simply the constant $\ell_{\mathrm{Pareto}}(x)=\alpha (x^\ast)^\alpha$, which does not vary at all.

There are two simple ways to turn a continuous regularly varying distribution into a integer-valued one, both of which again belong to the class of regularly varying distributions. In the first example we simply take the integer $k$ to be the floor of the continuous value $x$: $k = \lfloor x \rfloor$. If $x$ is Pareto-distributed, then since $P(k) = \overline{F}_{\mathrm{Pareto}}(k - 1) - \overline{F}_{\mathrm{Pareto}}(k)$, it follows that for all $k \ge \lfloor x^\ast \rfloor$
\begin{align*}
	P_{\mathrm{floorP}}(k) = \left(\frac{k - 1}{x^\ast}\right)^{-\alpha} 
	- \left(\frac{k}{x^\ast}\right)^{-\alpha}
	\hspace{-3pt}= \ell_{\mathrm{floorP}}(k)k^{-\gamma},
\end{align*}
where $\ell_{\mathrm{floorP}}(k)$ converges to $\alpha (x^\ast)^\alpha$ as $k \to \infty$. We note that in this example the slowly varying function is not a constant. Yet it approaches a constant asymptotically.

The second example is a mixed Poisson distribution~\cite{willmot2001mixed} with Pareto mixing. The easiest way to define a mixed Poisson distribution is via the procedure to sample from it: as its name suggests, first sample $x$ from the Pareto distribution, and then sample $k$ from the Poisson distribution with mean $x$. The resulting PDF of $k$ is thus 
\begin{align}
	P_{\mathrm{mPois}}(k)&=\int_{x^\ast}^\infty \frac{x^k{\mathrm e}^{-x}}{k!}P_{\mathrm{Pareto}}(x)\,dx\nonumber\\
	&= \alpha (x^\ast)^\alpha\frac{\Gamma(k + 1 - \gamma,x^\ast)}{\Gamma(k + 1)}=\ell_{\mathrm{mPois}}(k)\,k^{-\gamma},\label{eq:mixed-poisson}\\
    \ell_{\mathrm{mPois}}(k)&=\alpha (x^\ast)^\alpha k^\gamma \frac{\Gamma(k + 1 - \gamma,x^\ast)}{\Gamma(k + 1)},\nonumber
\end{align}
where $\Gamma(k,x)$ is the upper incomplete Gamma function. The function $\ell_{\mathrm{mPois}}(k)$ is slowly varying, and its $k\to\infty$ limit is, as in the previous example, $\alpha (x^\ast)^\alpha$.

Mixed Poisson distributions appear often as exact degree distributions in network models with hidden variables~\cite{boguna2003class}, also known in mathematics as inhomogeneous random graphs~\cite{bollobas2007phase}, or more generally, graphon-based $W$-random graphs~\cite{lovasz2006limits}. Both the expected value and the tail exponent of mixed Poisson $k$ are equal to those of Pareto $x$, versus floored Paretos in which the expected value of $k$ is $\langle k \rangle = \zeta\left(\langle x \rangle\right)$, where $\zeta$ is the Riemann zeta function.

\section{Consistent estimators for tail exponents of regularly varying distributions}\label{sec:estimators}

Here we give the definitions of the three consistent estimators of the tail of a regularly varying distribution that we use to infer the tail exponents in synthetic and real-world degree sequences. The two other consistent estimators that are also included in our software package~\cite{githubcode} are defined here as well.

Although we work only with regularly varying distributions, the used estimators are actually designed to estimate the \emph{index} of an \emph{extreme value distribution}. In fact, the consistency results are proven under the assumption that the distribution belongs to the \emph{maximum domain of attraction} of an extreme value distribution. It turns out that any regularly varying distribution satisfies this assumption. Therefore, we start with a brief review of extreme value distributions and their maximum domains of attraction, and then explain how these concepts are employed by the consistent estimators of tail exponents.

\subsection{Extreme value distributions and their maximum domains of attraction}\label{ssec:extreme_value}

Let $x_1, \dots, x_n$ be an i.i.d.\ sequence sampled from some distribution $P(x)$, and denote by $m_n = \max_{1 \le i \le n} x_i$ the largest value in the sequence. Extreme value theory is concerned with the properties of the distribution of $m_n$, whose CDF is given by the order statistics $F^n(x)$. The typical question is whether there is a non-degenerate limit law, i.e., a distribution which is not a delta function, for $\mu_n=(m_n - d_n)/c_n$ for some appropriately chosen $n$-dependent constants $c_n > 0$ and $d_n \in \mathbb{R}$. A degenerate limit for $\mu_n$ exists for any distribution as one can always select $d_n=0$ and any $c_n$ growing with $n$ faster than the expected value of $m_n$, in which case the distribution of $\mu_n$ would approach the delta-function distribution centered at zero. However, a non-degenerate limit exists~\cite[Theorem 3.1.3]{embrechts2013modelling} if and only if the CDF $F(x)$ of the distribution satisfies
\begin{equation}\label{eq:def_existence_mda}
	\lim_{x \to X_F} \frac{1 - F(x)}{F(x) - F(x-)} = 1 \quad \text{and} \quad F(X_F-) = 1,
\end{equation}
where $X_F = \sup\{x \, ; \, F(x) < 1\}$ is the right endpoint of the distribution, which can be infinite, and $F(x-) = \lim_{t \to \infty} F(x - 1/t)$ is the left limit of $F$ at $x$. In words, this requirement states that $F(x)$ must be sufficiently flat at its right end and must not jump there. Many distributions frequently appearing in practice do satisfy this requirement, but not all. Notable examples of distributions that do not satisfy it, are the Poisson~\cite[Example 3.1.4]{embrechts2013modelling} and geometric~\cite[Example 3.1.5]{embrechts2013modelling} distributions. Indeed, for a distribution with support on non-negative integers, the limit in~\eqref{eq:def_existence_mda} is equivalent to $\lim_{k \to \infty} {\overline{F}(k)}/{\overline{F}(k-1)} = 1$. For the Poisson distribution with mean $\lambda$, we have 
\[
	\frac{\overline{F}(k)}{\overline{F}(k-1)} \le 1 - \left(1 + \frac{\lambda}{k - \lambda}\right)^{-1}, 
\]
which tends to $0$ as $k \to \infty$, while for the geometric distribution with success probability $p$, $\overline{F}(k)/\overline{F}(k-1) \to 1 - p$, violating~\eqref{eq:def_existence_mda} in both cases.

If a distribution $P(x)$ does satisfy~\eqref{eq:def_existence_mda}, so that a non-degenerate limit distribution of $\mu_n=(m_n - d_n)/c_n$ does exist, this latter distribution $\mathcal{P}(\mu)$ is called an \emph{extreme value} distribution~\cite{resnick2013extreme,mikosch1999regular}. An important result~\cite{fisher1928limiting} (see also \cite[Proposition 0.3]{resnick2013extreme}) states that extreme value distributions are parameterized by an \emph{index} parameter $\xi \in \mathbb{R}$, and that the class of extreme value distributions consists of just three subclasses---Fr\'{e}chet, Gumbel, and Weibull distributions---corresponding, respectively, to $\xi > 0$, $\xi = 0$, and $\xi < 0$. The CDFs $\mathcal{F}(\mu)$ of these three distributions can be grouped into the CDF of the generalized extreme value distribution
\begin{align}
	\mathcal{F}(\lambda) &= {\mathrm e}^{-\lambda}, \text{ where}\nonumber\\
    \lambda &= \begin{cases}
          (1+\xi \nu)^{-1/\xi}, & \mbox{if } \xi\neq0, \\
          {\mathrm e}^{-\nu}, & \mbox{otherwise},
        \end{cases}\text{ where}\label{eq:def_frechet}\\
    \nu &= \frac{\mu-l}{s},\nonumber
\end{align}
where $l \in \mathbb{R}$ and $s > 0$ are known as, respectively, the location and scale parameters. The supports of the distributions are $\nu\geq-1/\xi$ for $\xi>0$, $\nu\leq-1/\xi$ for $\xi<0$, and $\nu\in\mathbb{R}$ for $\xi=0$.

A distribution $P(x)$ is said to belong to the \emph{maximum domain of attraction} (MDA) of an extreme value distribution $\mathcal{P}(\mu)$
if there exist $n$-sequences of constants $c_n > 0$ and $d_n \in \mathbb{R}$ such that the distribution of $\mu_n=(m_n - d_n)/c_n$ converges to $\mathcal{P}(\mu)$. The crucially important fact, originally proven in~\cite{gnedenko1943distribution}, is that the regularly varying distributions are exactly all the distributions comprising the MDA of the Fr\'{e}chet distribution, see also~\cite[Proposition 1.11]{resnick2013extreme} and \cite[Theorem 1.4.20]{mikosch1999regular}, so that any regularly varying distribution with PDF and CCDF tail exponents $\gamma$ and $\alpha$ belongs to the MDA of a Fr\'{e}chet distribution with index
\begin{equation}
\xi = \frac{1}{\alpha}=\frac{1}{\gamma - 1}.
\end{equation}

The sequences $d_n$ and $c_n$ in this regularly varying/Fr\'{e}chet case are, \cite[Proposition 1.11]{resnick2013extreme},
\begin{align*}
  d_n &= 0, \\
  c_n &= F^{-1}\left(1-\frac{1}{n}\right),
\end{align*}
where $F^{-1}$ is the inverse CDF of the distribution $P(x)$, while the location and scale parameters of the the Fr\'{e}chet distribution in~\eqref{eq:def_frechet} are
\begin{align*}
  l &= 1, \\
  s &= \xi,
\end{align*}
so that the distribution of the largest values $m_n$ among $n$ i.i.d.\ samples from any regularly varying distribution has the following limit upon rescaling by $c_n$:
\begin{equation}
	\lim_{n \to \infty} F^n(c_n \mu) = \mathcal{F}(\mu) = {\mathrm e}^{-\mu^{-1/\xi}}
\end{equation}
with support $\mu\geq0$.

If the distribution $P(x)$ is Pareto, for example, then
\begin{equation}
c_n = F^{-1}\left(1-\frac{1}{n}\right)=x^\ast n^\xi,
\end{equation}
related to the known observations that the expected maximum degree among $n$ samples from a power-law networks with exponent $\gamma$ is proportional to $n^{1/(\gamma-1)}$~\cite{boguna2004cut}.

We note that the expressions above specify not only the expected values but also the full limit distributions of such maxima. We also note that the mean of the limit Fr\'{e}chet distribution $\mathcal{F}(\mu)={\mathrm e}^{-\mu^{-1/\xi}}$ is $\langle\mu\rangle=\Gamma(1-\xi)$ if $\xi<1$ ($\gamma>2$), and that this mean is infinite if $\xi\geq1$ ($\gamma\leq2$), so that if $\gamma>2$ and $n$ is large, one can approximate the expected value of $m_n$ in Pareto by
\begin{equation}\label{eq:exact_expected_maximum}
  \langle m_n \rangle \approx \langle \mu \rangle c_n = \Gamma\left(\frac{\gamma-2}{\gamma-1}\right)x^\ast n^{1/(\gamma-1)}.
\end{equation}

To complete the picture of the classification of distributions based on their MDAs, the MDA of the Weibull distribution consists of all distributions with an upper-bounded support, $X_F < \infty$, and CCDFs satisfying $\overline{F}(X_F - 1/t) = \ell(t)t^{1/\xi}$ for $t\to\infty$, some slowly varying function $\ell(t)$, and $\xi<0$, which is the same $\xi$ as in~\eqref{eq:def_frechet}~\cite[Theorem 3.3.12]{embrechts2013modelling}. This requirement says that the CCDF approaches its right end as a regularly varying function. Examples are the uniform or beta distributions on $[0,1]$.

By exclusion, all other distributions satisfying~\eqref{eq:def_existence_mda} are in the MDA of the Gumbel distribution. However, there are more insightful characterizations of the Gumbel MDA (roughly, it consists of all von Mises functions and tail-equivalent distributions)~\cite[Theorems 3.3.26-3.3.27]{embrechts2013modelling}. Examples are the normal~\cite[Example 3.3.29]{embrechts2013modelling} and exponential~\cite[Example 3.3.19]{embrechts2013modelling} distributions, which are not heavy-tailed, but also heavy-tailed distributions that are not regularly varying---the subexponential lognormal distribution, for example~\cite[Example 3.3.31]{embrechts2013modelling}.

The key point, however, is that if a distribution is regularly varying with tail exponent $\gamma$, then it is in the MDA of the Fr\'{e}chet distribution with index $\xi=1/(\gamma-1)$ which all the following estimators actually estimate.

\subsection{Hill's estimator}

Hill's estimator~\cite{hill1975simple} was introduced to analyze the tail behavior of a distribution without any assumptions about its shape, other than that it belongs to the Fr\'{e}chet MDA.
Given an i.i.d.\ sample $x_i$, $i=1,\ldots,n$, and its order statistics $x_{(1)} \ge x_{(2)} \ge \dots \ge x_{(n)}$, the estimator is defined by
\begin{equation}\label{eq:def_hill_estimator}
	\widehat{\xi}^{\, \mathrm{Hill}}_{\kappa,n} = \frac{1}{\kappa} \sum_{i = 1}^{\kappa} \log\left(\frac{x_{(i)}}{x_{(\kappa+1)}}\right)
\end{equation}
Theorems~4.1 and~4.2 in~\cite{resnick2007heavy} prove that if $\kappa/n \to 0$ and $\kappa\to\infty$ as $n\to\infty$, then this estimator is statistically consistent, i.e., satisfies~\eqref{eq:def_consistency}, for any distribution in the MDA of the Fr\'{e}chet distribution. In other words, the estimator is statistically consistent for any regularly varying distribution with any tail exponent $\gamma>1$, or equivalently any index $\xi>0$.

\subsection{Moments estimator}

The moments estimator~\cite{dekkers1989moment} is a modification of Hill's estimator that is statistically consistent not only for distributions from the MDA of the Fr\'{e}chet distribution, but also for distributions from the MDAs of the Gumbel or Weibull distributions, i.e., for any $\xi \in \mathbb{R}$. To define it, denote
\[
	\widehat{\xi}^{\, \mathrm{Hill}, 2}_{\kappa,n} = \frac{1}{\kappa} \sum_{i = 1}^{\kappa} \left(\log\frac{x_{(i)}}{x_{(\kappa+1)}}\right)^2.
\]
	
With this notation, the Moments estimator is
\begin{equation}\label{eq:def_moments_estimator}
	\widehat{\xi}^{\, \mathrm{Moment}}_{\kappa,n} = \widehat{\xi}^{\, \mathrm{Hill}}_{\kappa,n} + 1 
	- \frac{1}{2} \left(1 - \frac{\left(\widehat{\xi}^{\, \mathrm{Hill}}_{\kappa,n}\right)^2}{\widehat{\xi}^{\, \mathrm{Hill}, 2}_{\kappa,n}}\right)^{-1}.
\end{equation}
Consistency of $\widehat{\xi}^{\, \mathrm{Moment}}_{\kappa,n}$ is proven in~\cite[Theorem 2.1]{dekkers1989moment}. It converges almost surely if $\kappa/n \to 0$ and $\kappa\to\infty$ as $n\to\infty$, and there exists a constant $\delta > 0$ such that $\log(n)^\delta/\kappa \to 0$.

\subsection{Kernel estimator}

Similar to the Moments estimator, the Kernel estimator~\cite{groeneboom2003kernel} is consistently applicable to distributions with any $\xi \in \mathbb{R}$. As its name suggests, the Kernel estimator uses a \emph{kernel}, which is a function $\phi \colon [0,1] \to [0,\infty)$ that can by chosen by the user, and that must satisfy a set of conditions for the estimator to be consistent~\cite{groeneboom2003kernel}. The estimator also employs a parameter $\lambda > 1/2$ to get rid of possible singularities. Finally, instead of using an integer-valued $\kappa$ to determine the range of the order statistics to consider for $\xi$-estimation, the estimator relies on a continuous bandwidth parameter $h > 0$ for that purpose. Thanks to this modification, as a function of $h$, the estimator tends to be smoother compared to the other estimators.

Given the chosen kernel $\phi$, denote $\phi_{h}(u) \coloneqq \phi(u/h)/h$. With this notation, the Kernel estimator is
\begin{align*}
	\widehat{\xi}_{h,n}^{\, \mathrm{Kernel}} &= \widehat{\xi}_{h,n}^{\, \mathrm{pos}} - 1 
	+ \frac{\widehat{q}_{h,n}^{\, (1)}}{\widehat{q}_{h,n}^{\, (2)}},\text{ where}\\
	\widehat{\xi}_{h,n}^{\, \mathrm{pos}} &= \sum_{i = 1}^{n-1} \frac{i}{n}\phi_{h}\left(\frac{i}{n}\right)
		\log\left(\frac{x_{(i)}}{x_{(i+1)}}\right),\\
	\widehat{q}_{h,n}^{\, (1)} &= \sum_{i = 1}^{n-1} \left(\frac{i}{n}\right)^{\lambda} \phi_{h}\left(\frac{i}{n}\right)
		\log\left(\frac{x_{(i)}}{x_{(i+1)}}\right),\\
	\widehat{q}_{h,n}^{\, (2)} &= \sum_{i = 1}^{n-1} \frac{\partial}{\partial u}[u^{\lambda + 1}\phi_{h}(u)]_{u=i/n}
		\log\left(\frac{x_{(i)}}{x_{(i+1)}}\right).
\end{align*}
The consistency of this estimator for $n\to\infty$, $h\to0$, and $hn\to\infty$ is proven in~\cite{groeneboom2003kernel}. 

For the experiments in this paper, which are also the default settings in~\cite{githubcode}, we prepare a list of fractions of order statistics $h_1,\ldots,h_s$. The estimator is then evaluated at each $h$-value $h_i$, $i=1,\ldots,s$. These fractions $h_i$ are logarithmically spaced in the interval $[1/n,1]$, where $n$ is the sequence length. The number of different $h$-values is set to $s=[0.3n]$. The logarithmic binning is chosen to scan the tail of the degree sequence more densely, while the choice of~$s$ guarantees that the sample sizes used in the double bootstrap procedure described in Appendix~\ref{ssec:finding_good_k} exceed~$s$, so that kernel smoothing is applied to both bootstrap samples as well. For~$\lambda$, we use the setting in~\cite{groeneboom2003kernel} where $\lambda = 0.6$.  In our software package~\cite{githubcode} the values of $\lambda$ and $s$ can be changed to any other values $\lambda>1/2$ and $s>0$. We note that the estimator is proven to be consistent for \emph{any} choice of $s$, $h_i$, and $\lambda$ satisfying the requirements above. Package~\cite{githubcode} also implements the bi- and tri-weight kernels from~\cite{groeneboom2003kernel}:
\begin{align*}
	\phi^{(1)}(u) &= \frac{15}{8}(1-u^2)^2,\\
	\phi^{(2)}(u) &= \frac{35}{16}(1-u^2)^3,
\end{align*}
where $\phi^{(1)}$ is used for the tail estimation, and the combination of $\phi^{(1)}$ and $\phi^{(2)}$ is used to find the optimal $h^\ast$ as described in Section~\ref{ssec:finding_good_k}. Once such an $h^\ast$ is found, the value of $\kappa^\ast$ is set to $\lfloor n h^{\ast} \rfloor$ in~\cite{githubcode}.

\subsection{Smooth Hill estimator}

Although Hill's estimator is consistent, it, as a function of the number of order statistics $\kappa$, can be highly irregular for finite-size data samples. The plots of such functions are even known as Hill Horror Plots, Section~4.4.2 in~\cite{resnick2007heavy}. These horrors make it essentially impossible to examine these plots in search of the stable regime of $\widehat{\xi}^{\, \mathrm{Hill}}_{\kappa,n}$, i.e., the region of $\kappa$s where $\widehat{\xi}^{\, \mathrm{Hill}}_{\kappa,n}$ is approximately constant. The value that the estimator yields in this constant regime is then one's best estimate of $\xi$, but if these plots are highly irregular, then this estimation procedure is unavoidably subjective. Even though no results presented in this paper rely on such subjective manipulations---instead we rely on the statistically consistent double bootstrap method to find~$\kappa^\ast$, Section~\ref{ssec:finding_good_k}---in practice one may wish to investigate such plots to get deeper insight into the data at hand. To this end, one usually uses either the smoothed version of Hill's estimator or Pickands estimator, which are both included in~\cite{githubcode}.

The smooth Hill estimator~\cite{resnick1997smoothing} is defined for any integer $r \ge 2$, which is a parameter, by
\begin{equation}\label{eq:def_smooth_hill_estimator}
	\widehat{\xi}^{\, \mathrm{smooH}}_{\kappa,n} = \frac{1}{(r - 1)\kappa} \sum_{j = \kappa + 1}^{r\kappa} \widehat{\xi}^{\, \mathrm{Hill}}_{j,n},
\end{equation}
which is just an average of Hill's estimators over the range $[\kappa+1, r\kappa]$. This estimator is also statistically consistent, for any $r$, as proven in~\cite{resnick1997smoothing}. The practical advantage of this smooth estimator compared to the original Hill's estimator is that by averaging the latter, the former suppresses its erratic behavior, making it easier to identify its stable region.

\subsection{Pickands estimator}

The Pickands estimator~\cite{pickands1975statistical} is defined by 
\begin{equation}\label{eq:def_pickands_estimator}
	\widehat{\xi}^{ \, \mathrm{Pickands}}_{\kappa,n} = \frac{1}{\log 2} \log\left(\frac{x_{(\kappa)} - x_{(2\kappa)}}{x_{(2\kappa)} - x_{(4\kappa)}}\right).
\end{equation}
Consistency of the estimator is proven also in~\cite{pickands1975statistical}. Similarly to the Moments and Kernel estimators, it is consistently applicable to distributions in the MDAs of extreme value distributions with any $\xi\in\mathbb{R}$. In practice this estimator provides a simple way to check whether the assumption that the data comes from a regularly varying distribution makes sense. Specifically, if the function $\widehat{\xi}^{ \, \mathrm{Pickands}}_{\kappa,n}$ of $\kappa$ is all negative, then this assumption can hardly be true. 

In contrast to the other estimators, the Pickands estimator has an issue dealing with integer-valued data containing ties. For instance, if there are many data points with the same value (many nodes with the same degree), it can happen that for some $\kappa$, $x_{(2\kappa)} = x_{(4\kappa)}$, in which case $\widehat{\xi}^{ \, \mathrm{Pickands}}_{\kappa,n}$ is undefined. This drawback can however be remedied, in a provably consistent manner, by adding uniform noise to the integer-valued data, as explained in Section~\ref{ssec:integer_data}.

It is known that in practice the Pickands estimator is quite volatile as a function of the number of order statistics $\kappa$, and that it has large asymptotic variance~\cite{segers2005generalized} and poor efficiency~\cite{groeneboom2003kernel}. Attempts to cure this poor behavior resulted in a number of different versions of generalized Pickands estimators~\cite{pereira1994second,falk1994efficiency,alves1995estimation,drees1995refined,yun2002generalized,segers2005generalized}, all of them using linear combinations of log-spacings of the order statistics $\kappa_{(1)},\ldots,\kappa_{(n)}$. Yet, to the best of our knowledge, the consistency of the double bootstrap method has been proven~\cite{draisma1999bootstrap} for only one version, the one defined in~\cite{pereira1994second}, so that we implemented only this version in~\cite{githubcode}.

\section{Estimating the tail exponent of an empirical degree sequence}\label{sec:real_estimation}

Here we discuss the technical details concerning the application of the consistent estimators discussed in the previous section to empirical degree sequences coming from either synthetic or real-world networks.

\subsection{Finding the optimal number of order statistics}\label{ssec:finding_good_k}

All the estimators in Section~\ref{sec:estimators} depend on the number of order statistics $\kappa$. That is, all the estimators operate only on the $\kappa$ largest-value data samples (degrees). The consistency of all the estimators is proven only in the limit of both $\kappa$ and $n$, the number of samples (nodes), tending to infinity. Therefore, when applied to a finite empirical degree sequence, these estimators have the value of $\kappa$ as a free parameter. The main focus of this section is the double bootstrap method that algorithmically identifies an optimal $\kappa$-value $\kappa^\ast$ in a statistically consistent manner, meaning that the value of $\hat{\xi}_{\kappa^\ast, n}$ estimated by these estimators with $\kappa=\kappa^\ast$ provably converges to the true value of $\xi$ as $n\to\infty$.

The identification of an optimal value $\kappa^{\ast}$ has been an active research topic in extreme value statistics for several decades~\cite{draisma1999bootstrap,qi2008bootstrap}. The existing methods for choosing $\kappa^{\ast}$ can be roughly split into two classes: (1)~heuristic approaches that propose to study tail index estimates plotted as functions of $\kappa$, and (2)~theoretical approaches based on the minimization of the asymptotic mean-squared-error (AMSE) of the estimator. 

The heuristic methods mainly consider various ways of identifying regions of $\kappa$ where estimators show \emph{stable behavior}, i.e., where the estimator plot is relatively flat as a function of $\kappa$. Examples of such approaches are the automated eyeball method~\cite{resnick1997smoothing}, or picking a fixed small percentage (typically $5\%$ or $10\%$) of the largest-value data samples. Such methods, involving (semi-)subjective ad-hoc choices, may not be robust.

The main idea behind the theoretical methods is as follows. Suppose $x_1, \dots, x_n$ is an i.i.d.\ sequence sampled from a distribution that belongs to the domain of attraction of the generalized extreme value distribution~\eqref{eq:def_frechet} with a given $\xi$. Denote by $\widehat{\xi}_{\kappa,n}$ the estimated value of $\xi$ returned by a given estimator applied to the $\kappa$ largest elements in this sequence. Observe that since the sequence is random, $\widehat{\xi}_{\kappa,n}$ is a random variable. Define the \emph{asymptotic mean squared error} between the true and estimated $\xi$s as
\begin{equation}\label{eq:amse}
    AMSE(n, \kappa) = \mathbb{E}\left[ (\widehat{\xi}_{\kappa,n} - \xi)^2 \right].
\end{equation}
The main goal is to find the optimal $\kappa$-value $\kappa^{\ast}$ that minimizes this error:
\begin{equation}\label{eq:AMSE-minimum}
    \kappa^{\ast} = \argmin_{\kappa} AMSE(n, \kappa).
\end{equation}

To estimate $\kappa^{\ast}$ in this paper we use the AMSE-based double bootstrap method developed in~\cite{danielsson2001using,draisma1999bootstrap,qi2008bootstrap,groeneboom2003kernel} because of its proved consistency, stability, and applicability to the considered estimators. The method finds a consistent optimal value $\kappa^\ast$ for a given consistent estimator by employing not only this estimator, but also another consistent estimator. The two estimators are applied to two collections of bootstrap samples from the original data, estimating $\xi$ at all possible values of $\kappa$ in these collections, and the value $\kappa^\ast$ is then determined as the value of $\kappa$ at which the two estimators agree most in their estimation of $\xi$ according to the empirical AMSE evaluated on the bootstrap collections.

Specifically, the double bootstrap method operates using the following steps with two parameters: $r$ denotes the number of bootstrap samples, and $t \in (0,1)$ defines the first and second bootstrap sample sizes as $n_1 = \sqrt{t}n$ and $n_2 = tn$, where $n$ is the original sequence length. In all the experiments in this paper, and in the software package~\cite{githubcode}, these parameters are set to $r=500$ and $t=1/2$ by default, so that the size of the second bootstrap sample is $n_2 = n/2$.
\begin{enumerate}
    \item{Sample $r > 1$ bootstrap samples of size $n_1 = \left[\sqrt{t}n\right]$ from the original data.}
    \item{Using the two consistent estimators, estimate $\xi^{(1)}_{\kappa_1, j}$ and $\xi^{(2)}_{\kappa_1, j}$ for each value of $\kappa_1 = 1,\ldots,n_1$ in each bootstrap sample $j = 1,\ldots,r$.}
    \item{Find $\kappa_1^{\ast}$ that minimizes the empirical AMSE between the two estimates with respect to the $r$ bootstrap samples, i.e.,
    \begin{equation*}
        \kappa_1^{\ast} = \argmin_{\kappa_1} \frac{1}{r} \sum_{j = 1}^{r} (\xi^{(1)}_{\kappa_1, j} 
        - \xi^{(2)}_{\kappa_1, j})^2.
    \end{equation*}}
    \item{Repeat the same procedure for a smaller bootstrap sample size $n_2 = \left[tn\right]$ and find $\kappa_2^{\ast}$ in the same manner.}
    \item{The optimal value of $\kappa$ for the original data is given by:
    \begin{equation}\label{eq:kappa-ast-dbs}
        \kappa^{\ast} = A(\kappa_1^\ast, n_1, n) \frac{(\kappa_1^\ast)^2}{\kappa_2^\ast},
    \end{equation}
    where $A(\kappa_1^\ast, n_1, n)$ is a pre-factor that depends on $\kappa_1^\ast, n_1, n$, and whose exact form depends on the two estimators used.}
\end{enumerate}
Following the derivations in~\cite{danielsson2001using,draisma1999bootstrap,qi2008bootstrap,groeneboom2003kernel}, we use the following combinations of consistent estimators for the double bootstrap procedure applied to the Hill, Kernel, and Moments estimators: (1)~the 1st (Hill) and 2nd moment estimators for the Hill double bootstrap; (2)~the 2nd and 3rd moment estimators for the Moments double bootstrap; (3)~the bi-weight and tri-weight kernel estimators for the Kernel double bootstrap. We note that in principle any combination of consistent estimators can be used in the double bootstrap method, but proofs of the consistency of such combinations must be carried out for each combination, so that we use the combinations that are already proven to be consistent and optimal.

We also note that these proofs are based on an additional assumption that the regularly varying distribution of the samples satisfies the \emph{second order condition}~\cite{haan1996generalized}, \cite[Definition 2.3.1]{haan2007extreme}. This condition is often invoked to prove asymptotic normality of estimators~\cite{haan1996second,gomes2015extreme}, but it is known to be either difficult or impossible to check in real-world data~\cite{jordanova2015flexible}. To define it for a given distribution with CDF $F(x)$, let $U(x) = F^{-1}(1-1/x)$ be the inverse of the CCDF $1 - F(x)$. If the distribution is in an MDA of some extreme value distribution, then it is known~\cite[Theorem 1.1.6]{haan2007extreme} that there exists a positive function $a(x)$ such that, for any $t>0$,
\[
	\lim_{x \to \infty} \frac{U(tx) - U(x)}{a(x)} = b_\xi(t) \coloneqq
\begin{cases}
  \frac{t^{\xi} - 1}{\xi}, & \mbox{if } \xi\neq0, \\
  \log t, & \mbox{otherwise}.
\end{cases}
\]
The second order condition concerns the scaling of $(U(tx) - U(x))/a(x) - b_\xi(t)$ as $x \to \infty$. The distribution is said to satisfy the second order condition if there exist functions $A(x)$ with $\lim_{x \to \infty} A(x) = 0$ and a non-degenerate $H(t)\neq cb_\xi(t)$ for any $c\neq0$, such that for any $t>0$
\begin{equation}\label{eq:def_second_order_condition}
	\lim_{x \to \infty} \left(\frac{U(tx) - U(x)}{a(x)} - \frac{t^{\xi} - 1}{\xi}\right)/A(x) = H(t).
\end{equation}
A simple example of a regularly varying CDF that satisfies the second order condition is
\[
	F(x) = 1 - x^{-\alpha} - d x^{-\delta}, 
\]
where $d > 0$, $\delta > \alpha > 0$, and $x\geq x^\ast$, where $x^\ast$ is the root of $F(x)=0$. The simplest example of a distribution that does not satisfy the second order condition is a Pareto distribution. To see this, note that in case of Pareto $U(x) = x^\ast x^{1/\alpha}$, so that $a(x) = \alpha^{-1}x^\ast x^{1/\alpha}$, and
\[
	\frac{U(tx) - U(x)}{a(x)} = \alpha (t^{1/\alpha} - 1) = \frac{t^\xi - 1}{\xi}.
\] 
Hence the left hand side in~\eqref{eq:def_second_order_condition} is always zero, meaning that no non-degenerate function $H(t)$ exists. 

Since the proofs of consistency of the double bootstrap method rely on the second order condition, nothing can be said regarding the convergence and consistency of the considered estimators equipped with the double bootstrap method, if they are applied to sequences sampled from distributions that do not satisfy the second order condition. However, in our experiments we find that even in these cases the double bootstrap procedure performs well, and the values of $\widehat{\xi}_{\kappa^\ast,n}$ quickly converge to the true~$\xi$s as $n \to \infty$ in most such cases, Section~\ref{ssec:toy-models}.

Further technical details on the double bootstrap procedure for the Hill estimator can be found in~\cite{danielsson2001using,qi2008bootstrap}, for the Moments estimator in~\cite{draisma1999bootstrap}, and for the Kernel estimator in~\cite{groeneboom2003kernel}, where the consistency of double bootstrapping applied to these estimators is also proven.

\subsection{Working with integer data}\label{ssec:integer_data}

A common issue with all the known consistent estimators is their instability, i.e., erratic behavior of $\widehat{\xi}_{\kappa,n}$s as functions of sampled sequences and the number of order statistics $\kappa$, on integer-valued sequences~\cite{velthoen2014estimation,matsui2013estimation}, which is the case with degree sequences. For instance, just rounding samples in sequences sampled from a continuous regularly varying distribution makes the estimators unstable~\cite{matsui2013estimation}, even though such rounded sequences are still regularly varying with the same exponent, Section~\ref{ssec:examples_regularly_varying}. In other words, the estimators remain consistent on integer-valued regularly varying distributions, but they tend to be unstable and 
exhibit slow convergence in such cases.

\begin{figure}[ht!]
  \centering
  \includegraphics[width=.49\textwidth]{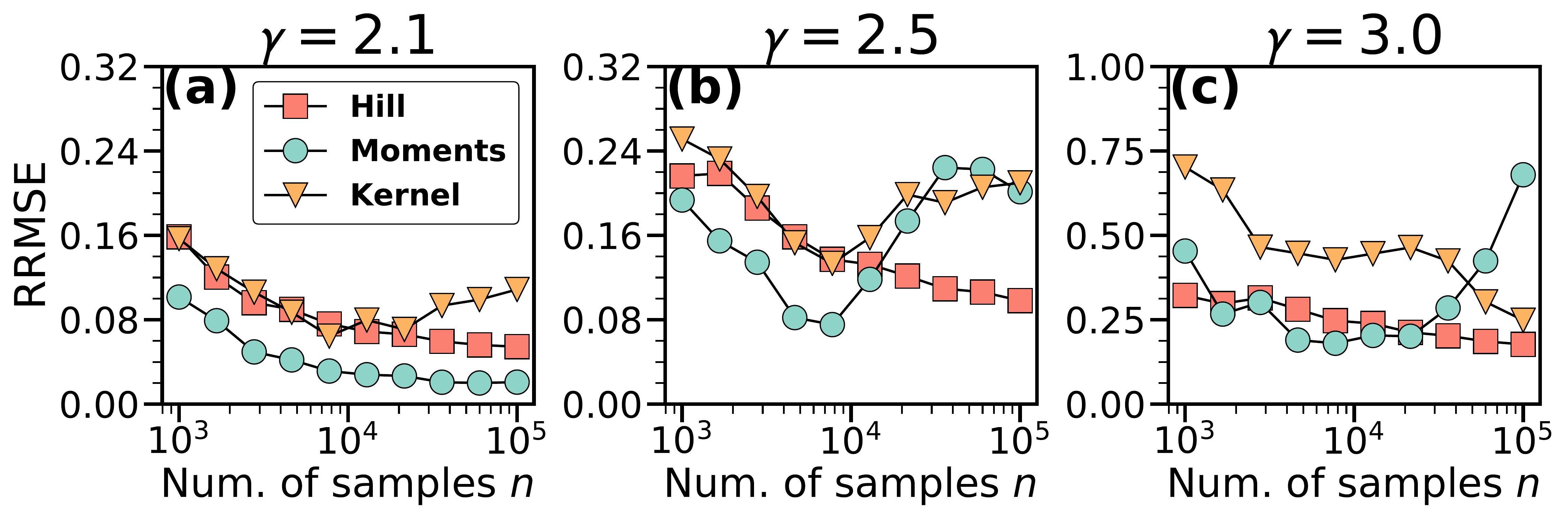}
  \caption{The relative root mean squared error (RRMSE)~\eqref{eq:rrmse-xi} of the three estimators for the i.i.d.\ sequences of random integers $k$ of varying length $n$ sampled from the Zeta distributions with different values of exponent $\gamma$, Section~\ref{ssec:toy-models}. The integers are fed to the estimators as is, without adding the uniform symmetric noise. The RRMSE is larger than with noise, cf.~Section~\ref{ssec:toy-models}.}
  \label{fig:zeta-no-noise}
\end{figure}

To resolve this issue we add uniform symmetric noise to the integer-valued sequences $x_i$, $i=1,\ldots,n$, that is, to all sequences considered in this paper. Specifically, instead of applying the estimators to $x_i$, we apply them to $y_i=x_i+u_i$, where $u_i$s are i.i.d.\ samples from the uniform distribution on $[-1/2,1/2]$. This does not affect the tail exponent: if $x$ is a regularly varying random variable with tail index $\xi > 0$, and $u$ is a uniform random variable on $[-1/2\cdot10^{-p}, 1/2\cdot10^{-p}]$, where $p \geq 0$, then $x + u$ is also regularly varying with the same exponent~\cite[Theorem~5.3.1]{velthoen2014estimation}. Adding such noise greatly improves the stability and convergence of the estimators, see Fig.~\ref{fig:zeta-no-noise} and compare it with Section~\ref{ssec:toy-models}.

\subsection{Example of the estimator operation using the double bootstrap method}\label{ssec:discrepancies-explained}

\begin{figure*}[t]
  \centering
  \includegraphics[width=.85\textwidth]{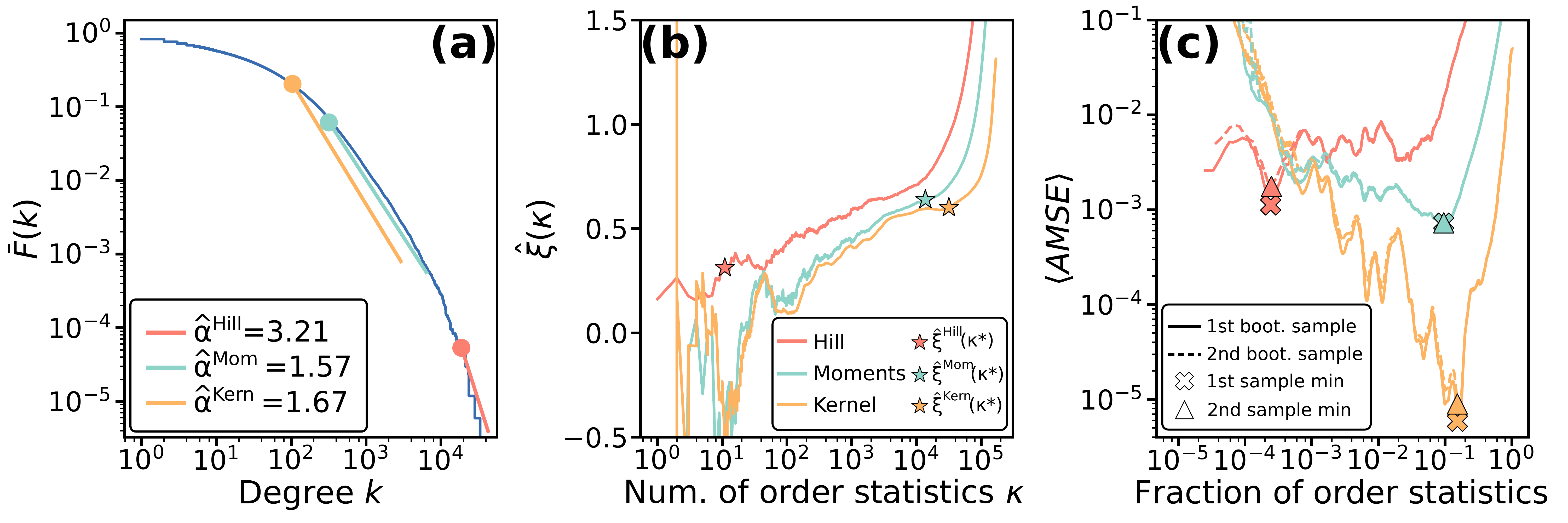}
  \caption{An example of the estimator operation on the in-degree sequence of the \textit{Libimseti} network, an online dating social network.
  Panel~\textbf{(a)} follows the same style and notations as in Fig.~\ref{fig:examples}.
  Panel~\textbf{(b)} shows the estimated values $\hat{\xi}(\kappa)$ of the extreme value index $\xi$ as a function of the number of order statistics $\kappa$. The filled symbols correspond to $\hat{\xi}(\kappa^{\ast})$, where $\kappa^{\ast}$ is the optimal value of $\kappa$ found by the double bootstrap algorithm.
  Panel~\textbf{(c)} shows the averaged asymptotic mean squared error (AMSE), defined in~\eqref{eq:amse}, as a function of the fraction $f=\kappa/n$ of the number of order statistics used in the two bootstrap samples of sizes $[n/\sqrt{2}]$ and $[n/2]$, corresponding to the ``1st boot.''\ and ``2nd boot.''\ curves in the figure. Their minima, given by~\eqref{eq:AMSE-minimum}, are shown by the triangle and cross markers. The values of the number of order statistics $\kappa$ corresponding to these minima---$\kappa_1^{\ast}$ and $\kappa_2^{\ast}$ for the 1st and 2nd bootstrap samples, respectively---are then used to identify the optimal value $\kappa^{\ast}$ via~\eqref{eq:kappa-ast-dbs}.
  }
  \label{fig:diagn-plot}
\end{figure*}

\begin{figure*}[ht!]
  \centering
  \includegraphics[width=.99\textwidth]{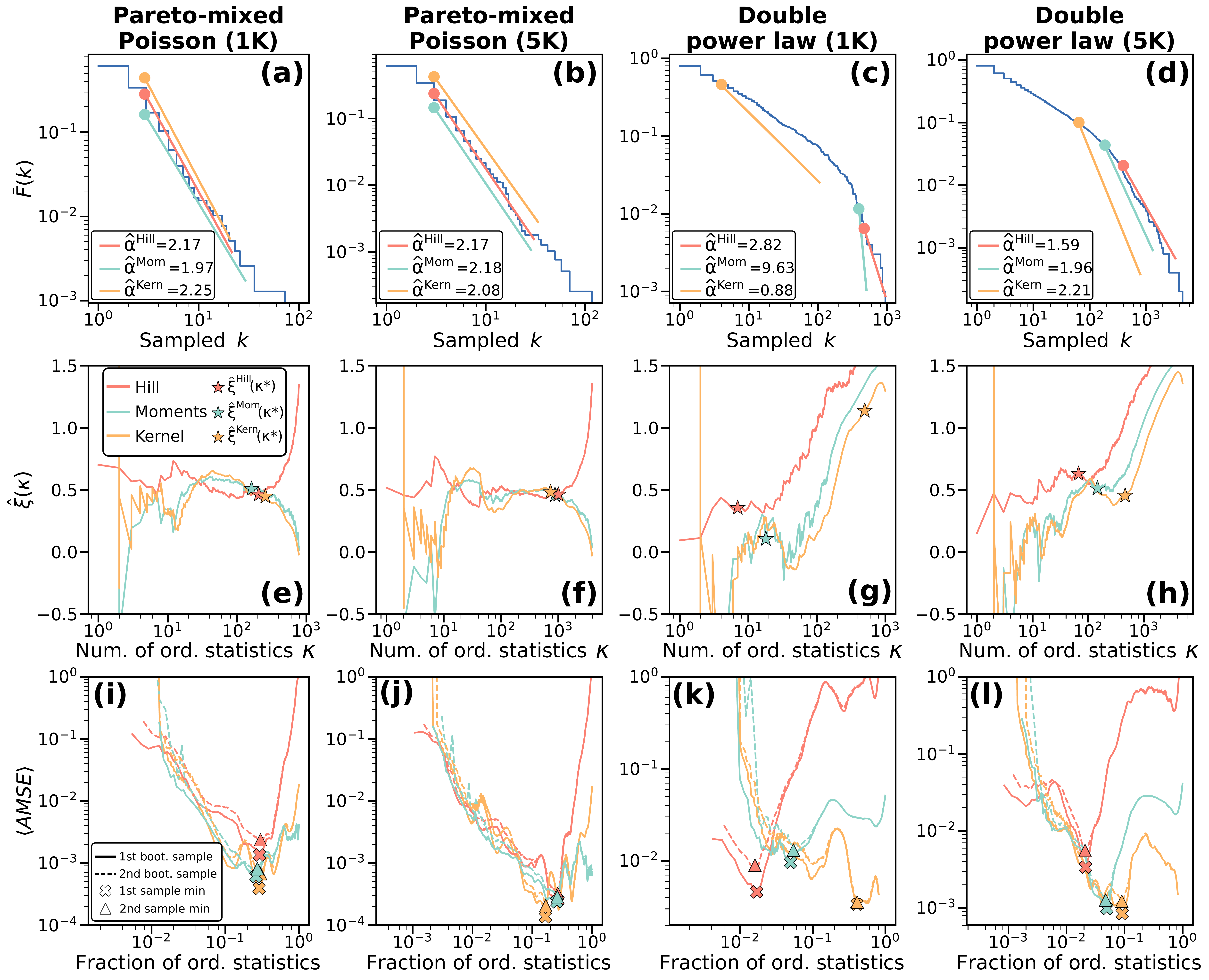}
  \caption{An example of convergence of the estimators on the i.i.d.\ sequences of two different sizes, ``small'' $n_{s}=1,000$ and ``large'' $n_{l}=5,000$, sampled from 
  two regularly varying distributions with the same tail exponent $\gamma=3$, but with different slowly varying functions $\ell(k)$: the Pareto-mixed Poisson distribution and double power law, see Section~\ref{ssec:toy-models} for details. The parameters for the Pareto-mixed Poisson are $\gamma = 3$ and $x_0=1$, while for the double power law, they are: $\gamma = 3$, $\gamma_0 = 1.5$, $c = 500$, $r = 0.1$, and $x_0=1$. The first, second, and third rows follow the same style and notations as panels~(a,b,c) in Fig.~\ref{fig:diagn-plot}, respectively. One can see that while all the estimators yield similar estimates for the sequences sampled from the Pareto-mixed Poisson distribution for both small and large sequence sizes, the estimates of different estimators for the small sequence size $n_s$ are far apart in the case of the double power law distribution with ``uglier" $\ell(k)$. In the latter case, the estimators start to agree on their estimates only for the large sequence size $n_l$. One can also see that the main reason for this effect is that the AMSE minima occur at far-apart locations if the sample size is small and $\ell(k)$ is not so ``nice.''
  }
  \label{fig:l-k-example}
\end{figure*}

To emphasize the importance of using as many consistent estimators as possible in application to degree sequences in real-world networks, here we consider an example of how the estimators work in conjunction with the double bootstrap method, showing that different estimators may explore different parts of the empirical degree distribution for any finite sequence, thus explaining why they may return different estimations on such sequences, especially if the slowly varying function $\ell(k)$ is not trivial.

Figure~\ref{fig:diagn-plot} shows that the Hill estimator yields a higher estimation of $\alpha=1/\xi$ than the other two estimators applied to the in-degree sequence of the \textit{Libimseti} network. This happens because the value of the optimal number of order statistics $\kappa^\ast$ returned by Hill's double bootstrap is substantially lower than for the other two estimators, so that the Hill estimator considers a smaller part of the distribution tail. The value of Hill's $\kappa^\ast$ is smaller because it is based on finding the minimum of the AMSE as a function of the number of order statistics $\kappa$, and as we can see in the figure these minima occur at quite different values of $\kappa$ for the Hill versus the two other estimators.

This effect is actually expected in small-sized sequences sampled from regularly varying distributions with nontrivial slowly varying functions $\ell(k)$. Figure~\ref{fig:l-k-example} shows the details behind estimator convergence in two different cases, with a ``nice'' and ``not so nice'' slowly varying function $\ell(k)$. The figure illustrates the point that the farther the $\ell(k)$ is from a constant, the larger the network size must be for all the estimators to yield similar values of $\kappa^\ast$ and $\widehat{\xi}$. The estimators are guaranteed to converge to the true value of $\xi$ for any $\ell(k)$, but only in the infinite sample limit $n\to\infty$,~and, to the best of our knowledge, there are no results (for the bounds) on the speed of this convergence, partly because this speed may depend in an unknown way on some properties of $\ell(k)$. That is why using as many consistent estimators as possible in application to real-world data is the best strategy one can follow.

\section{Evaluation on synthetic sequences and network models}\label{sec:experiments}

Here we show that the estimators based on extreme value (EV) theory---the Hill, Moments, and Kernel estimators equipped with the double bootstrap procedure, the code in~\cite{githubcode}---yield the expected results when applied to synthetic degree sequences and to network models. We also compare the estimations that these estimators produce with the ones by the PLFit~\cite{clauset2009power}, which is based on a combination of techniques inspired by maximum-likelihood estimation (MLE) and Kolmogorov-Smirnov (KS) distance minimization. We use the \texttt{plfit.m} code version~1.0.11 by Aaron Clauset, which is widely used and publicly available at~\cite{plfiturl}. As stated in the code comments of the \texttt{fit.py} script in~\cite{plfitpythonurl}---the Python implementation of the PLFit used in~\cite{broido2018scale}---this implementation is based on the original MATLAB code~\cite{plfiturl}, so that the results obtained using any of these two implementations~\cite{plfiturl,plfitpythonurl} should be identical.

\subsection{Synthetic sequences}\label{ssec:toy-models}

\begin{figure*}[ht]
  \centering
  \includegraphics[width=.99\textwidth]{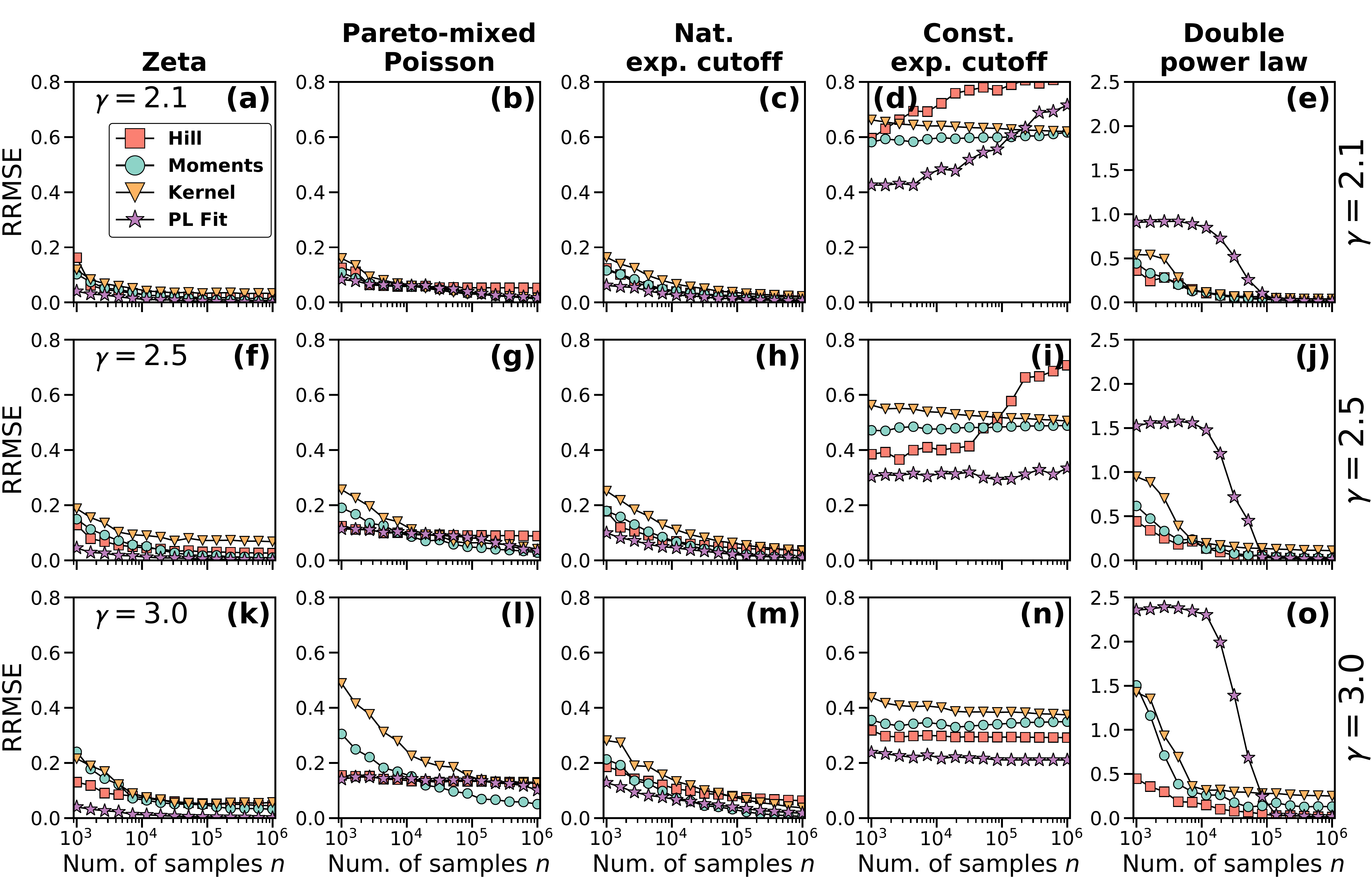}
  \caption{The relative root mean squared error (RRMSE)~\eqref{eq:rrmse-xi} of the three EV-based estimators and the MLE-based PLFit algorithm~\cite{clauset2009power} applied to i.i.d.\ sequences of random numbers sampled from the distributions described in Section~\ref{ssec:toy-models}. The columns correspond to these distributions, while the rows are for the three values of $\gamma$---$2.1$, $2.5$, and $3.0$---used in these distributions to sample the number sequences. The sequence length varies in the range $n\in[10^3,10^6]$ everywhere.
  }
  \label{fig:toys}
\end{figure*}

Here we sample different numbers $n$ of positive integers $k\in\N_+$ from the distributions listed below, so that the sampled sequence length is always $n$. The set of chosen distributions is intended to be diverse and representative of distributions claimed to be observed in real-world networks. In cases where the distribution has support on non-negative integers $k\in\N$, we discard all the zero entries from the sequence since they would correspond to nodes of degree $k = 0$ in networks. The parameter $\gamma$ in all the distributions below can be any real number greater than~$1$.

  {\emph{Zeta distribution:}
  The distribution PDF (or PMF, to be precise) is
    \begin{equation}
      P(k) = \frac{k^{-\gamma}}{\zeta(\gamma)}, \quad k\in\N_+,
    \end{equation}
    where $\zeta(\gamma)$ is the Riemann zeta function. This is the ``clean'' integer-valued power-law distribution with constant slowly varying function $\ell(k)=1/\zeta(\gamma)$.
  }

  {\emph{Pareto-mixed Poisson distribution:}
  For each sample, we first sample a real number $x$ from the Pareto distribution, and then sample an integer $k$ from the Poisson distribution with mean $x$:
  \begin{align}
    P(k|x) &= \frac{x^k {\mathrm e}^{-x}}{k!},\quad k\in\N, \\
    P(x) &= \alpha x_{0}^{\alpha} x^{-\gamma},\quad x\geq x_0 > 0,
  \end{align}
  where $\alpha=\gamma-1$, and we set $x_0=1$ in the experiments. The Pareto-mixed Poisson distribution is ubiquitous in network models with hidden variables~\cite{boguna2003class}, also known in mathematics as inhomogeneous random graphs~\cite{bollobas2007phase}, and more generally, as graphon-based $W$-random graphs~\cite{lovasz2006limits}. This is one of the simplest regularly varying distribution with non-constant $\ell(k)$ that converges to a constant, $\ell(k)\to\alpha x_0^\alpha$, Section~\ref{ssec:examples_regularly_varying}.
  }

  {\emph{Pareto distribution with 
  natural exponential cutoff}. We sample a random number $x$ from the Pareto distribution with the exponential cutoff at $n^\xi$,
  \begin{equation}\label{eq:pareto_exponentail_cut_off}
    P_n(x) = \frac{x_0^\alpha}{E_{\gamma}\left(x_{0}/n^\xi\right)} x^{-\gamma} {\mathrm e}^{-x / n^\xi},\quad x\geq x_0 >0,
  \end{equation}
  where $E_{\gamma}$ is the exponential integral function, $\xi=1/\alpha$, and $\alpha=\gamma-1$. We then round 
  $x$ to the closest integer $k = \left[x\right]$. We set $x_0=1$. The value $n^\xi$ of where the exponential decay becomes prominent corresponds to the natural cutoff~\cite{boguna2004cut}, which is proportional to the exact expected maximum value~\eqref{eq:exact_expected_maximum} among $n$ i.i.d.\ samples from the Pareto distribution with exponent $\gamma$. This is an example of not a fixed distribution, but of an $n$-dependent family of distributions. For any fixed $n$, the distribution is not regularly varying since it has an exponential tail instead of a power-law tail. Yet as $n$ increases, the location $n^\xi$ of the ``soft beginning'' of the exponential tail diverges, so that in the $n\to\infty$ limit the distributions in this family converge to the pure Pareto distribution with exponent~$\gamma$, which is regularly varying.
  }
  
  {\emph{Pareto distribution with a constant exponential cutoff}. The sampling is the same as in the previous example, except that the location of the exponential cutoff does not depend on $n$, and is fixed to be $10$ instead of $n^\xi$. This is an example of a distribution which is not regularly varying.
  }
  
  {\emph{Double power law}. We sample a random number $x$ from the double power-law distribution with the PDF
  \begin{align}\label{eq:double-pl}
    P(x) &= \beta x^{-\gamma_0} \left( 1 + \left( \frac{x}{c} \right)^{\alpha_0/r} \right)^{-r},\quad x\geq x_0 >0,\\
    \alpha_0&=\gamma-\gamma_0,\quad \gamma \geq \gamma_0 > 1,
  \end{align}
  where $c$ is the location of the switch between the two power laws with exponents $\gamma_0$ for $x\ll c$ and $\gamma$ for $x\gg c$, $r$ is the parameter that controls how smooth this switch is, and $\beta$ is the normalizing constant given by
  \begin{equation*}
    \beta = \frac{\alpha x_0^\alpha}{c^{\alpha_0} {}_{2}F_{1}\left(r,\, r\alpha/\alpha_0,\, 1 + r\alpha/\alpha_0,\, -\left(c/x_0\right)^{\alpha_0/r} \right)},
  \end{equation*}
  where $\alpha = \gamma-1$ and $_{2}F_{1}$ is the Gauss hypergeometric function. As in all other examples, given this random $x$, we round it to integer $k = \left[x\right]$. In our experiments, we set $\gamma_0 = 1.5$, $c = 500$, $r = 0.1$, and $x_0 = 1$. This distribution is regularly varying with exponent $\gamma$, which we vary in the experiments. Yet, as discussed in Section~\ref{sec:power-law-networks}, it may be difficult for the estimators to see that it is indeed $\gamma$ and not $\gamma_0$ if $n$ is small. Distributions of this form characterize the degree distribution in the causal set of the universe~\cite{krioukov2012network}, and they also frequently appear in astrophysics~\cite{huppenkothen2013quasi}.
  }

To assess the accuracy of the estimators, we sample $s=100$ random sequences for each combination of the distributions listed above, the values of $\gamma$, and the numbers of samples $n$ in a sequence. On each sampled sequence $j$, $j=1,\ldots,s$, each estimator returns an estimated value $\widehat{\xi}_j$ of $\xi$. Given a collection of these $\widehat{\xi}_j$s, we compute the relative root-mean-squared-error (RRMSE), a standard measure used to assess the accuracy of the tail index estimation~\cite{brzezinski2016robust,jordanova2016weak}, defined as
\begin{equation}
    \label{eq:rrmse-xi}
    RRMSE = \frac{\sqrt{\frac{1}{s}\sum_{j=1}^{s} (\widehat{\xi}_j - \xi)^2}}{\xi},
\end{equation}
and show the results in Fig.~\ref{fig:toys} both for the extreme value (EV) estimators (Hill, Kernel, Moments), and for the MLE-based PLFit~\cite{clauset2009power}.

We observe that all the results are as expected. On sequences sampled from distributions that are regularly varying, all the EV estimators converge. They also converge in the case where the distributions are not regularly varying for any finite sample size $n$, but where they converge to a regularly varying distribution at $n\to\infty$---the Pareto distribution with the diverging natural cutoff. No estimator converges in the case of a fixed distribution which is not regularly varying---the Pareto distribution with a fixed exponential cutoff.

Also as expected, the PLFit yields a lower estimation error in case of the zeta distribution. This is because the zeta distribution satisfies PLFit's main assumption of a clean power law with constant $\ell(k)$, but does not satisfy the second order condition, thus affecting the optimality of the double bootstrap, Section~\ref{ssec:finding_good_k}. In other cases with reasonably ``nice'' regularly varying distributions with $\ell(k)$ quickly converging to a constant, the accuracy and convergence rates of the EV and PLFit estimators are comparable. However, as soon as the regularly varying distribution is not really nice---the double power law case with non-constant $\ell(k)$ over a wide range of degrees~$k$---the PLFit estimations are completely off, as opposed to the EV estimators. This is also expected for the reasons discussed in Section~\ref{ssec:plfit}.

\subsection{Network models}\label{ssec:network_models}

The main motivation to test the performance of the EV estimators not only on synthetic sequences of numbers sampled from various distributions, but also on degree sequences in network models, is to see whether and how their performance is affected by possible non-i.i.d.-ness of the latter sequences. To
this end we consider three paradigmatic network models in which the degree distributions have been proven to converge to a regularly varying distribution, and in which the degree sequences are not i.i.d: 1)~the erased configuration model (ECM)~\cite{britton2006generating}, 2)~preferential attachment (PA)~\cite{barabasi1999emergence}, and 3) hyperbolic random graphs (HRG)~\cite{krioukov2010hyperbolic}. 

\emph{Erased configuration model.}
We sample varying-length i.i.d.\ sequences of random integers from the zeta distributions with different values of the exponent, and then either accept or reject the sequence based on whether the sum of its elements is even or odd. Each number in the sequence is the number of stubs attached to a node in a network to be formed. We match pairs of stubs uniformly at random, and then delete loops and multi-edges. For any finite sample size, the degree sequence in the resulting network is neither zeta-distributed nor i.i.d., but it converges to the original zeta distribution as the sample size tends to infinity~\cite[Theorem 2.1]{britton2006generating}.

\emph{Preferential attachment.} We use the redirection implementation in~\cite{krapivsky2001organization}: starting with the first node of degree~$0$, nodes arrive one by one, and each new node picks an already existing node uniformly at random, and then connects either to it with probability $1-r$, or to its random neighbor with probability $r$. The only exception is the second node that connects to the first node with probability~$1$. We use this redirection probability to control the exponent of the power-law tail of the degree distribution, because this distribution converges to the following regularly varying distribution with exponent $\gamma=1+1/r$~\cite{krapivsky2001organization}:
\begin{equation}\label{eq:DD-PA}
	P(k) =  (\gamma-1)\frac{\Gamma(2\gamma - 3)}{\Gamma(\gamma-2)} \, \frac{\Gamma(k + \gamma - 3)}{\Gamma(k + 2\gamma - 3)}.
\end{equation}

\emph{Hyperbolic random graphs.} The degree distribution in random geometric graphs in hyperbolic spaces converges to regularly varying Pareto-mixed Poisson distributions~\eqref{eq:mixed-poisson}, and, as opposed to the previous two models, these graphs also have non-vanishing average local clustering coefficients~\cite{krioukov2010hyperbolic}. We use the software package developed in~\cite{aldecoa2015hyperbolic} and available at~\cite{hggenerator} to generate these graphs. We fix the average degree parameter to $\bar{k} = 10$, the temperature parameter to $T = 0$ corresponding to strongest clustering, and vary the $\gamma$ parameter.

\begin{figure}[t]
  \centering
  \includegraphics[width=.49\textwidth]{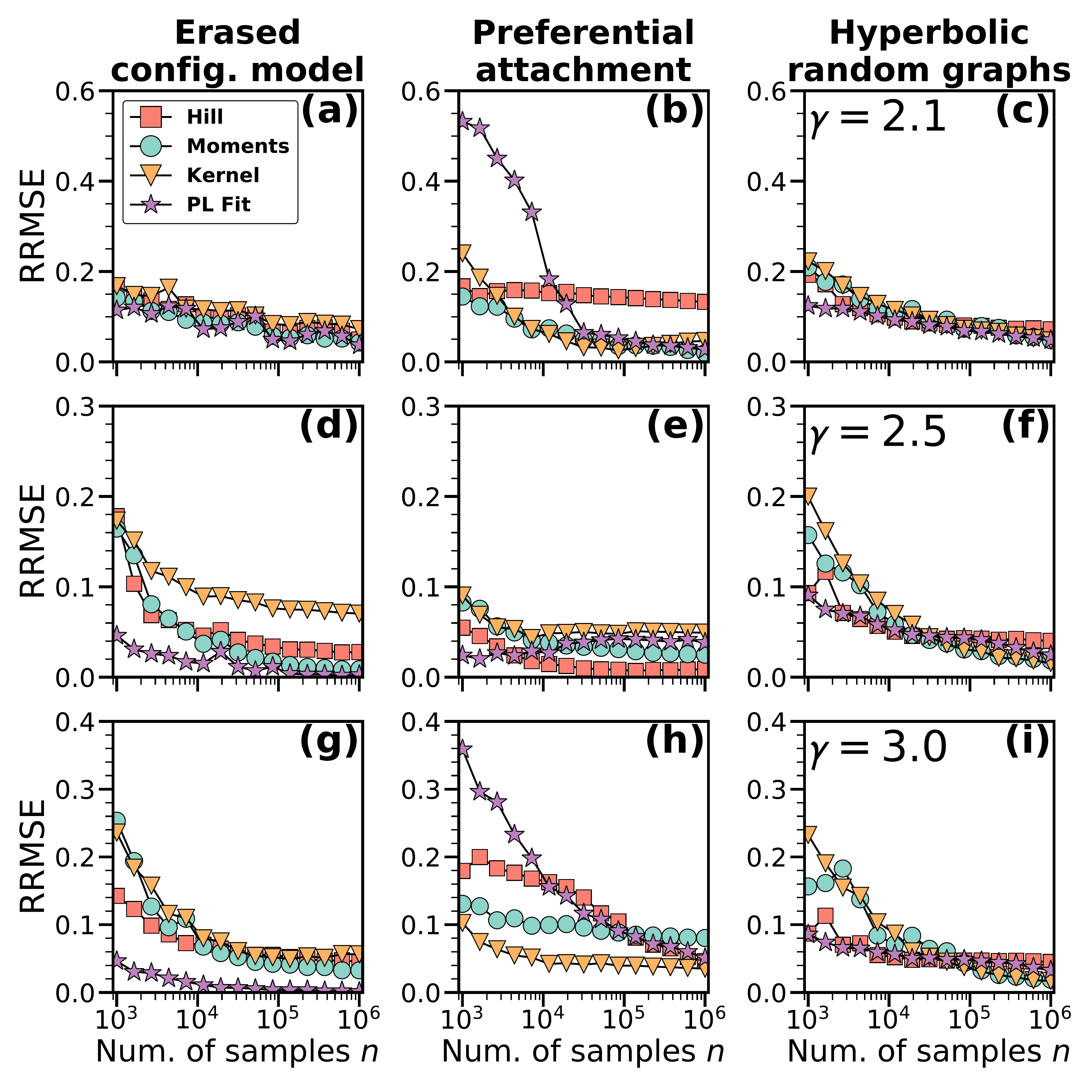}
  \caption{The relative root mean squared error (RRMSE)~\eqref{eq:rrmse-xi} of the three EV-based estimators and the MLE-based PLFit algorithm~\cite{clauset2009power} applied to the network models described in Section~\ref{ssec:network_models}. The first, second, and third columns show the results for the erased configuration model, preferential attachment, and hyperbolic random graphs, respectively. The first, second, and third rows show the results for $\gamma=2.1$, $\gamma=2.5$, and $\gamma=3.0$, respectively. The network size varies in the range $n \in [10^3,10^6]$ everywhere.
  }
  \label{fig:nets-rrmse}
\end{figure}

For each model, we vary the $\gamma$ over the three values $\gamma = 2.1$, $2.5$, and $3.0$, and vary the network size $n$ from $10^3$ to $10^6$. For each combination of the model, $\gamma$, and $n$, we generate $100$ random networks, read off their degree sequences, and feed them to all the considered estimators. We then compute the RRMSE~\eqref{eq:rrmse-xi}, and show the results in Fig.~\ref{fig:nets-rrmse}.

We observe that in all the considered cases, all the EV estimators converge, even though the degree sequences are not i.i.d. The slow convergence in some cases is explained by the slow convergence of the degree distributions in these finite-sized networks to their limits. This is the case, for example, in the HRGs with $\gamma=2.1$: the degree distribution in HRGs converges to its Pareto-mixed Poisson limit 
the more slowly, the closer the $\gamma$ is to~$2$~\cite{krioukov2010hyperbolic}.

The most notable results are for PA. Here the EV estimators clearly outperform the PLFit if $\gamma = 2.1$ or $\gamma= 3$, while all the estimators are on par if $\gamma = 2.5$ for the reasons that we discuss in the next section.

\subsection{Anatomy of the PLFit}\label{ssec:plfit}

To better understand the slow convergence of the PLFit in the double power law and preferential attachment cases in the previous two sections, it is instructive to recall first how exactly the PLFit algorithm works.
The algorithm is a variation of estimators in~\cite{goldstein2004problems,bauke2007parameter} based on maximum-likelihood estimation (MLE). The starting point of the PLFit operations is a sequence of possible $\gamma$-values~$\gamma_s$ to experiment with. By default, this sequence is linearly spaced in the region $[1.5,3.5]$ with step size~$0.01$ in the code~\cite{plfiturl} released with~\cite{clauset2009power}. These default settings have to be manually changed for the code to be applicable to degree sequences coming from distributions with $\gamma > 3.5$. The default values of $\gamma_s$ in the code~\cite{plfitpythonurl} used in~\cite{broido2018scale} are linearly spaced in $[1.01,6.50]$ with step size~$0.01$.

Given the sequence $\gamma_s$ and a degree sequence of length~$n$ supplied as
input data, the PLFit algorithm first finds the sequence of unique degree values $k_t$ appearing in the degree sequence. For each value $k_t$, the algorithm computes the vector of log-likelihood values
\begin{equation}\label{eq:appendix-plfit-mle}
    {\mathcal L}_{ts} = -n_{t}\log{\zeta(\gamma_s, k_{t})} - \gamma_s \sum_{i=1}^n \mathbbm{1}\{k_i \geq k_t\}\log{k_{i}},
\end{equation}
where $n_t$ is the number of nodes with degrees $k_i\geq k_t$, $\mathbbm{1}$~the indicator function, and
\begin{equation}
  \zeta(\gamma_s, k_{t}) = \sum_{k=k_t}^{\infty} k^{-\gamma_s}
\end{equation}
is the Hurwitz zeta function. This likelihood is based on the assumption that the degrees that are greater than or equal to~$k_t$ form a sequence of i.i.d.\ samples from a pure power law with exponent~$\gamma_s$, i.e., from the generalized zeta distribution~\eqref{eq:zeta_def} with parameters $\gamma_s$, $k_{\min}=k_t$, and the normalization constant $c=1/\zeta(\gamma_s,k_t)$. Among all the considered values $\gamma_s$, the algorithm then identifies the one, $\gamma_t^*$, that corresponds to the maximum value of ${\mathcal L}_{ts}$ for the given~$k_t$. This $\gamma_t^*$ serves as an approximation of the MLE of $\gamma$ for the degrees that are greater than or equal to~$k_t$.
For the same $k_t$, the algorithm then computes the Kolmogorov-Smirnov (KS) distance $D_t^{KS}$ between the generalized zeta distribution with parameters~$\gamma_t^*$ and $k_{\min}=k_t$, and the empirical CDF of degrees $k_i \geq k_t$. This procedure is then repeated for each $k_t$ observed in the sequence, and the estimates $\hat{\gamma}$ and $\hat{k}_{\min}$ that the algorithm eventually returns are those that correspond to the minimum $D_{t^*}^{KS}$ of $D_t^{KS}$ across all possible values of~$k_t$, i.e., $\hat{k}_{\min}=k_{t^*}$ and $\hat{\gamma}=\gamma_{t^*}^*$.

The algorithm is thus a mixture of two optimization strategies: one is based on likelihood maximization, while the other one deals with the KS distance minimization. We note that since the algorithm does not implement MLE exactly, it trivially cannot be consistent if the true value of $\gamma$ does not belong to the finite set of $\gamma_s$ values, because it can never report any $\gamma$-estimate~$\hat{\gamma}$ that does not belong to the finite set of $\gamma_s$s. More importantly, even though the correct implementation of MLE with a fixed and known $k_{\min}$ had long been proven to be consistent~\cite{clauset2009power}, the consistency of MLE in combination with KS-distance minimization has been proven only very recently in~\cite{drees2018minimum}, and only for pure power laws, i.e., for the Pareto or generalized zeta distributions. If the distribution is not a pure power law but a general regularly varying distribution, the consistency of the algorithm is a question that has not been rigorously explored at all, except the conjectures in~\cite{drees2018minimum} that this MLE-KS combination appears to be consistent for regularly varying distributions satisfying the second order condition, and for regularly varying distributions whose slowly varying functions $\ell(k)$ converge to constants, and that the algorithm is likely not to be consistent for all other classes of regularly varying distributions. That is, in all these other cases the algorithm may be consistent, or it may not be.

The problem is that there is the following logical inconsistency in the algorithm: it operates under the assumption that above a certain $k_{\min}$, the distribution of degrees $k$ is a pure power law, but then it recognizes that the distribution may be not a pure power law, and tries to account for that by finding a reasonable value of $k_{\min}$ such that above this value the assumption would hold ``approximately.'' If the distribution was a pure power law, then the search for this $k_{\min}$ would not be necessary, since the value of $k_{\min}$ would be equal with high probability to the smallest value observed in the sequence. But if the distribution is not a pure power law, then such a value of $k_{\min}$ simply does not exist, since for any $k_{\min}$ the distribution of $k>k_{\min}$ is \emph{not} a pure power law. Yet the distribution of such $k$s may converge to a pure power law, but only if the $k_{\min}$ value that the algorithm finds diverges with the sample size~$n$, and only if the slowly varying function $\ell(k)$ converges to a constant. Therefore, for this subclass of regularly varying distributions with $\ell(k)$s converging to constants, the algorithm is likely to be consistent, yet the full proof is currently lacking~\cite{drees2018minimum}. If $\ell(k)$ does not converge to a constant, then the consistency of the algorithm is quite unclear at present.

\begin{figure}[t]
  \centering
  \includegraphics[width=.48\textwidth]{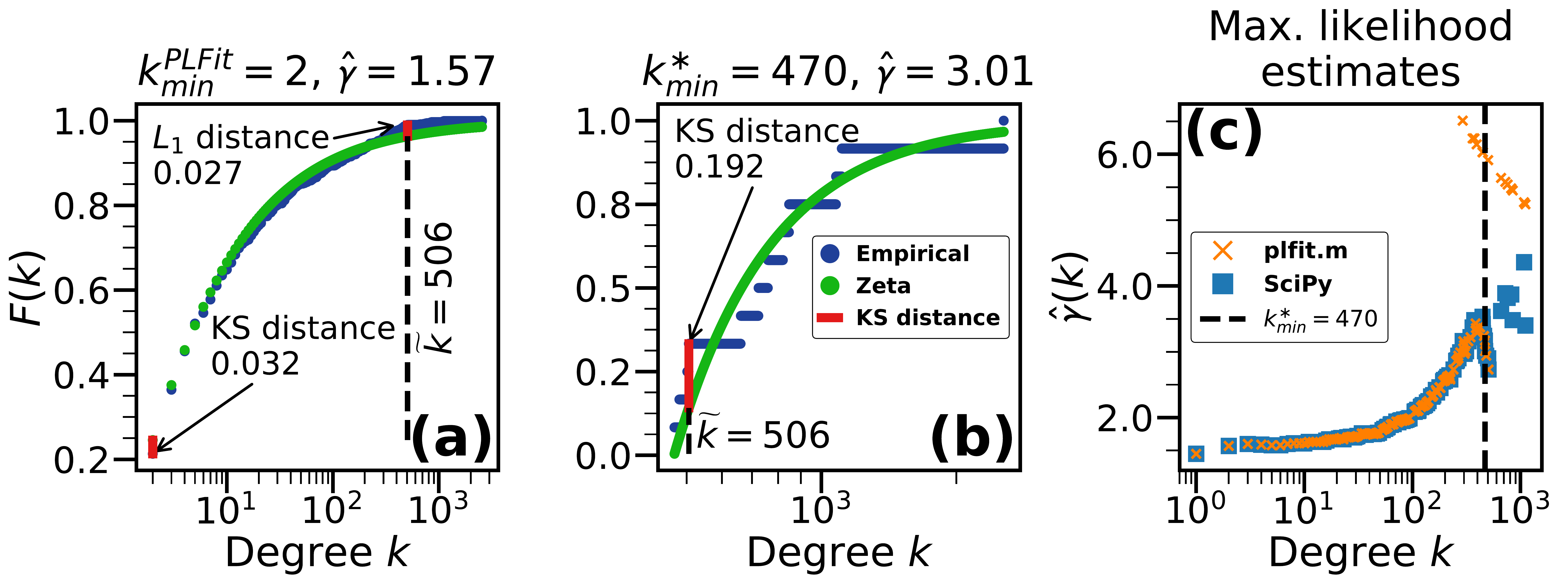}
  \caption{Anatomy of Kolmogorov-Smirnov (KS) distance minimization on a sequence of random numbers sampled from the double power-law distribution with the parameters as in Section~\ref{ssec:toy-models}, tail exponent $\gamma = 3$ and size $n=1,000$. Applied to this degree sequence, the PLFit algorithm estimates $k_{\min}^{PLFit}=2$ and $\hat{\gamma}\left(k_{\min}^{PLFit}\right)=1.57$, the MLE estimate of $\gamma$ with $k_{\min}=2$. Panel~(a) shows the empirical and theoretical (generalized zeta) cumulative distribution functions (CDFs) with these parameters. The KS distance ($0.032$) is marked. Panel~(b) shows the same CDFs but for different $k_{\min}^{\ast} = 470$ and $\hat{\gamma}\left(k_{\min}^{\ast}\right) = 3.01$, the MLE estimate of $\gamma$ with $k_{\min}=470$. This $k_{\min}$ is optimal in this sequence, in the sense that the MLE value of $\gamma$ with this $k_{\min}$ is closest to the true $\gamma$ across all other degree values present in the sequence to which $k_{\min}$ can be set. Yet the marked KS distance ($0.192$), achieved at $\tilde{k}=506$, is greater than the KS distance achieved at a different location in panel~(a), where the $L^1$ distance ($0.027$) between the two CDFs at $\tilde{k}$ is also shown. Panel~(c) shows a collection of numerical errors produced by the PLFit if modified to compute the MLE values of $\gamma$ for large $k_{\min}$s. The errors are due to the numerically incorrect computations of the Hurwitz zeta function in the \texttt{plfit.m} code~\cite{plfiturl}.
  }
  \label{fig:plfit-figure1}
\end{figure}

\begin{figure}[t]
  \centering
  \includegraphics[width=.45\textwidth]{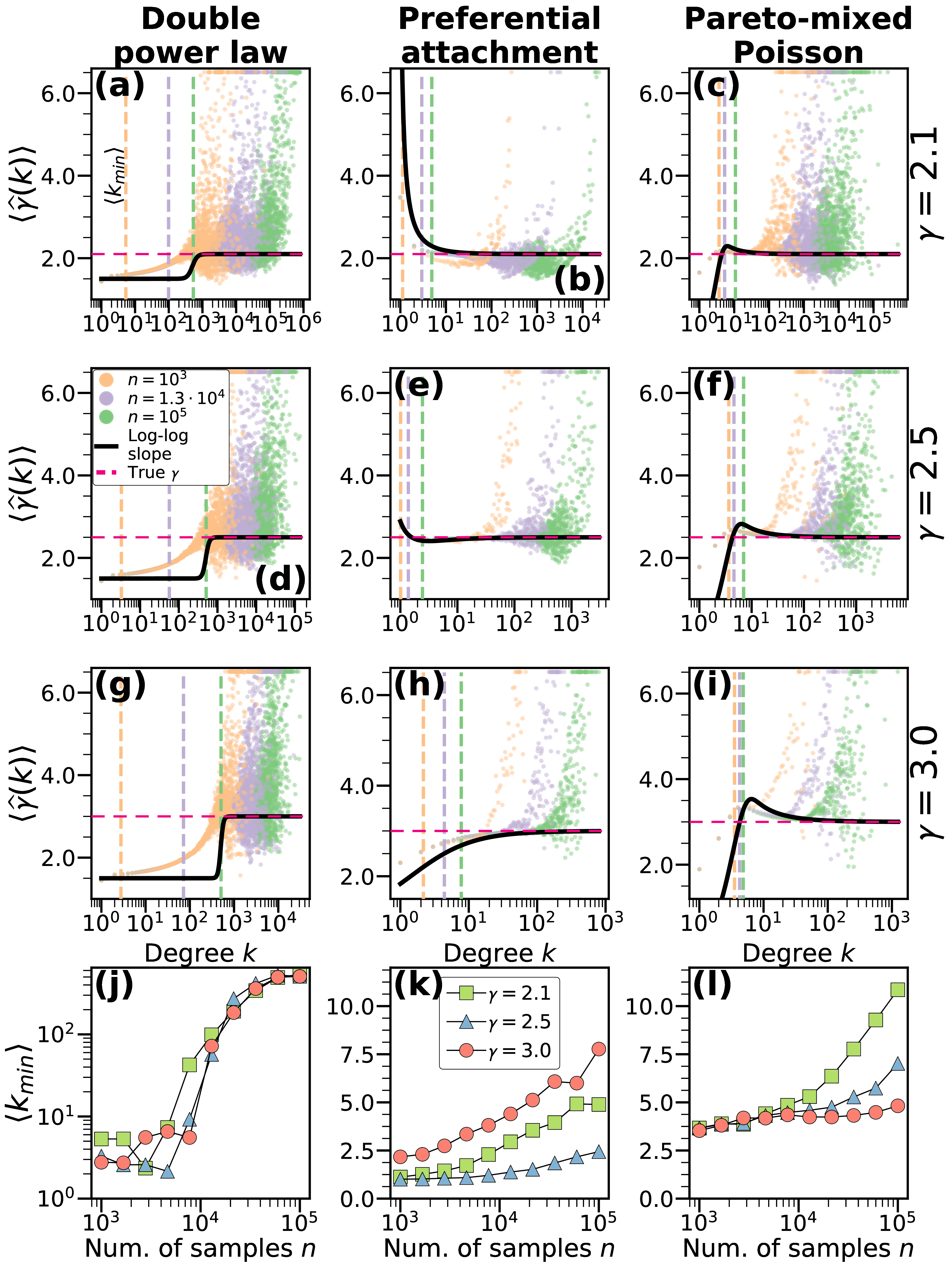}
  \caption{Anatomy of maximum-likelihood estimation (MLE) of $\gamma$ on different sequences of different sizes. The columns, left to right, correspond to: 1)~number sequences sampled from the double power-law distribution~\eqref{eq:double-pl} with the same parameters as in Section~\ref{ssec:toy-models}; 2)~degree sequences of random graphs in the preferential attachment model with the degree distribution given by~\eqref{eq:DD-PA}; 3)~number sequences sampled from the Pareto-mixed Poisson distribution~\eqref{eq:mixed-poisson}. The first three rows show the results for $\gamma=2.1$, $2.5$, and $3.0$, while the last row show the average values of $k_{\min}$ found by the PLFit algorithm for different $\gamma$s as functions of sequence length $n$. The first three rows show the results only for three different sequence sizes $n$, marked by different colors. The colored dots are the average values, computed using the \texttt{SciPy} package~\cite{scipy}, of the MLE $\hat{\gamma}$ estimates obtained by the minimization of~\eqref{eq:appendix-plfit-mle} over $\gamma_s$ linearly spaced in $[1.01,6.50]$ with step $0.01$, shown as functions of $k_{\min}=k_t$, where $k_t$ is a degree value appearing at least once in the
  collection of sequences. For any combination of the model, $\gamma$, and $n$, the results are averaged over $100$ random sequences. The solid black curves in the first three rows are the loglog slopes of the corresponding theoretical PDFs considered as functions of continuous~$k$. The horizontal dashed lines are the true $\gamma$s, while the vertical dashed lines color-correspond to the $n$s, showing the average $k_{\min}$s for these $n$s from the bottom row.
  }
  \label{fig:plfit-figure2}
\end{figure}

In all the regularly varying distributions considered in the previous two sections, the function $\ell(k)$ does converge to a constant, and indeed in all these cases the PLFit appears to eventually converge. Yet in two of these cases, namely, double power laws and preferential attachment, its convergence is worse than that of any of the considered EV estimators. To see why, we analyze the two components of the PLFit, KS distance minimization and likelihood maximization, separately---in Figs.~\ref{fig:plfit-figure1} and~\ref{fig:plfit-figure2}, respectively.

Figure~\ref{fig:plfit-figure1} illustrates that the KS distance minimization component of the PLFit drives the values of $k_{\min}$ that the PLFit attempts to estimate to erroneously low values, in full agreement with more recent and in-depth investigations in~\cite{drees2018minimum}. This happens because the smaller the $k_{\min}$, the smaller the deviations of the empirical CDF at degrees $k$ right above $k_{\min}$ from the theoretical CDF, because if the distribution is regularly varying, there are more nodes with smaller degrees. The larger the $k_{\min}$, the larger are these deviations caused by ``sparser statistics'' in the distribution tail, and as a consequence the KS distance grows larger. If $k_{\min}$ is set to a small value, the deviations in the tail are suppressed as they are getting ``squished'' in the high-degree region of the CDF close to~$1$, cf.~panels~(a) and~(b) in the figure.

Panel~(c) in Fig.~\ref{fig:plfit-figure1} shows that the \texttt{plfit.m} code~\cite{plfiturl}, both originally released with~\cite{clauset2009power} as well as its current version, cannot be used to compute the MLE values of~$\gamma$ for large $k_{\min}$s, because it contains errors in computing the Hurwitz zeta function with the required accuracy~\cite{bauke2007parameter}, leading to numerical errors. Therefore we use a \texttt{SciPy}~\cite{scipy} implementation instead in Fig.~\ref{fig:plfit-figure2}.

Figure~\ref{fig:plfit-figure2} shows that if $k_{\min}$ is small---and it is, thanks to the KS distance minimization part of the PLFit---then the MLE component of the PLFit does very little other than trying to fit the loglog slope of the PDF evaluated at this $k_{\min}$. The reasons behind this problem are the same as those behind the KS distance minimization problems discussed above: since there is a lot of data with degrees right above a small value of~$k_{\min}$, and since the MLE is primarily concerned with fitting as much data as possible, it tries to fit the part of the distribution with degrees~$k$ right above~$k_{\min}$, versus the true tail of the distribution with large~$k$s, thus getting bad estimates of the tail exponent.

In other words, the errors in PLFit's estimates are due to the combination of the following two factors related, ironically, to the two key ideas behind the PLFit: (1)~the small values of $k_{\min}$ returned by the KS distance minimization part of the algorithm, and (2)~the MLE part of the PLFit that estimates not the tail exponent but, roughly, the loglog slope of the PDF evaluated close to this~$k_{\min}$. If this slope is different from the slope at large~$k$s, i.e., the true tail exponent, then the PLFit does not really fit any power-law tail. However, if the distribution is such that at least one of these conditions is not satisfied, then the PLFit estimates are more accurate, cf.~the Pareto-mixed Poisson or the $\gamma=2.5$ preferential attachment cases in Fig.~\ref{fig:plfit-figure2}.

If these two conditions are satisfied, which is the case with the double power law and preferential attachment with $\gamma=2.1$ and $\gamma=3.0$ in Fig.~\ref{fig:plfit-figure2}, then the PLFit estimates of the tail exponent are quite off. But if they are off, and if one then performs KS hypothesis testing using these inaccurate estimates, then the hypothesis that the degree sequence comes from a pure power law with the estimated exponent will be rejected with high probability, simply because the true tail exponent is different. For these reasons, if one applies the PLFit to preferential attachment networks with these exponents and then deploys the KS hypothesis tests, one will likely find that these networks are not power-law~\cite{broido2018scale}.

Finally, Figure~\ref{fig:plfit-figure2} also shows that the whole idea of using MLE to estimate tail exponents of regularly varying distributions is quite problematic to begin with, explaining why it has not been seriously explored in statistics. Indeed, for such an estimation procedure to be accurate, the values of $k_{\min}$ must be large and diverging in the $n\to\infty$ limit for the reasons discussed above. However, the larger the $k_{\min}$, the smaller the second term in~\eqref{eq:appendix-plfit-mle}. On the other hand, as a function of $\gamma_s$, the first term in~\eqref{eq:appendix-plfit-mle} grows monotonically at a much higher rate than the linear rate of growth of the second term. Therefore, if $k_{\min}$ is above a certain threshold, then the MLE will do nothing but select the largest possible value of $\gamma_s$ to maximize the likelihood via the first term. This is exactly what we see in Fig.~\ref{fig:plfit-figure2}, where for many instances of large $k_{\min}$, the MLE-selected values of $\gamma$ are the largest possible values within the range that we offer the MLE to operate with. Therefore, the correctness of MLE depends on whether there exists a ``sweet-spot'' range of values of $k_{\min}$ that are not too large and not too small. It might be the case that such a range simply does not exist for some regularly varying distributions. Worse, even if it can be proven to always exist, which is unclear at present, we have seen above that the KS distance minimization procedure is quite unlikely to be a correct, statistically consistent, procedure to identify this range. At least, the KS distance minimization has not been proven to be such a procedure for general regularly varying distributions. Whether a required procedure exists at all, is also unclear. After all, for the reasons mentioned above, even the required sweet-spot range of $k_{\min}$s is quite unlikely to exist for regularly varying distributions whose slowly varying functions do not converge to constants.

\onecolumngrid

\clearpage
\twocolumngrid
\begin{longtable*}[ht!]{|c | c | c | c c c|}
    \caption{The tail exponent estimation results for the 35 real-world undirected networks collected from the KONECT database~\cite{kunegis2013konect}. Each network name is followed by its KONECT code in braces. The estimators return estimates $\hat{\xi}$ of $\xi$ that are translated to $\hat{\gamma}=1+1/\hat{\xi}$. If $\hat{\xi} \leq 0$, then $\hat{\gamma}$ is set to $\infty$. The estimates are colored according to the definitions in Section~\ref{sec:power-law-networks}: 1)~{\color{tablered}\textit{not power-law networks} (red)} -- at least one estimate is nonpositive $\hat{\xi} \leq 0$; 2)~{\color{tableorange}\textit{hardly power-law networks} (yellow)} -- all the estimates are positive $\hat{\xi} > 0$, and at least one estimate is $\hat{\xi} \leq \xiThreshold$; 3)~{\color{tablegreen}\textit{power-law networks with a divergent second moment} (green)} -- all the estimates are $\hat{\xi} > 1/2$; and 4)~{\color{tableblue}\textit{other power-law networks} (blue)} -- the rest of the cases.
    }\label{tab:real-undir}\\
    \hline
    {\bf Network Name} & $\boldsymbol{n}$ & $\boldsymbol{\bar{k}}$ & $\boldsymbol{\hat{\gamma}^{Hill}}$ & $\boldsymbol{\hat{\gamma}^{Mom}}$ & $\boldsymbol{\hat{\gamma}^{Kern}}$ \\
    \hhline{======}
    \endfirsthead
    \caption{(Continued).}\\
    \hline
    {\bf Network Name} & $\boldsymbol{n}$ & $\boldsymbol{\bar{k}}$ & $\boldsymbol{\hat{\gamma}^{Hill}}$ & $\boldsymbol{\hat{\gamma}^{Mom}}$ & $\boldsymbol{\hat{\gamma}^{Kern}}$ \\
    \hhline{======}
    \endhead
    
    CAIDA (IN) & 26,475 & 4.03 & \boldsymbol{\textcolor{tablegreen}{ 2.1 }} & \boldsymbol{\textcolor{tablegreen}{ 2.11 }} & \boldsymbol{\textcolor{tablegreen}{ 2.11 }} \\ \hline
    Skitter (SK) & 1,696,415 & 13.08 & \boldsymbol{\textcolor{tablegreen}{ 2.38 }} & \boldsymbol{\textcolor{tablegreen}{ 2.36 }} & \boldsymbol{\textcolor{tablegreen}{ 2.43 }} \\ \hline
    Actor collaborations (CL) & 382,219 & 173.28 & \boldsymbol{\textcolor{tableorange}{ 3.71 }} & \boldsymbol{\textcolor{tableorange}{ $6.7\cdot10^3$ }} & \boldsymbol{\textcolor{tableorange}{ 2.36 }} \\ \hline
    Amazon (CA) & 334,863 & 5.53 & \boldsymbol{\textcolor{tableblue}{ 3.99 }} & \boldsymbol{\textcolor{tableblue}{ 3.48 }} & \boldsymbol{\textcolor{tableblue}{ 3.44 }} \\ \hline
    arXiv (AP) & 18,771 & 21.1 & \boldsymbol{\textcolor{tableorange}{ 4.41 }} & \boldsymbol{\textcolor{tableorange}{ 5.78 }} & \boldsymbol{\textcolor{tableorange}{ 7.29 }} \\ \hline
    Bible names (MN) & 1,773 & 10.3 & \boldsymbol{\textcolor{tableblue}{ 3.09 }} & \boldsymbol{\textcolor{tableblue}{ 3.36 }} & \boldsymbol{\textcolor{tableblue}{ 2.88 }} \\ \hline
    Brightkite (BK) & 58,228 & 7.35 & \boldsymbol{\textcolor{tableblue}{ 3.51 }} & \boldsymbol{\textcolor{tableblue}{ 3.8 }} & \boldsymbol{\textcolor{tableblue}{ 2.96 }} \\ \hline
    Catster (Sc) & 149,684 & 72.8 & \boldsymbol{\textcolor{tablegreen}{ 2.09 }} & \boldsymbol{\textcolor{tablegreen}{ 2.06 }} & \boldsymbol{\textcolor{tablegreen}{ 1.98 }} \\ \hline
    Catster/Dogster (Scd) & 623,748 & 50.33 & \boldsymbol{\textcolor{tablegreen}{ 2.1 }} & \boldsymbol{\textcolor{tablegreen}{ 2.11 }} & \boldsymbol{\textcolor{tablegreen}{ 2.04 }} \\ \hline
    Chicago roads (CR) & 1,467 & 1.77 & \boldsymbol{\textcolor{tablered}{ 77.92 }} & \boldsymbol{\textcolor{tablered}{ $\infty$ }} & \boldsymbol{\textcolor{tablered}{ $\infty$ }} \\ \hline
    DBLP (CD) & 317,080 & 6.62 & \boldsymbol{\textcolor{tableorange}{ 6.59 }} & \boldsymbol{\textcolor{tableorange}{ 13.99 }} & \boldsymbol{\textcolor{tableorange}{ 3.06 }} \\ \hline
    Dogster (Sd) & 426,816 & 40.03 & \boldsymbol{\textcolor{tablegreen}{ 2.15 }} & \boldsymbol{\textcolor{tablegreen}{ 2.15 }} & \boldsymbol{\textcolor{tablegreen}{ 2.12 }} \\ \hline
    Douban (DB) & 154,908 & 4.22 & \boldsymbol{\textcolor{tableorange}{ 4.42 }} & \boldsymbol{\textcolor{tableorange}{ 6.88 }} & \boldsymbol{\textcolor{tableorange}{ 1.86 }} \\ \hline
    U. Rovira I Virgili (A$@$) & 1,133 & 9.62 & \boldsymbol{\textcolor{tablered}{ 6.49 }} & \boldsymbol{\textcolor{tablered}{ $\infty$ }} & \boldsymbol{\textcolor{tablered}{ $\infty$ }} \\ \hline
    Euro roads (ET) & 1,174 & 2.41 & \boldsymbol{\textcolor{tableorange}{ 4.73 }} & \boldsymbol{\textcolor{tableorange}{ 44.48 }} & \boldsymbol{\textcolor{tableorange}{ 29.57 }} \\ \hline
    Flickr (LF) & 1,715,254 & 18.13 & \boldsymbol{\textcolor{tableorange}{ 3.94 }} & \boldsymbol{\textcolor{tableorange}{ 4.29 }} & \boldsymbol{\textcolor{tableorange}{ 5.02 }} \\ \hline
    Flickr (FI) & 105,938 & 43.74 & \boldsymbol{\textcolor{tableorange}{ 6.18 }} & \boldsymbol{\textcolor{tableorange}{ 1.79 }} & \boldsymbol{\textcolor{tableorange}{ 1.65 }} \\ \hline
    Flixster (FX) & 2,523,386 & 6.28 & \boldsymbol{\textcolor{tableorange}{ 53.63 }} & \boldsymbol{\textcolor{tableorange}{ 1.93 }} & \boldsymbol{\textcolor{tableorange}{ 1.95 }} \\ \hline
    Gowalla (GW) & 196,591 & 9.67 & \boldsymbol{\textcolor{tablegreen}{ 2.8 }} & \boldsymbol{\textcolor{tablegreen}{ 2.8 }} & \boldsymbol{\textcolor{tablegreen}{ 2.86 }} \\ \hline
    Hamsterster (Shf) & 1,858 & 13.49 & \boldsymbol{\textcolor{tableorange}{ 4.45 }} & \boldsymbol{\textcolor{tableorange}{ 8.09 }} & \boldsymbol{\textcolor{tableorange}{ 3.51 }} \\ \hline
    Hamsterster (Sh) & 2,426 & 13.71 & \boldsymbol{\textcolor{tableorange}{ 4.57 }} & \boldsymbol{\textcolor{tableorange}{ 25.39 }} & \boldsymbol{\textcolor{tableorange}{ 6.32 }} \\ \hline
    Hyves (HY) & 1,402,673 & 3.96 & \boldsymbol{\textcolor{tablegreen}{ 2.98 }} & \boldsymbol{\textcolor{tablegreen}{ 2.23 }} & \boldsymbol{\textcolor{tablegreen}{ 1.99 }} \\ \hline
    LiveJournal (Lj) & 5,203,763 & 18.72 & \boldsymbol{\textcolor{tableblue}{ 3.86 }} & \boldsymbol{\textcolor{tableblue}{ 4.04 }} & \boldsymbol{\textcolor{tableblue}{ 3.15 }} \\ \hline
    Livemocha (LM) & 104,103 & 42.13 & \boldsymbol{\textcolor{tablered}{ 9.13 }} & \boldsymbol{\textcolor{tablered}{ $\infty$ }} & \boldsymbol{\textcolor{tablered}{ 2.39 }} \\ \hline
    Orkut (OR) & 3,072,441 & 76.28 & \boldsymbol{\textcolor{tableblue}{ 3.58 }} & \boldsymbol{\textcolor{tableblue}{ 2.65 }} & \boldsymbol{\textcolor{tableblue}{ 3.35 }} \\ \hline
    Power grid (UG) & 4,941 & 2.67 & \boldsymbol{\textcolor{tableorange}{ 6.62 }} & \boldsymbol{\textcolor{tableorange}{ 7.76 }} & \boldsymbol{\textcolor{tableorange}{ 9.2 }} \\ \hline
    Proteins (Mp) & 1,846 & 2.39 & \boldsymbol{\textcolor{tableblue}{ 3.09 }} & \boldsymbol{\textcolor{tableblue}{ 3.31 }} & \boldsymbol{\textcolor{tableblue}{ 3.87 }} \\ \hline
    Reactome (RC) & 6,229 & 46.93 & \boldsymbol{\textcolor{tablered}{ 4.86 }} & \boldsymbol{\textcolor{tablered}{ 34.33 }} & \boldsymbol{\textcolor{tablered}{ $\infty$ }} \\ \hline
    Roads CA (RO) & 1,965,206 & 2.82 & \boldsymbol{\textcolor{tablered}{ 18.86 }} & \boldsymbol{\textcolor{tablered}{ $\infty$ }} & \boldsymbol{\textcolor{tablered}{ $\infty$ }} \\ \hline
    Roads PA (RD) & 1,088,092 & 2.83 & \boldsymbol{\textcolor{tablered}{ 18.24 }} & \boldsymbol{\textcolor{tablered}{ $\infty$ }} & \boldsymbol{\textcolor{tablered}{ $\infty$ }} \\ \hline
    Roads TX (R1) & 1,379,917 & 2.79 & \boldsymbol{\textcolor{tablered}{ 21.83 }} & \boldsymbol{\textcolor{tablered}{ $\infty$ }} & \boldsymbol{\textcolor{tablered}{ $\infty$ }} \\ \hline
    Route views (AS) & 6,474 & 3.88 & \boldsymbol{\textcolor{tablegreen}{ 2.13 }} & \boldsymbol{\textcolor{tablegreen}{ 2.16 }} & \boldsymbol{\textcolor{tablegreen}{ 2.14 }} \\ \hline
    WordNet (WO) & 146,005 & 9.00 & \boldsymbol{\textcolor{tablegreen}{ 2.86 }} & \boldsymbol{\textcolor{tablegreen}{ 2.68 }} & \boldsymbol{\textcolor{tablegreen}{ 2.61 }} \\ \hline
    Youtube (CY) & 1,134,890 & 5.27 & \boldsymbol{\textcolor{tablegreen}{ 2.48 }} & \boldsymbol{\textcolor{tablegreen}{ 2.58 }} & \boldsymbol{\textcolor{tablegreen}{ 2.17 }} \\ \hline
    Human PPI (MV) & 3,023 & 4.07 & \boldsymbol{\textcolor{tableblue}{ 3.04 }} & \boldsymbol{\textcolor{tableblue}{ 3.4 }} & \boldsymbol{\textcolor{tableblue}{ 3.03 }} \\ \hline

\end{longtable*}
\hphantom{1pt}

\clearpage
\begin{longtable*}[ht!]{|c | c c | c c | c c c | c c c|}
    \caption{The tail exponent estimation results for the 49 real-world directed networks collected from the KONECT database~\cite{kunegis2013konect}. The style and notations are the same as in Table~\ref{tab:real-undir}. The estimators and coloring are applied to the in- and out-degree sequences separately. Since the operation of the estimators is undefined on zeros, the zero entries in the in- and out-degree sequences are removed, explaining the different lengths of the in- and out-degree sequences $n_{in}$ and $n_{out}$ for the same network, as well as the different values of the in and out average degrees $\bar{k}_{in}$ and $\bar{k}_{out}$.}\label{tab:real-dir}\\
    \hline
    {\bf Network Name} & $\boldsymbol{n_{in}}$ & $\boldsymbol{n_{out}}$ & $\boldsymbol{\bar{k}_{in}}$ & $\boldsymbol{\bar{k}_{out}}$ & $\boldsymbol{\hat{\gamma}_{in}^{Hill}}$ & $\boldsymbol{\hat{\gamma}_{in}^{Mom}}$ & $\boldsymbol{\hat{\gamma}_{in}^{Kern}}$ & $\boldsymbol{\hat{\gamma}_{out}^{Hill}}$ & $\boldsymbol{\hat{\gamma}_{out}^{Mom}}$ & $\boldsymbol{\hat{\gamma}_{out}^{Kern}}$ \\
    \hhline{===========}
    \endfirsthead
    \caption{(Continued).}\\
    \hline
    {\bf Network Name} & $\boldsymbol{n_{in}}$ & $\boldsymbol{n_{out}}$ & $\boldsymbol{\bar{k}_{in}}$ & $\boldsymbol{\bar{k}_{out}}$ & $\boldsymbol{\hat{\gamma}_{in}^{Hill}}$ & $\boldsymbol{\hat{\gamma}_{in}^{Mom}}$ & $\boldsymbol{\hat{\gamma}_{in}^{Kern}}$ & $\boldsymbol{\hat{\gamma}_{out}^{Hill}}$ & $\boldsymbol{\hat{\gamma}_{out}^{Mom}}$ & $\boldsymbol{\hat{\gamma}_{out}^{Kern}}$ \\
    \hhline{===========}
    \endhead

    Adolescent health (ME) & 2,441 & 2,313 & 5.31 & 5.61 & \boldsymbol{\textcolor{tablered}{ 6.68 }} & \boldsymbol{\textcolor{tablered}{ $\infty$ }} & \boldsymbol{\textcolor{tablered}{ $\infty$ }} & \boldsymbol{\textcolor{tablered}{ 501.0 }} & \boldsymbol{\textcolor{tablered}{ $\infty$ }} & \boldsymbol{\textcolor{tablered}{ $\infty$ }} \\ \hline
    Advogato (AD) & 4,422 & 4,009 & 10.66 & 11.76 & \boldsymbol{\textcolor{tableblue}{ 3.16 }} & \boldsymbol{\textcolor{tableblue}{ 3.35 }} & \boldsymbol{\textcolor{tableblue}{ 3.0 }} & \boldsymbol{\textcolor{tableblue}{ 3.29 }} & \boldsymbol{\textcolor{tableblue}{ 3.43 }} & \boldsymbol{\textcolor{tableblue}{ 4.37 }} \\ \hline
    Air traffic control (TC) & 1,210 & 1,083 & 2.16 & 2.41 & \boldsymbol{\textcolor{tablered}{ 6.0 }} & \boldsymbol{\textcolor{tablered}{ $\infty$ }} & \boldsymbol{\textcolor{tablered}{ 7.58 }} & \boldsymbol{\textcolor{tableorange}{ 4.28 }} & \boldsymbol{\textcolor{tableorange}{ 23.22 }} & \boldsymbol{\textcolor{tableorange}{ 8.3 }} \\ \hline
    Amazon (Am) & 403,312 & 402,439 & 8.4 & 8.42 & \boldsymbol{\textcolor{tableblue}{ 3.04 }} & \boldsymbol{\textcolor{tableblue}{ 3.58 }} & \boldsymbol{\textcolor{tableblue}{ 3.14 }} & \boldsymbol{\textcolor{tablered}{ $1.2\cdot10^7$ }} & \boldsymbol{\textcolor{tablered}{ $\infty$ }} & \boldsymbol{\textcolor{tablered}{ $\infty$ }} \\ \hline
    arXiv (PHc) & 28,226 & 32,153 & 14.93 & 13.11 & \boldsymbol{\textcolor{tableblue}{ 3.51 }} & \boldsymbol{\textcolor{tableblue}{ 4.16 }} & \boldsymbol{\textcolor{tableblue}{ 4.18 }} & \boldsymbol{\textcolor{tableorange}{ 3.79 }} & \boldsymbol{\textcolor{tableorange}{ 4.89 }} & \boldsymbol{\textcolor{tableorange}{ 6.46 }} \\ \hline
    arXiv (THc) & 23,176 & 25,055 & 15.22 & 14.08 & \boldsymbol{\textcolor{tableblue}{ 2.9 }} & \boldsymbol{\textcolor{tableblue}{ 2.98 }} & \boldsymbol{\textcolor{tableblue}{ 3.11 }} & \boldsymbol{\textcolor{tableorange}{ 4.09 }} & \boldsymbol{\textcolor{tableorange}{ 9.13 }} & \boldsymbol{\textcolor{tableorange}{ 3.84 }} \\ \hline
    Baidu (BAi) & 1,241,374 & 1,654,404 & 14.33 & 10.75 & \boldsymbol{\textcolor{tablegreen}{ 2.47 }} & \boldsymbol{\textcolor{tablegreen}{ 2.51 }} & \boldsymbol{\textcolor{tablegreen}{ 2.57 }} & \boldsymbol{\textcolor{tableorange}{ 4.58 }} & \boldsymbol{\textcolor{tableorange}{ 5.22 }} & \boldsymbol{\textcolor{tableorange}{ 5.57 }} \\ \hline
    Baidu (BAr) & 277,991 & 394,482 & 11.81 & 8.33 & \boldsymbol{\textcolor{tableblue}{ 2.72 }} & \boldsymbol{\textcolor{tableblue}{ 2.79 }} & \boldsymbol{\textcolor{tableblue}{ 3.26 }} & \boldsymbol{\textcolor{tableorange}{ 19.52 }} & \boldsymbol{\textcolor{tableorange}{ 2.73 }} & \boldsymbol{\textcolor{tableorange}{ 2.59 }} \\ \hline
    Berkley Stanford (BS) & 617,094 & 680,486 & 12.32 & 11.17 & \boldsymbol{\textcolor{tablegreen}{ 2.09 }} & \boldsymbol{\textcolor{tablegreen}{ 2.08 }} & \boldsymbol{\textcolor{tablegreen}{ 2.06 }} & \boldsymbol{\textcolor{tableorange}{ 334.33 }} & \boldsymbol{\textcolor{tableorange}{ 3.25 }} & \boldsymbol{\textcolor{tableorange}{ 3.36 }} \\ \hline
    Blogs (Mg) & 990 & 1,064 & 19.21 & 17.88 & \boldsymbol{\textcolor{tableorange}{ 3.27 }} & \boldsymbol{\textcolor{tableorange}{ 4.11 }} & \boldsymbol{\textcolor{tableorange}{ 6.1 }} & \boldsymbol{\textcolor{tableorange}{ 5.15 }} & \boldsymbol{\textcolor{tableorange}{ 14.7 }} & \boldsymbol{\textcolor{tableorange}{ 7.21 }} \\ \hline
    CiteSeer (CS) & 194,959 & 336,024 & 8.95 & 5.19 & \boldsymbol{\textcolor{tableblue}{ 3.46 }} & \boldsymbol{\textcolor{tableblue}{ 3.84 }} & \boldsymbol{\textcolor{tableblue}{ 3.31 }} & \boldsymbol{\textcolor{tableorange}{ 4.08 }} & \boldsymbol{\textcolor{tableorange}{ 5.55 }} & \boldsymbol{\textcolor{tableorange}{ 4.23 }} \\ \hline
    Cora (CC) & 13,879 & 21,201 & 6.59 & 4.32 & \boldsymbol{\textcolor{tableblue}{ 3.03 }} & \boldsymbol{\textcolor{tableblue}{ 3.2 }} & \boldsymbol{\textcolor{tableblue}{ 2.95 }} & \boldsymbol{\textcolor{tableorange}{ 5.1 }} & \boldsymbol{\textcolor{tableorange}{ 4.94 }} & \boldsymbol{\textcolor{tableorange}{ 6.68 }} \\ \hline
    DBLP (Pi) & 11,564 & 3,158 & 4.3 & 15.75 & \boldsymbol{\textcolor{tableblue}{ 3.1 }} & \boldsymbol{\textcolor{tableblue}{ 3.4 }} & \boldsymbol{\textcolor{tableblue}{ 2.98 }} & \boldsymbol{\textcolor{tableblue}{ 3.42 }} & \boldsymbol{\textcolor{tableblue}{ 2.81 }} & \boldsymbol{\textcolor{tableblue}{ 2.84 }} \\ \hline
    Edinburgh thesaurus (EA) & 22,675 & 8,210 & 13.75 & 37.98 & \boldsymbol{\textcolor{tableorange}{ 6.13 }} & \boldsymbol{\textcolor{tableorange}{ 3.94 }} & \boldsymbol{\textcolor{tableorange}{ 4.97 }} & \boldsymbol{\textcolor{tablered}{ 30.41 }} & \boldsymbol{\textcolor{tablered}{ $\infty$ }} & \boldsymbol{\textcolor{tablered}{ $\infty$ }} \\ \hline
    Ego Google Plus (GP) & 23,591 & 131 & 1.66 & 199.56 & \boldsymbol{\textcolor{tableorange}{ 5.83 }} & \boldsymbol{\textcolor{tableorange}{ 4.21 }} & \boldsymbol{\textcolor{tableorange}{ 4.83 }} & \boldsymbol{\textcolor{tablered}{ 3.91 }} & \boldsymbol{\textcolor{tablered}{ 11.75 }} & \boldsymbol{\textcolor{tablered}{ $\infty$ }} \\ \hline
    Ego Twitter (TL) & 22,964 & 938 & 1.44 & 35.29 & \boldsymbol{\textcolor{tableblue}{ 4.46 }} & \boldsymbol{\textcolor{tableblue}{ 3.4 }} & \boldsymbol{\textcolor{tableblue}{ 3.79 }} & \boldsymbol{\textcolor{tablered}{ 7.45 }} & \boldsymbol{\textcolor{tablered}{ $\infty$ }} & \boldsymbol{\textcolor{tablered}{ 48.62 }} \\ \hline
    Epinions (ES) & 51,957 & 60,341 & 9.79 & 8.43 & \boldsymbol{\textcolor{tableblue}{ 3.43 }} & \boldsymbol{\textcolor{tableblue}{ 4.38 }} & \boldsymbol{\textcolor{tableblue}{ 3.62 }} & \boldsymbol{\textcolor{tableblue}{ 3.54 }} & \boldsymbol{\textcolor{tableblue}{ 3.77 }} & \boldsymbol{\textcolor{tableblue}{ 3.98 }} \\ \hline
    FOLDOC (FO) & 13,309 & 13,356 & 9.03 & 9.0 & \boldsymbol{\textcolor{tableblue}{ 3.16 }} & \boldsymbol{\textcolor{tableblue}{ 2.67 }} & \boldsymbol{\textcolor{tableblue}{ 2.5 }} & \boldsymbol{\textcolor{tableorange}{ 5.42 }} & \boldsymbol{\textcolor{tableorange}{ 7.9 }} & \boldsymbol{\textcolor{tableorange}{ 10.01 }} \\ \hline
    Gnutella (GN) & 62,283 & 16,387 & 2.37 & 9.02 & \boldsymbol{\textcolor{tablered}{ 5.85 }} & \boldsymbol{\textcolor{tablered}{ $\infty$ }} & \boldsymbol{\textcolor{tablered}{ 3.67 }} & \boldsymbol{\textcolor{tableorange}{ 8.14 }} & \boldsymbol{\textcolor{tableorange}{ 3.15 }} & \boldsymbol{\textcolor{tableorange}{ 2.3 }} \\ \hline
    Google (GO) & 714,545 & 739,454 & 7.14 & 6.9 & \boldsymbol{\textcolor{tablegreen}{ 2.5 }} & \boldsymbol{\textcolor{tablegreen}{ 2.55 }} & \boldsymbol{\textcolor{tablegreen}{ 2.68 }} & \boldsymbol{\textcolor{tableorange}{ 5.1 }} & \boldsymbol{\textcolor{tableorange}{ 3.34 }} & \boldsymbol{\textcolor{tableorange}{ 2.97 }} \\ \hline
    Google (GC) & 15,762 & 12,447 & 10.81 & 13.68 & \boldsymbol{\textcolor{tablegreen}{ 2.16 }} & \boldsymbol{\textcolor{tablegreen}{ 2.17 }} & \boldsymbol{\textcolor{tablegreen}{ 2.04 }} & \boldsymbol{\textcolor{tableblue}{ 3.12 }} & \boldsymbol{\textcolor{tableblue}{ 2.71 }} & \boldsymbol{\textcolor{tableblue}{ 2.23 }} \\ \hline
    Hudong (HUi) & 798,202 & 1,725,741 & 18.39 & 8.51 & \boldsymbol{\textcolor{tableblue}{ 3.23 }} & \boldsymbol{\textcolor{tableblue}{ 3.65 }} & \boldsymbol{\textcolor{tableblue}{ 2.34 }} & \boldsymbol{\textcolor{tableblue}{ 3.72 }} & \boldsymbol{\textcolor{tableblue}{ 2.53 }} & \boldsymbol{\textcolor{tableblue}{ 2.63 }} \\ \hline
    Hudong (HUr) & 991,745 & 2,232,238 & 19.01 & 8.45 & \boldsymbol{\textcolor{tableblue}{ 3.29 }} & \boldsymbol{\textcolor{tableblue}{ 3.46 }} & \boldsymbol{\textcolor{tableblue}{ 3.07 }} & \boldsymbol{\textcolor{tableorange}{ 126.0 }} & \boldsymbol{\textcolor{tableorange}{ 3.62 }} & \boldsymbol{\textcolor{tableorange}{ 2.64 }} \\ \hline
    JDK dependencies (DJ) & 2,375 & 6,369 & 63.57 & 23.71 & \boldsymbol{\textcolor{tablegreen}{ 2.07 }} & \boldsymbol{\textcolor{tablegreen}{ 2.07 }} & \boldsymbol{\textcolor{tablegreen}{ 1.71 }} & \boldsymbol{\textcolor{tableorange}{ 10.62 }} & \boldsymbol{\textcolor{tableorange}{ 2.36 }} & \boldsymbol{\textcolor{tableorange}{ 2.52 }} \\ \hline
    JUNG/JAVAX dependencies (Dj) & 2,208 & 6,055 & 62.82 & 22.91 & \boldsymbol{\textcolor{tablegreen}{ 2.0 }} & \boldsymbol{\textcolor{tablegreen}{ 2.07 }} & \boldsymbol{\textcolor{tablegreen}{ 1.71 }} & \boldsymbol{\textcolor{tableorange}{ 10.62 }} & \boldsymbol{\textcolor{tableorange}{ 2.4 }} & \boldsymbol{\textcolor{tableorange}{ 2.54 }} \\ \hline
    Libimseti (LI) & 168,791 & 135,359 & 102.85 & 128.25 & \boldsymbol{\textcolor{tableblue}{ 4.21 }} & \boldsymbol{\textcolor{tableblue}{ 2.57 }} & \boldsymbol{\textcolor{tableblue}{ 2.67 }} & \boldsymbol{\textcolor{tablegreen}{ 2.54 }} & \boldsymbol{\textcolor{tablegreen}{ 2.49 }} & \boldsymbol{\textcolor{tablegreen}{ 2.56 }} \\ \hline
    Linux (LX) & 12,013 & 25,619 & 17.77 & 8.33 & \boldsymbol{\textcolor{tablegreen}{ 1.97 }} & \boldsymbol{\textcolor{tablegreen}{ 1.96 }} & \boldsymbol{\textcolor{tablegreen}{ 2.03 }} & \boldsymbol{\textcolor{tableorange}{ 4.42 }} & \boldsymbol{\textcolor{tableorange}{ 11.75 }} & \boldsymbol{\textcolor{tableorange}{ 5.39 }} \\ \hline
    Notre Dame (ND) & 325,729 & 136,934 & 4.51 & 10.73 & \boldsymbol{\textcolor{tablegreen}{ 2.05 }} & \boldsymbol{\textcolor{tablegreen}{ 2.57 }} & \boldsymbol{\textcolor{tablegreen}{ 1.99 }} & \boldsymbol{\textcolor{tablegreen}{ 2.55 }} & \boldsymbol{\textcolor{tablegreen}{ 2.73 }} & \boldsymbol{\textcolor{tablegreen}{ 2.26 }} \\ \hline
    Open flights (OF) & 3,418 & 3,409 & 19.8 & 19.85 & \boldsymbol{\textcolor{tableorange}{ 6.38 }} & \boldsymbol{\textcolor{tableorange}{ 3.63 }} & \boldsymbol{\textcolor{tableorange}{ 1.84 }} & \boldsymbol{\textcolor{tableorange}{ 6.21 }} & \boldsymbol{\textcolor{tableorange}{ 4.72 }} & \boldsymbol{\textcolor{tableorange}{ 1.84 }} \\ \hline
    Pokec (PL) & 1,519,452 & 1,432,693 & 20.15 & 21.37 & \boldsymbol{\textcolor{tableorange}{ 4.97 }} & \boldsymbol{\textcolor{tableorange}{ 6.95 }} & \boldsymbol{\textcolor{tableorange}{ 7.76 }} & \boldsymbol{\textcolor{tableorange}{ 4.1 }} & \boldsymbol{\textcolor{tableorange}{ 4.94 }} & \boldsymbol{\textcolor{tableorange}{ 6.52 }} \\ \hline
    Slashdot Zoo (SZ) & 65,220 & 45,598 & 7.9 & 11.3 & \boldsymbol{\textcolor{tableblue}{ 2.99 }} & \boldsymbol{\textcolor{tableblue}{ 3.34 }} & \boldsymbol{\textcolor{tableblue}{ 4.02 }} & \boldsymbol{\textcolor{tablered}{ 29.57 }} & \boldsymbol{\textcolor{tablered}{ $\infty$ }} & \boldsymbol{\textcolor{tablered}{ $\infty$ }} \\ \hline
    Standford (SF) & 261,588 & 281,731 & 8.84 & 8.21 & \boldsymbol{\textcolor{tablegreen}{ 2.15 }} & \boldsymbol{\textcolor{tablegreen}{ 2.15 }} & \boldsymbol{\textcolor{tablegreen}{ 2.15 }} & \boldsymbol{\textcolor{tableorange}{ 63.5 }} & \boldsymbol{\textcolor{tableorange}{ 5.55 }} & \boldsymbol{\textcolor{tableorange}{ 3.09 }} \\ \hline
    TREC WT10g (WT) & 1,295,841 & 1,532,051 & 6.22 & 5.26 & \boldsymbol{\textcolor{tableblue}{ 2.98 }} & \boldsymbol{\textcolor{tableblue}{ 3.67 }} & \boldsymbol{\textcolor{tableblue}{ 2.53 }} & \boldsymbol{\textcolor{tablegreen}{ 2.19 }} & \boldsymbol{\textcolor{tablegreen}{ 2.17 }} & \boldsymbol{\textcolor{tablegreen}{ 2.27 }} \\ \hline
    Twitter ICWSM (Ws) & 465,016 & 2,502 & 1.8 & 333.65 & \boldsymbol{\textcolor{tablegreen}{ 2.54 }} & \boldsymbol{\textcolor{tablegreen}{ 2.57 }} & \boldsymbol{\textcolor{tablegreen}{ 2.64 }} & \boldsymbol{\textcolor{tablered}{ $1.6\cdot10^4$ }} & \boldsymbol{\textcolor{tablered}{ $\infty$ }} & \boldsymbol{\textcolor{tablered}{ $\infty$ }} \\ \hline
    Twitter MPI (TF) & 49,395,940 & 43,983,853 & 39.75 & 44.64 & \boldsymbol{\textcolor{tablegreen}{ 2.02 }} & \boldsymbol{\textcolor{tablegreen}{ 2.31 }} & \boldsymbol{\textcolor{tablegreen}{ 2.01 }} & \boldsymbol{\textcolor{tablegreen}{ 1.91 }} & \boldsymbol{\textcolor{tablegreen}{ 1.99 }} & \boldsymbol{\textcolor{tablegreen}{ 2.34 }} \\ \hline
    Twitter WWW (TW) & 35,689,148 & 40,103,281 & 41.14 & 36.61 & \boldsymbol{\textcolor{tablegreen}{ 1.93 }} & \boldsymbol{\textcolor{tablegreen}{ 2.01 }} & \boldsymbol{\textcolor{tablegreen}{ 2.43 }} & \boldsymbol{\textcolor{tablegreen}{ 2.04 }} & \boldsymbol{\textcolor{tablegreen}{ 2.03 }} & \boldsymbol{\textcolor{tablegreen}{ 1.98 }} \\ \hline
    US airports (AF) & 1,504 & 1,478 & 18.77 & 19.1 & \boldsymbol{\textcolor{tablered}{ 13.2 }} & \boldsymbol{\textcolor{tablered}{ $\infty$ }} & \boldsymbol{\textcolor{tablered}{ 2.46 }} & \boldsymbol{\textcolor{tablered}{ 14.89 }} & \boldsymbol{\textcolor{tablered}{ $\infty$ }} & \boldsymbol{\textcolor{tablered}{ 2.26 }} \\ \hline
    US patents (PC) & 3,258,983 & 2,089,345 & 5.07 & 7.91 & \boldsymbol{\textcolor{tableblue}{ 4.28 }} & \boldsymbol{\textcolor{tableblue}{ 4.6 }} & \boldsymbol{\textcolor{tableblue}{ 4.52 }} & \boldsymbol{\textcolor{tableblue}{ 3.43 }} & \boldsymbol{\textcolor{tableblue}{ 3.6 }} & \boldsymbol{\textcolor{tableblue}{ 3.82 }} \\ \hline
    Wikipedia links DE (Wde) & 2,262,745 & 3,221,527 & 36.06 & 25.33 & \boldsymbol{\textcolor{tablegreen}{ 2.42 }} & \boldsymbol{\textcolor{tablegreen}{ 2.4 }} & \boldsymbol{\textcolor{tablegreen}{ 2.15 }} & \boldsymbol{\textcolor{tableorange}{ 5.17 }} & \boldsymbol{\textcolor{tableorange}{ 4.25 }} & \boldsymbol{\textcolor{tableorange}{ 3.22 }} \\ \hline
    Wikipedia links EN (Wen) & 7,549,312 & 12,114,964 & 50.08 & 31.21 & \boldsymbol{\textcolor{tableblue}{ 2.78 }} & \boldsymbol{\textcolor{tableblue}{ 2.79 }} & \boldsymbol{\textcolor{tableblue}{ 3.03 }} & \boldsymbol{\textcolor{tableorange}{ 4.94 }} & \boldsymbol{\textcolor{tableorange}{ 8.58 }} & \boldsymbol{\textcolor{tableorange}{ 5.9 }} \\ \hline
    Wikipedia links FR (Wfr) & 2,127,693 & 2,993,436 & 48.11 & 34.2 & \boldsymbol{\textcolor{tablegreen}{ 2.4 }} & \boldsymbol{\textcolor{tablegreen}{ 2.37 }} & \boldsymbol{\textcolor{tablegreen}{ 2.54 }} & \boldsymbol{\textcolor{tableorange}{ 4.0 }} & \boldsymbol{\textcolor{tableorange}{ 5.24 }} & \boldsymbol{\textcolor{tableorange}{ 3.61 }} \\ \hline
    Wikipedia links IT (Wit) & 1,488,860 & 1,855,986 & 61.47 & 49.31 & \boldsymbol{\textcolor{tableblue}{ 2.88 }} & \boldsymbol{\textcolor{tableblue}{ 2.76 }} & \boldsymbol{\textcolor{tableblue}{ 3.17 }} & \boldsymbol{\textcolor{tableorange}{ 6.81 }} & \boldsymbol{\textcolor{tableorange}{ 5.46 }} & \boldsymbol{\textcolor{tableorange}{ 2.93 }} \\ \hline
    Wikipedia links JA (Wja) & 1,253,659 & 1,609,718 & 56.67 & 44.14 & \boldsymbol{\textcolor{tablegreen}{ 2.48 }} & \boldsymbol{\textcolor{tablegreen}{ 2.47 }} & \boldsymbol{\textcolor{tablegreen}{ 2.68 }} & \boldsymbol{\textcolor{tableorange}{ 5.37 }} & \boldsymbol{\textcolor{tableorange}{ 3.73 }} & \boldsymbol{\textcolor{tableorange}{ 4.27 }} \\ \hline
    Wikipedia links PL (Wpl) & 1,196,546 & 1,528,795 & 48.04 & 37.6 & \boldsymbol{\textcolor{tablegreen}{ 2.59 }} & \boldsymbol{\textcolor{tablegreen}{ 2.59 }} & \boldsymbol{\textcolor{tablegreen}{ 2.95 }} & \boldsymbol{\textcolor{tableblue}{ 3.82 }} & \boldsymbol{\textcolor{tableblue}{ 4.46 }} & \boldsymbol{\textcolor{tableblue}{ 4.12 }} \\ \hline
    Wikipedia links PT (Wpt) & 1,137,929 & 1,591,426 & 43.07 & 30.8 & \boldsymbol{\textcolor{tablegreen}{ 2.58 }} & \boldsymbol{\textcolor{tablegreen}{ 2.51 }} & \boldsymbol{\textcolor{tablegreen}{ 2.72 }} & \boldsymbol{\textcolor{tableorange}{ 14.7 }} & \boldsymbol{\textcolor{tableorange}{ 4.03 }} & \boldsymbol{\textcolor{tableorange}{ 4.5 }} \\ \hline
    Wikipedia links RU (Wru) & 1,834,424 & 2,852,544 & 44.72 & 28.76 & \boldsymbol{\textcolor{tablegreen}{ 2.48 }} & \boldsymbol{\textcolor{tablegreen}{ 2.46 }} & \boldsymbol{\textcolor{tablegreen}{ 2.62 }} & \boldsymbol{\textcolor{tableorange}{ 7.8 }} & \boldsymbol{\textcolor{tableorange}{ 251.0 }} & \boldsymbol{\textcolor{tableorange}{ 4.89 }} \\ \hline
    Yahoo ads (YD) & 194,317 & 653,260 & 15.09 & 4.49 & \boldsymbol{\textcolor{tablegreen}{ 2.2 }} & \boldsymbol{\textcolor{tablegreen}{ 2.18 }} & \boldsymbol{\textcolor{tablegreen}{ 2.23 }} & \boldsymbol{\textcolor{tablered}{ 4.02 }} & \boldsymbol{\textcolor{tablered}{ 3.98 }} & \boldsymbol{\textcolor{tablered}{ $\infty$ }} \\ \hline
    Human PPI (MF) & 2,033 & 338 & 3.17 & 19.09 & \boldsymbol{\textcolor{tablered}{ 27.32 }} & \boldsymbol{\textcolor{tablered}{ $\infty$ }} & \boldsymbol{\textcolor{tablered}{ $\infty$ }} & \boldsymbol{\textcolor{tablegreen}{ 2.4 }} & \boldsymbol{\textcolor{tablegreen}{ 2.86 }} & \boldsymbol{\textcolor{tablegreen}{ 2.04 }} \\ \hline
    Human PPI (MS) & 1,698 & 1,597 & 3.63 & 3.86 & \boldsymbol{\textcolor{tableblue}{ 4.62 }} & \boldsymbol{\textcolor{tableblue}{ 2.44 }} & \boldsymbol{\textcolor{tableblue}{ 2.57 }} & \boldsymbol{\textcolor{tableorange}{ 6.75 }} & \boldsymbol{\textcolor{tableorange}{ 2.42 }} & \boldsymbol{\textcolor{tableorange}{ 2.63 }} \\ \hline

\end{longtable*}

\clearpage
\begin{longtable*}[ht!]{|c | c c | c c | c c c | c c c|}
    \caption{The tail exponent estimation results for the 31 real-world bipartite networks collected from the KONECT database~\cite{kunegis2013konect}. The style and notations are the same as in Table~\ref{tab:real-dir}. The estimators and coloring are applied to the degree sequences of nodes of types 1 and 2 (domains 1 and 2) separately.}\label{tab:real-bip}\\
    \hline
    {\bf Network Name} & $\boldsymbol{n_{d_1}}$ & $\boldsymbol{n_{d_2}}$ & $\boldsymbol{\bar{k}_{d_1}}$ & $\boldsymbol{\bar{k}_{d_2}}$ & $\boldsymbol{\hat{\gamma}_{d_1}^{Hill}}$ & $\boldsymbol{\hat{\gamma}_{d_1}^{Mom}}$ & $\boldsymbol{\hat{\gamma}_{d_1}^{Kern}}$ & $\boldsymbol{\hat{\gamma}_{d_2}^{Hill}}$ & $\boldsymbol{\hat{\gamma}_{d_2}^{Mom}}$ & $\boldsymbol{\hat{\gamma}_{d_2}^{Kern}}$ \\
    \hhline{===========}
    \endfirsthead
    \caption{(Continued).}\\
    \hline
    {\bf Network Name} & $\boldsymbol{n_{d_1}}$ & $\boldsymbol{n_{d_2}}$ & $\boldsymbol{\bar{k}_{d_1}}$ & $\boldsymbol{\bar{k}_{d_2}}$ & $\boldsymbol{\hat{\gamma}_{d_1}^{Hill}}$ & $\boldsymbol{\hat{\gamma}_{d_1}^{Mom}}$ & $\boldsymbol{\hat{\gamma}_{d_1}^{Kern}}$ & $\boldsymbol{\hat{\gamma}_{d_2}^{Hill}}$ & $\boldsymbol{\hat{\gamma}_{d_2}^{Mom}}$ & $\boldsymbol{\hat{\gamma}_{d_2}^{Kern}}$ \\
    \hhline{===========}
    \endhead

    \normalsize{ Movies/Actors (AM) } & 383,640 & 127,823 & 3.83 & 11.5 & \boldsymbol{\textcolor{tableorange}{ 4.92 }} & \boldsymbol{\textcolor{tableorange}{ 7.67 }} & \boldsymbol{\textcolor{tableorange}{ 5.24 }} & \boldsymbol{\textcolor{tableorange}{ 6.13 }} & \boldsymbol{\textcolor{tableorange}{ 11.99 }} & \boldsymbol{\textcolor{tableorange}{ 6.13 }} \\ \hline
    \normalsize{ arXiv (AC) } & 22,015 & 16,726 & 2.66 & 3.5 & \boldsymbol{\textcolor{tableorange}{ 9.33 }} & \boldsymbol{\textcolor{tableorange}{ 12.9 }} & \boldsymbol{\textcolor{tableorange}{ 13.99 }} & \boldsymbol{\textcolor{tableorange}{ 5.33 }} & \boldsymbol{\textcolor{tableorange}{ 6.05 }} & \boldsymbol{\textcolor{tableorange}{ 6.75 }} \\ \hline
    \normalsize{ Book Crossing implicit (BX) } & 340,523 & 105,278 & 3.38 & 10.92 & \boldsymbol{\textcolor{tableblue}{ 3.54 }} & \boldsymbol{\textcolor{tableblue}{ 2.26 }} & \boldsymbol{\textcolor{tableblue}{ 2.31 }} & \boldsymbol{\textcolor{tableblue}{ 3.08 }} & \boldsymbol{\textcolor{tableblue}{ 3.37 }} & \boldsymbol{\textcolor{tableblue}{ 1.88 }} \\ \hline
    \normalsize{ Book Crossing ratings (Bx) } & 185,955 & 77,802 & 2.33 & 5.57 & \boldsymbol{\textcolor{tablegreen}{ 2.35 }} & \boldsymbol{\textcolor{tablegreen}{ 2.39 }} & \boldsymbol{\textcolor{tablegreen}{ 2.43 }} & \boldsymbol{\textcolor{tablegreen}{ 2.83 }} & \boldsymbol{\textcolor{tablegreen}{ 2.82 }} & \boldsymbol{\textcolor{tablegreen}{ 2.86 }} \\ \hline
    \normalsize{ Countries DBPedia (CN) } & 2,302 & 590,112 & 276.77 & 1.08 & \boldsymbol{\textcolor{tablegreen}{ 1.43 }} & \boldsymbol{\textcolor{tablegreen}{ 1.44 }} & \boldsymbol{\textcolor{tablegreen}{ 1.93 }} & \boldsymbol{\textcolor{tablered}{ 9.77 }} & \boldsymbol{\textcolor{tablered}{ 3.03 }} & \boldsymbol{\textcolor{tablered}{ $\infty$ }} \\ \hline
    \normalsize{ DBLP (PA) } & 4,000,150 & 1,425,813 & 6.07 & 2.16 & \boldsymbol{\textcolor{tableorange}{ 5.22 }} & \boldsymbol{\textcolor{tableorange}{ 4.3 }} & \boldsymbol{\textcolor{tableorange}{ 6.75 }} & \boldsymbol{\textcolor{tablegreen}{ 2.22 }} & \boldsymbol{\textcolor{tablegreen}{ 2.23 }} & \boldsymbol{\textcolor{tablegreen}{ 2.14 }} \\ \hline
    \normalsize{ Discogs labels/artists (Dl) } & 270,771 & 1,754,823 & 53.24 & 8.21 & \boldsymbol{\textcolor{tablegreen}{ 2.08 }} & \boldsymbol{\textcolor{tablegreen}{ 2.15 }} & \boldsymbol{\textcolor{tablegreen}{ 1.93 }} & \boldsymbol{\textcolor{tableblue}{ 3.12 }} & \boldsymbol{\textcolor{tableblue}{ 3.11 }} & \boldsymbol{\textcolor{tableblue}{ 3.29 }} \\ \hline
    \normalsize{ Discogs genres/artists (Da) } & 15 & 1,754,823 & $1.3\cdot10^6$ & 10.85 & \boldsymbol{\textcolor{tableorange}{ 7.67 }} & \boldsymbol{\textcolor{tableorange}{ 2.72 }} & \boldsymbol{\textcolor{tableorange}{ 1.0 }} & \boldsymbol{\textcolor{tableblue}{ 3.12 }} & \boldsymbol{\textcolor{tableblue}{ 3.07 }} & \boldsymbol{\textcolor{tableblue}{ 3.21 }} \\ \hline
    \normalsize{ Discogs genres/labels (Dr) } & 15 & 270,771 & $2.8\cdot10^5$ & 15.32 & \boldsymbol{\textcolor{tableorange}{ 1.9 }} & \boldsymbol{\textcolor{tableorange}{ 2.56 }} & \boldsymbol{\textcolor{tableorange}{ 7.76 }} & \boldsymbol{\textcolor{tablegreen}{ 2.2 }} & \boldsymbol{\textcolor{tablegreen}{ 2.2 }} & \boldsymbol{\textcolor{tablegreen}{ 2.17 }} \\ \hline
    \normalsize{ Flickr (FG) } & 103,631 & 395,979 & 82.46 & 21.58 & \boldsymbol{\textcolor{tableblue}{ 2.66 }} & \boldsymbol{\textcolor{tableblue}{ 3.7 }} & \boldsymbol{\textcolor{tableblue}{ 2.85 }} & \boldsymbol{\textcolor{tableorange}{ 5.22 }} & \boldsymbol{\textcolor{tableorange}{ 6.52 }} & \boldsymbol{\textcolor{tableorange}{ 6.29 }} \\ \hline
    \normalsize{ Genres DBPedia (GE) } & 7,783 & 258,769 & 59.55 & 1.79 & \boldsymbol{\textcolor{tablered}{ 2.16 }} & \boldsymbol{\textcolor{tablered}{ 2.57 }} & \boldsymbol{\textcolor{tablered}{ $\infty$ }} & \boldsymbol{\textcolor{tablered}{ 5.61 }} & \boldsymbol{\textcolor{tablered}{ 12.49 }} & \boldsymbol{\textcolor{tablered}{ $\infty$ }} \\ \hline
    \normalsize{ Github (GH) } & 120,867 & 56,519 & 3.64 & 7.79 & \boldsymbol{\textcolor{tablegreen}{ 2.08 }} & \boldsymbol{\textcolor{tablegreen}{ 2.07 }} & \boldsymbol{\textcolor{tablegreen}{ 2.07 }} & \boldsymbol{\textcolor{tableblue}{ 2.77 }} & \boldsymbol{\textcolor{tableblue}{ 4.07 }} & \boldsymbol{\textcolor{tableblue}{ 3.31 }} \\ \hline
    \normalsize{ Jester100 (J1) } & 100 & 73,421 & $4.1\cdot10^4$ & 56.34 & \boldsymbol{\textcolor{tablered}{ $4.3\cdot10^4$ }} & \boldsymbol{\textcolor{tablered}{ $\infty$ }} & \boldsymbol{\textcolor{tablered}{ $\infty$ }} & \boldsymbol{\textcolor{tablered}{ $5.1\cdot10^5$ }} & \boldsymbol{\textcolor{tablered}{ $\infty$ }} & \boldsymbol{\textcolor{tablered}{ $\infty$ }} \\ \hline
    \normalsize{ Jester150 (J2) } & 140 & 50,692 & $1.2\cdot10^4$ & 34.1 & \boldsymbol{\textcolor{tableorange}{ 3.26 }} & \boldsymbol{\textcolor{tableorange}{ 3.32 }} & \boldsymbol{\textcolor{tableorange}{ 5.46 }} & \boldsymbol{\textcolor{tableorange}{ 251.0 }} & \boldsymbol{\textcolor{tableorange}{ 3.29 }} & \boldsymbol{\textcolor{tableorange}{ 2.73 }} \\ \hline
    \normalsize{ LiveJournal (LG) } & 7,489,073 & 3,201,203 & 15.0 & 35.08 & \boldsymbol{\textcolor{tablegreen}{ 1.78 }} & \boldsymbol{\textcolor{tablegreen}{ 1.8 }} & \boldsymbol{\textcolor{tablegreen}{ 1.8 }} & \boldsymbol{\textcolor{tablered}{ $1.0\cdot10^3$ }} & \boldsymbol{\textcolor{tablered}{ $\infty$ }} & \boldsymbol{\textcolor{tablered}{ 4.32 }} \\ \hline
    \normalsize{ Locations DBPedia (LO) } & 53,407 & 172,079 & 5.5 & 1.71 & \boldsymbol{\textcolor{tablegreen}{ 2.02 }} & \boldsymbol{\textcolor{tablegreen}{ 2.02 }} & \boldsymbol{\textcolor{tablegreen}{ 2.05 }} & \boldsymbol{\textcolor{tableorange}{ 6.52 }} & \boldsymbol{\textcolor{tableorange}{ 3.9 }} & \boldsymbol{\textcolor{tableorange}{ 38.04 }} \\ \hline
    \normalsize{ Movies DBPedia (ST) } & 81,085 & 76,098 & 3.47 & 3.7 & \boldsymbol{\textcolor{tableorange}{ 3.94 }} & \boldsymbol{\textcolor{tableorange}{ 4.85 }} & \boldsymbol{\textcolor{tableorange}{ 5.98 }} & \boldsymbol{\textcolor{tableorange}{ 5.52 }} & \boldsymbol{\textcolor{tableorange}{ 13.35 }} & \boldsymbol{\textcolor{tableorange}{ 8.69 }} \\ \hline
    \normalsize{ Occupations DBPedia (OC) } & 101,730 & 127,571 & 2.47 & 1.97 & \boldsymbol{\textcolor{tablegreen}{ 1.73 }} & \boldsymbol{\textcolor{tablegreen}{ 1.74 }} & \boldsymbol{\textcolor{tablegreen}{ 1.73 }} & \boldsymbol{\textcolor{tableorange}{ 6.56 }} & \boldsymbol{\textcolor{tableorange}{ 19.18 }} & \boldsymbol{\textcolor{tableorange}{ $1.0\cdot10^3$ }} \\ \hline
    \normalsize{ Orkut (OG) } & 8,730,857 & 2,783,196 & 37.46 & 117.5 & \boldsymbol{\textcolor{tablegreen}{ 1.88 }} & \boldsymbol{\textcolor{tablegreen}{ 1.89 }} & \boldsymbol{\textcolor{tablegreen}{ 1.89 }} & \boldsymbol{\textcolor{tableorange}{ 53.63 }} & \boldsymbol{\textcolor{tableorange}{ 2.4 }} & \boldsymbol{\textcolor{tableorange}{ 2.63 }} \\ \hline
    \normalsize{ Producers DBPedia (PR) } & 138,839 & 48,833 & 1.49 & 4.24 & \boldsymbol{\textcolor{tableblue}{ 3.69 }} & \boldsymbol{\textcolor{tableblue}{ 3.89 }} & \boldsymbol{\textcolor{tableblue}{ 4.13 }} & \boldsymbol{\textcolor{tableblue}{ 4.09 }} & \boldsymbol{\textcolor{tableblue}{ 4.58 }} & \boldsymbol{\textcolor{tableblue}{ 2.12 }} \\ \hline
    \normalsize{ Labels DBPedia (RL) } & 18,421 & 168,268 & 12.66 & 1.39 & \boldsymbol{\textcolor{tablegreen}{ 2.05 }} & \boldsymbol{\textcolor{tablegreen}{ 2.08 }} & \boldsymbol{\textcolor{tablegreen}{ 2.09 }} & \boldsymbol{\textcolor{tableorange}{ 5.95 }} & \boldsymbol{\textcolor{tableorange}{ 5.18 }} & \boldsymbol{\textcolor{tableorange}{ 6.05 }} \\ \hline
    \normalsize{ Reuters (RE) } & 283,911 & 781,265 & 213.34 & 77.53 & \boldsymbol{\textcolor{tableorange}{ 6.38 }} & \boldsymbol{\textcolor{tableorange}{ 1.55 }} & \boldsymbol{\textcolor{tableorange}{ 1.56 }} & \boldsymbol{\textcolor{tablered}{ 6.46 }} & \boldsymbol{\textcolor{tablered}{ $\infty$ }} & \boldsymbol{\textcolor{tablered}{ $\infty$ }} \\ \hline
    \normalsize{ TREC (TR) } & 1,173,225 & 551,787 & 71.28 & 151.56 & \boldsymbol{\textcolor{tablegreen}{ 1.61 }} & \boldsymbol{\textcolor{tablegreen}{ 1.63 }} & \boldsymbol{\textcolor{tablegreen}{ 1.6 }} & \boldsymbol{\textcolor{tableorange}{ 3.93 }} & \boldsymbol{\textcolor{tableorange}{ 8.87 }} & \boldsymbol{\textcolor{tableorange}{ 2.94 }} \\ \hline
    \normalsize{ TV tropes DBPedia (DBT) } & 87,678 & 64,415 & 36.86 & 50.18 & \boldsymbol{\textcolor{tableblue}{ 3.18 }} & \boldsymbol{\textcolor{tableblue}{ 3.73 }} & \boldsymbol{\textcolor{tableblue}{ 3.37 }} & \boldsymbol{\textcolor{tableblue}{ 3.34 }} & \boldsymbol{\textcolor{tableblue}{ 3.07 }} & \boldsymbol{\textcolor{tableblue}{ 2.97 }} \\ \hline
    \normalsize{ Teams DBPedia (TM) } & 34,461 & 901,130 & 39.65 & 1.52 & \boldsymbol{\textcolor{tableorange}{ 8.46 }} & \boldsymbol{\textcolor{tableorange}{ 12.76 }} & \boldsymbol{\textcolor{tableorange}{ 2.2 }} & \boldsymbol{\textcolor{tablered}{ 21.83 }} & \boldsymbol{\textcolor{tablered}{ $\infty$ }} & \boldsymbol{\textcolor{tablered}{ $\infty$ }} \\ \hline
    \normalsize{ vi.sualize.us images/tags (Vti) } & 495,402 & 82,035 & 4.64 & 28.02 & \boldsymbol{\textcolor{tableblue}{ 3.15 }} & \boldsymbol{\textcolor{tableblue}{ 3.34 }} & \boldsymbol{\textcolor{tableblue}{ 2.97 }} & \boldsymbol{\textcolor{tablegreen}{ 1.81 }} & \boldsymbol{\textcolor{tablegreen}{ 1.85 }} & \boldsymbol{\textcolor{tablegreen}{ 1.86 }} \\ \hline
    \normalsize{ vi.sualize.us tags/users (Vut) } & 82,035 & 17,122 & 28.02 & 134.26 & \boldsymbol{\textcolor{tablegreen}{ 1.81 }} & \boldsymbol{\textcolor{tablegreen}{ 1.85 }} & \boldsymbol{\textcolor{tablegreen}{ 1.85 }} & \boldsymbol{\textcolor{tablegreen}{ 2.54 }} & \boldsymbol{\textcolor{tablegreen}{ 2.57 }} & \boldsymbol{\textcolor{tablegreen}{ 2.13 }} \\ \hline
    \normalsize{ vi.sualize.us images/users (Vui) } & 495,402 & 17,122 & 4.64 & 134.26 & \boldsymbol{\textcolor{tableblue}{ 3.15 }} & \boldsymbol{\textcolor{tableblue}{ 3.31 }} & \boldsymbol{\textcolor{tableblue}{ 2.98 }} & \boldsymbol{\textcolor{tablegreen}{ 2.49 }} & \boldsymbol{\textcolor{tablegreen}{ 2.58 }} & \boldsymbol{\textcolor{tablegreen}{ 2.51 }} \\ \hline
    \normalsize{ Web trackers (WT) } & 12,756,244 & 27,665,730 & 11.02 & 5.08 & \boldsymbol{\textcolor{tablegreen}{ 2.03 }} & \boldsymbol{\textcolor{tablegreen}{ 2.03 }} & \boldsymbol{\textcolor{tablegreen}{ 2.04 }} & \boldsymbol{\textcolor{tablegreen}{ 2.62 }} & \boldsymbol{\textcolor{tablegreen}{ 2.1 }} & \boldsymbol{\textcolor{tablegreen}{ 2.05 }} \\ \hline
    \normalsize{ Writers DBPedia (WR) } & 46,213 & 89,355 & 3.12 & 1.62 & \boldsymbol{\textcolor{tableblue}{ 3.77 }} & \boldsymbol{\textcolor{tableblue}{ 4.1 }} & \boldsymbol{\textcolor{tableblue}{ 4.06 }} & \boldsymbol{\textcolor{tableorange}{ 5.88 }} & \boldsymbol{\textcolor{tableorange}{ 7.62 }} & \boldsymbol{\textcolor{tableorange}{ 15.08 }} \\ \hline
    \normalsize{ Youtube (YG) } & 30,087 & 94,238 & 9.75 & 3.11 & \boldsymbol{\textcolor{tablegreen}{ 2.31 }} & \boldsymbol{\textcolor{tablegreen}{ 2.36 }} & \boldsymbol{\textcolor{tablegreen}{ 2.45 }} & \boldsymbol{\textcolor{tablegreen}{ 2.79 }} & \boldsymbol{\textcolor{tablegreen}{ 2.87 }} & \boldsymbol{\textcolor{tablegreen}{ 2.56 }} \\ \hline

\end{longtable*}

\hphantom{1pt}

\clearpage
\balance

\onecolumngrid
\twocolumngrid

\end{document}